\documentclass[longauth]{aa}

\usepackage{graphicx}
\usepackage{txfonts}
\usepackage{comment} 
\usepackage[normalem]{ulem}
\usepackage{amsmath}
\usepackage[]{hyperref}
\usepackage{graphicx}
\usepackage{caption}
\usepackage{subcaption}

\usepackage{natbib}
\defcitealias{Wong2020}{W20}
\defcitealias{Mansfield2020}{M20}
\defcitealias{Ahlers2020}{A20}

\hypersetup{
    colorlinks = true,
    linkcolor = {blue}, 
    citecolor = {blue},
    urlcolor = {blue}
}

\usepackage{natbib}
\usepackage{xcolor}
\usepackage{float}
\usepackage{xspace}

\newcommand{\Spitzer}{\textsl{Spitzer}\xspace}

\newcommand{\lmax}{\ensuremath{\ell_\mathrm{max}}}

\begin{document}

\title{The stable climate of KELT-9b\thanks{The 
photometric time series data are only available in electronic form
at the CDS via anonymous ftp to cdsarc.u-strasbg.fr (130.79.128.5) or via 
\url{http://cdsweb.u-strasbg.fr/cgi-bin}. The CHEOPS program ID is CH\_PR100036.}}

\author{
K. D. Jones\inst{1} $^{\href{https://orcid.org/0000-0002-2316-6850}{\includegraphics[scale=0.5]{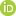}}}$, 
B. M. Morris\inst{1}$^{\href{https://orcid.org/0000-0003-2528-3409}{\includegraphics[scale=0.5]{figures/orcid.jpg}}}$, 
B.-O. Demory\inst{1} $^{\href{https://orcid.org/0000-0002-9355-5165}{\includegraphics[scale=0.5]{figures/orcid.jpg}}}$,
K. Heng\inst{1,2} $^{\href{https://orcid.org/0000-0003-1907-5910}{\includegraphics[scale=0.5]{figures/orcid.jpg}}}$,
M. J. Hooton\inst{3,4} $^{\href{https://orcid.org/0000-0003-0030-332X}{\includegraphics[scale=0.5]{figures/orcid.jpg}}}$,
N. Billot\inst{5} $^{\href{https://orcid.org/0000-0003-3429-3836}{\includegraphics[scale=0.5]{figures/orcid.jpg}}}$,
D. Ehrenreich\inst{5} $^{\href{https://orcid.org/0000-0001-9704-5405}{\includegraphics[scale=0.5]{figures/orcid.jpg}}}$,
S. Hoyer\inst{6} $^{\href{https://orcid.org/0000-0003-3477-2466}{\includegraphics[scale=0.5]{figures/orcid.jpg}}}$,
A. E. Simon\inst{4} $^{\href{https://orcid.org/0000-0001-9773-2600}{\includegraphics[scale=0.5]{figures/orcid.jpg}}}$,
M. Lendl\inst{5} $^{\href{https://orcid.org/0000-0001-9699-1459}{\includegraphics[scale=0.5]{figures/orcid.jpg}}}$,
O. D. S. Demangeon\inst{7,8} $^{\href{https://orcid.org/0000-0001-7918-0355}{\includegraphics[scale=0.5]{figures/orcid.jpg}}}$,
S. G. Sousa\inst{7} $^{\href{https://orcid.org/0000-0001-9047-2965}{\includegraphics[scale=0.5]{figures/orcid.jpg}}}$,
A. Bonfanti\inst{9} $^{\href{https://orcid.org/0000-0002-1916-5935}{\includegraphics[scale=0.5]{figures/orcid.jpg}}}$,
T. G. Wilson\inst{10} $^{\href{https://orcid.org/0000-0001-8749-1962}{\includegraphics[scale=0.5]{figures/orcid.jpg}}}$,
S. Salmon\inst{5} $^{\href{https://orcid.org/0000-0002-1714-3513}{\includegraphics[scale=0.5]{figures/orcid.jpg}}}$,
Sz. Csizmadia\inst{11} $^{\href{https://orcid.org/0000-0001-6803-9698}{\includegraphics[scale=0.5]{figures/orcid.jpg}}}$,
H. Parviainen\inst{12} $^{\href{https://orcid.org/0000-0001-5519-1391}{\includegraphics[scale=0.5]{figures/orcid.jpg}}}$,
G. Bruno\inst{13} $^{\href{https://orcid.org/0000-0002-3288-0802}{\includegraphics[scale=0.5]{figures/orcid.jpg}}}$,
Y. Alibert\inst{4} $^{\href{https://orcid.org/0000-0002-4644-8818}{\includegraphics[scale=0.5]{figures/orcid.jpg}}}$,
R. Alonso\inst{12,14} $^{\href{https://orcid.org/0000-0001-8462-8126}{\includegraphics[scale=0.5]{figures/orcid.jpg}}}$,
G. Anglada\inst{15,16} $^{\href{https://orcid.org/0000-0002-3645-5977}{\includegraphics[scale=0.5]{figures/orcid.jpg}}}$,
T. Bárczy\inst{17} $^{\href{https://orcid.org/0000-0002-7822-4413}{\includegraphics[scale=0.5]{figures/orcid.jpg}}}$,
D. Barrado y Navascues\inst{18} $^{\href{https://orcid.org/0000-0002-5971-9242}{\includegraphics[scale=0.5]{figures/orcid.jpg}}}$,
S. C. C. Barros\inst{7,8} $^{\href{https://orcid.org/0000-0003-2434-3625}{\includegraphics[scale=0.5]{figures/orcid.jpg}}}$,
W. Baumjohann\inst{9} $^{\href{https://orcid.org/0000-0001-6271-0110}{\includegraphics[scale=0.5]{figures/orcid.jpg}}}$,
M. Beck\inst{5} $^{\href{https://orcid.org/0000-0003-3926-0275}{\includegraphics[scale=0.5]{figures/orcid.jpg}}}$,
T. Beck\inst{4}, 
W. Benz\inst{4,1} $^{\href{https://orcid.org/0000-0001-7896-6479}{\includegraphics[scale=0.5]{figures/orcid.jpg}}}$,
X. Bonfils\inst{19} $^{\href{https://orcid.org/0000-0001-9003-8894}{\includegraphics[scale=0.5]{figures/orcid.jpg}}}$,
A. Brandeker\inst{20} $^{\href{https://orcid.org/0000-0002-7201-7536}{\includegraphics[scale=0.5]{figures/orcid.jpg}}}$,
C. Broeg\inst{4,1} $^{\href{https://orcid.org/0000-0001-5132-2614}{\includegraphics[scale=0.5]{figures/orcid.jpg}}}$,
J. Cabrera\inst{11}, 
S. Charnoz\inst{21} $^{\href{https://orcid.org/0000-0002-7442-491X}{\includegraphics[scale=0.5]{figures/orcid.jpg}}}$,
A. Collier Cameron\inst{10} $^{\href{https://orcid.org/0000-0002-8863-7828}{\includegraphics[scale=0.5]{figures/orcid.jpg}}}$,
M. B. Davies\inst{22} $^{\href{https://orcid.org/0000-0001-6080-1190}{\includegraphics[scale=0.5]{figures/orcid.jpg}}}$,
M. Deleuil\inst{6} $^{\href{https://orcid.org/0000-0001-6036-0225}{\includegraphics[scale=0.5]{figures/orcid.jpg}}}$,
A. Deline\inst{5}, 
L. Delrez\inst{23,24} $^{\href{https://orcid.org/0000-0001-6108-4808}{\includegraphics[scale=0.5]{figures/orcid.jpg}}}$,
A. Erikson\inst{11}, 
A. Fortier\inst{4,1} $^{\href{https://orcid.org/0000-0001-8450-3374}{\includegraphics[scale=0.5]{figures/orcid.jpg}}}$,
L. Fossati\inst{9} $^{\href{https://orcid.org/0000-0003-4426-9530}{\includegraphics[scale=0.5]{figures/orcid.jpg}}}$,
M. Fridlund\inst{25,26} $^{\href{https://orcid.org/0000-0002-0855-8426}{\includegraphics[scale=0.5]{figures/orcid.jpg}}}$,
D. Gandolfi\inst{27} $^{\href{https://orcid.org/0000-0001-8627-9628}{\includegraphics[scale=0.5]{figures/orcid.jpg}}}$,
M. Gillon\inst{23} $^{\href{https://orcid.org/0000-0003-1462-7739}{\includegraphics[scale=0.5]{figures/orcid.jpg}}}$,
M. Güdel\inst{28}, 
K. G. Isaak\inst{29} $^{\href{https://orcid.org/0000-0001-8585-1717}{\includegraphics[scale=0.5]{figures/orcid.jpg}}}$,
L. L. Kiss\inst{30,31}, 
J. Laskar\inst{32} $^{\href{https://orcid.org/0000-0003-2634-789X}{\includegraphics[scale=0.5]{figures/orcid.jpg}}}$,
A. Lecavelier des Etangs\inst{33} $^{\href{https://orcid.org/0000-0002-5637-5253}{\includegraphics[scale=0.5]{figures/orcid.jpg}}}$,
C. Lovis\inst{5} $^{\href{https://orcid.org/0000-0001-7120-5837}{\includegraphics[scale=0.5]{figures/orcid.jpg}}}$,
D. Magrin\inst{34} $^{\href{https://orcid.org/0000-0003-0312-313X}{\includegraphics[scale=0.5]{figures/orcid.jpg}}}$,
P. F. L. Maxted\inst{35} $^{\href{https://orcid.org/0000-0003-3794-1317}{\includegraphics[scale=0.5]{figures/orcid.jpg}}}$,
V. Nascimbeni\inst{34} $^{\href{https://orcid.org/0000-0001-9770-1214}{\includegraphics[scale=0.5]{figures/orcid.jpg}}}$,
G. Olofsson\inst{20} $^{\href{https://orcid.org/0000-0003-3747-7120}{\includegraphics[scale=0.5]{figures/orcid.jpg}}}$,
R. Ottensamer\inst{36}, 
I. Pagano\inst{13} $^{\href{https://orcid.org/0000-0001-9573-4928}{\includegraphics[scale=0.5]{figures/orcid.jpg}}}$,
E. Pallé\inst{12} $^{\href{https://orcid.org/0000-0003-0987-1593}{\includegraphics[scale=0.5]{figures/orcid.jpg}}}$,
G. Peter\inst{37} $^{\href{https://orcid.org/0000-0001-6101-2513}{\includegraphics[scale=0.5]{figures/orcid.jpg}}}$,
G. Piotto\inst{34,38} $^{\href{https://orcid.org/0000-0002-9937-6387}{\includegraphics[scale=0.5]{figures/orcid.jpg}}}$,
D. Pollacco\inst{2}, 
D. Queloz\inst{39,3} $^{\href{https://orcid.org/0000-0002-3012-0316}{\includegraphics[scale=0.5]{figures/orcid.jpg}}}$,
R. Ragazzoni\inst{34,38} $^{\href{https://orcid.org/0000-0002-7697-5555}{\includegraphics[scale=0.5]{figures/orcid.jpg}}}$,
N. Rando\inst{40}, 
F. Ratti\inst{40}, 
H. Rauer\inst{11,41,42} $^{\href{https://orcid.org/0000-0002-6510-1828}{\includegraphics[scale=0.5]{figures/orcid.jpg}}}$,
C. Reimers\inst{36}, 
I. Ribas\inst{15,16} $^{\href{https://orcid.org/0000-0002-6689-0312}{\includegraphics[scale=0.5]{figures/orcid.jpg}}}$,
N. C. Santos\inst{7,8} $^{\href{https://orcid.org/0000-0003-4422-2919}{\includegraphics[scale=0.5]{figures/orcid.jpg}}}$,
G. Scandariato\inst{13} $^{\href{https://orcid.org/0000-0003-2029-0626}{\includegraphics[scale=0.5]{figures/orcid.jpg}}}$,
D. Ségransan\inst{5} $^{\href{https://orcid.org/0000-0003-2355-8034}{\includegraphics[scale=0.5]{figures/orcid.jpg}}}$,
A. M. S. Smith\inst{11} $^{\href{https://orcid.org/0000-0002-2386-4341}{\includegraphics[scale=0.5]{figures/orcid.jpg}}}$,
M. Steller\inst{9} $^{\href{https://orcid.org/0000-0003-2459-6155}{\includegraphics[scale=0.5]{figures/orcid.jpg}}}$,
Gy. M. Szabó\inst{43,44}, 
N. Thomas\inst{4}, 
S. Udry\inst{5} $^{\href{https://orcid.org/0000-0001-7576-6236}{\includegraphics[scale=0.5]{figures/orcid.jpg}}}$,
V. Van Grootel\inst{24} $^{\href{https://orcid.org/0000-0003-2144-4316}{\includegraphics[scale=0.5]{figures/orcid.jpg}}}$,
I. Walter\inst{45} $^{\href{https://orcid.org/0000-0002-5839-1521}{\includegraphics[scale=0.5]{figures/orcid.jpg}}}$,
N. A. Walton\inst{46} $^{\href{https://orcid.org/0000-0003-3983-8778}{\includegraphics[scale=0.5]{figures/orcid.jpg}}}$,
W. Wang Jungo\inst{1}, 
}

\institute{
\label{inst:1} Center for Space and Habitability, Gesellsschaftstrasse 6, 3012 Bern, Switzerland \and
\label{inst:2} Department of Physics, University of Warwick, Gibbet Hill Road, Coventry CV4 7AL, United Kingdom \and
\label{inst:3} Cavendish Laboratory, JJ Thomson Avenue, Cambridge CB3 0HE, UK \and
\label{inst:4} Physikalisches Institut, University of Bern, Gesellsschaftstrasse 6, 3012 Bern, Switzerland \and
\label{inst:5} Observatoire Astronomique de l'Université de Genève, Chemin Pegasi 51, Versoix, Switzerland \and
\label{inst:6} Aix Marseille Univ, CNRS, CNES, LAM, 38 rue Frédéric Joliot-Curie, 13388 Marseille, France \and
\label{inst:7} Instituto de Astrofisica e Ciencias do Espaco, Universidade do Porto, CAUP, Rua das Estrelas, 4150-762 Porto, Portugal \and
\label{inst:8} Departamento de Fisica e Astronomia, Faculdade de Ciencias, Universidade do Porto, Rua do Campo Alegre, 4169-007 Porto, Portugal \and
\label{inst:9} Space Research Institute, Austrian Academy of Sciences, Schmiedlstrasse 6, A-8042 Graz, Austria \and
\label{inst:10} Centre for Exoplanet Science, SUPA School of Physics and Astronomy, University of St Andrews, North Haugh, St Andrews KY16 9SS, UK \and
\label{inst:11} Institute of Planetary Research, German Aerospace Center (DLR), Rutherfordstrasse 2, 12489 Berlin, Germany \and
\label{inst:12} Instituto de Astrofisica de Canarias, 38200 La Laguna, Tenerife, Spain \and
\label{inst:13} INAF, Osservatorio Astrofisico di Catania, Via S. Sofia 78, 95123 Catania, Italy \and
\label{inst:14} Departamento de Astrofisica, Universidad de La Laguna, 38206 La Laguna, Tenerife, Spain \and
\label{inst:15} Institut de Ciencies de l'Espai (ICE, CSIC), Campus UAB, Can Magrans s/n, 08193 Bellaterra, Spain \and
\label{inst:16} Institut d'Estudis Espacials de Catalunya (IEEC), 08034 Barcelona, Spain \and
\label{inst:17} Admatis, 5. Kandó Kálmán Street, 3534 Miskolc, Hungary \and
\label{inst:18} Depto. de Astrofisica, Centro de Astrobiologia (CSIC-INTA), ESAC campus, 28692 Villanueva de la Cañada (Madrid), Spain \and
\label{inst:19} Université Grenoble Alpes, CNRS, IPAG, 38000 Grenoble, France \and
\label{inst:20} Department of Astronomy, Stockholm University, AlbaNova University Center, 10691 Stockholm, Sweden \and
\label{inst:21} Université de Paris, Institut de physique du globe de Paris, CNRS, F-75005 Paris, France \and
\label{inst:22} Centre for Mathematical Sciences, Lund University, Box 118, 221 00 Lund, Sweden \and
\label{inst:23} Astrobiology Research Unit, Université de Liège, Allée du 6 Août 19C, B-4000 Liège, Belgium \and
\label{inst:24} Space sciences, Technologies and Astrophysics Research (STAR) Institute, Université de Liège, Allée du 6 Août 19C, 4000 Liège, Belgium \and
\label{inst:25} Leiden Observatory, University of Leiden, PO Box 9513, 2300 RA Leiden, The Netherlands \and
\label{inst:26} Department of Space, Earth and Environment, Chalmers University of Technology, Onsala Space Observatory, 439 92 Onsala, Sweden \and
\label{inst:27} Dipartimento di Fisica, Universita degli Studi di Torino, via Pietro Giuria 1, I-10125, Torino, Italy \and
\label{inst:28} University of Vienna, Department of Astrophysics, Türkenschanzstrasse 17, 1180 Vienna, Austria \and
\label{inst:29} Science and Operations Department - Science Division (SCI-SC), Directorate of Science, European Space Agency (ESA), European Space Research and Technology Centre (ESTEC),
Keplerlaan 1, 2201-AZ Noordwijk, The Netherlands \and
\label{inst:30} Konkoly Observatory, Research Centre for Astronomy and Earth Sciences, 1121 Budapest, Konkoly Thege Miklós út 15-17, Hungary \and
\label{inst:31} ELTE E\"otv\"os Lor\'and University, Institute of Physics, P\'azm\'any P\'eter s\'et\'any 1/A, 1117 \and
\label{inst:32} IMCCE, UMR8028 CNRS, Observatoire de Paris, PSL Univ., Sorbonne Univ., 77 av. Denfert-Rochereau, 75014 Paris, France \and
\label{inst:33} Institut d'astrophysique de Paris, UMR7095 CNRS, Université Pierre \& Marie Curie, 98bis blvd. Arago, 75014 Paris, France \and
\label{inst:34} INAF, Osservatorio Astronomico di Padova, Vicolo dell'Osservatorio 5, 35122 Padova, Italy \and
\label{inst:35} Astrophysics Group, Keele University, Staffordshire, ST5 5BG, United Kingdom \and
\label{inst:36} Department of Astrophysics, University of Vienna, Tuerkenschanzstrasse 17, 1180 Vienna, Austria \and
\label{inst:37} Institute of Optical Sensor Systems, German Aerospace Center (DLR), Rutherfordstrasse 2, 12489 Berlin, Germany \and
\label{inst:38} Dipartimento di Fisica e Astronomia "Galileo Galilei", Universita degli Studi di Padova, Vicolo dell'Osservatorio 3, 35122 Padova, Italy \and
\label{inst:39} ETH Zurich, Department of Physics, Wolfgang-Pauli-Strasse 2, CH-8093 Zurich, Switzerland \and
\label{inst:40} ESTEC, European Space Agency, 2201AZ, Noordwijk, NL \and
\label{inst:41} Center for Astronomy and Astrophysics, Technical University Berlin, Hardenberstrasse 36, 10623 Berlin, Germany \and
\label{inst:42} Institut für Geologische Wissenschaften, Freie UniversitÃ¤t Berlin, 12249 Berlin, Germany \and
\label{inst:43} ELTE E\"otv\"os Lor\'and University, Gothard Astrophysical Observatory, 9700 Szombathely, Szent Imre h. u. 112, Hungary \and
\label{inst:44} MTA-ELTE Exoplanet Research Group, 9700 Szombathely, Szent Imre h. u. 112, Hungary \and
\label{inst:45} German Aerospace Center (DLR), Institute of Optical Sensor Systems, Rutherfordstraße 2, 12489 Berlin \and
\label{inst:46} Institute of Astronomy, University of Cambridge, Madingley Road, Cambridge, CB3 0HA, United Kingdom 
}

 \authorrunning{K. D. Jones et al.}
 
   \date{Received 2022}
 
  \abstract
  {Even among the most irradiated gas giants, so-called ultra-hot Jupiters, KELT-9b stands out as the hottest planet thus far discovered with a dayside temperature of over 4500\,K. At these extreme irradiation levels, we expect an increase in heat redistribution efficiency and a low Bond albedo owed to an extended atmosphere with molecular hydrogen dissociation occurring on the planetary dayside. We present new photometric observations of the KELT-9 system throughout 4 full orbits and 9 separate occultations obtained by the 30\,cm space telescope CHEOPS. The CHEOPS bandpass, located at optical wavelengths, captures the peak of the thermal emission spectrum of KELT-9b. In this work we simultaneously analyse CHEOPS phase curves along with public phase curves from TESS and \Spitzer to infer joint constraints on the phase curve variation, gravity-darkened transits, and occultation depth in three bandpasses, as well as derive 2D temperature maps of the atmosphere at three different depths. We find a day-night heat redistribution efficiency of $\sim\,$0.3 which confirms expectations of enhanced energy transfer to the planetary nightside due to dissociation and recombination of molecular hydrogen. We also calculate a Bond albedo consistent with zero. We find no evidence of variability of the brightness temperature of the planet, excluding variability greater than 1\% (1$\sigma$).}


   \maketitle
%

\section{Introduction}
\label{sec:introduction}

Understanding the climate of an exoplanet involves quantifying its global thermal and chemical structure. To date there have been a plethora of methods used to aid in this characterisation, including observing a phase curve. This is a lightcurve of the system taken as the planet orbits the star. The planet can add to the total flux of the system with either thermal emission or reflected light from the star. Similar to phases of the Moon, different fractions of the dayside of the exoplanet will be visible to us at different times in its orbit. There are three main observables within a phase curve: phase curve amplitude and shape, the occultation depth (when the planet goes behind the star) and the planet's hotspot phase offset (from the substellar point). From this information, one can derive 2-dimensional temperature maps of the planet’s photosphere. For planets with a rotation period significantly less than its orbital period, it could be assumed that the planet would absorb the light from the star and then re-radiate it as thermal emission relatively uniformly across the entire surface. However, for tidally locked planets, where a single hemisphere is always facing the star, it can be seen that the heat from the star is not so evenly distributed. The level of this ‘heat redistribution’ can provide insights into the dynamics and chemistry of the planet’s atmosphere.

Ultra-hot Jupiters (UHJs) are a newly emerging class of short-period exoplanets with temperatures exceeding $\sim$2500\,K. Crucially, several chemical properties distinguish them from regular hot Jupiters. Firstly, their dayside atmospheres are hot enough that hydrogen is predicted to exist in its atomic form \citep{Bell2018}. The photon energies involved in H$^-$ opacities become the dominant source of the spectral continuum over Rayleigh scattering \citep{Arcangeli2018}, resulting in a low optical geometric albedo. As these planets are tidally locked with their host star, the hydrogen will recombine to molecular hydrogen at cooler longitudes, assisting with the heat transport around the planet. The prediction of higher heat redistribution in UHJs as opposed to moderately hot Jupiters follows this \citep{Bell2018}. Secondly, the high dayside temperatures result in the thermal dissociation of most molecules, leaving only water and carbon monoxide as the main opacity sources \citep[see also, e.g.][]{Lothringer2018,Kitzmann2018b, Parmentier2018}. Following this, metals are predicted to exist in their atomic form rather than being bound in molecules, a prediction that has been observationally verified for KELT-9b \citep{Hoeijmakers2018,Hoeijmakers2019, Pino2020, Bello-Arufe2022}. Due to the high levels of irradiation, UHJs are expected to have extended atmospheres which also lends itself to hydrogen atmospheric escape, a phenomenon also detected for KELT-9b \citep{Yan2018, Wyttenbach2020}. Chemically, UHJs are objects in between gas-giant exoplanets and stars.

Not only is it important to characterise the thermal and reflective properties of a planet, but also any transient departures from the mean global state (weather). Observed variability may be caused by dynamical processes in the atmosphere of the exoplanet \citep[see review by][]{Heng2015}; for example, baroclinic instabilities in the Earth's atmosphere occur on timescales of 3 to 7 days \citep{Peixoto1984}. To date, quantifying the climate of an exoplanet and any associated transient phenomena has only been attempted for hot Jupiters. This is due to their short orbital periods (which facilitate repeated observations) and large atmospheric pressure scale heights (due to their hot, hydrogen-dominated atmospheres). \cite{Armstrong2016} claimed variability from HAT-P-7b in the form of a shift in the peak offset of its Kepler phase curve from either side of the substellar point. However this was reassessed in \cite{Lally2022} and they concluded that the apparent variations were artifacts most likely caused by stellar supergranulation. Furthermore, \cite{Agol2010} analysed 7 transits and  7 secondary eclipses of HD 189733b, measured by the \Spitzer Space Telescope, and set an upper limit of 2.7\% on the variability of flux from its dayside. In addition, \cite{Owens2021} scrutinised 27 days of TESS data and found no detectable variability from WASP-12b (another UHJ). Other works focussed on the aspects of a climate that can be probed with phase curves include \cite{Kreidberg2018}, who used Hubble-WFC3 and \Spitzer phase curves of the UHJ WASP-103b to infer inefficient dayside-to-nightside heat redistribution, an absence of water and a carbon-to-oxygen ratio less than 0.9. \cite{Arcangeli2019} used Hubble-WFC3 phase curves of the UHJ WASP-18b to find that atmospheric drag is needed to explain the observed dayside-nightside flux contrast and they speculated that this drag may be magnetic in nature.

KELT-9b is the hottest known exoplanet and the most extreme member of the UHJ class with an equilibrium temperature of $4050\pm180$\,K \citep{Gaudi2017}, It has a near-polar, 1.48-day orbit around a rapidly rotating A0/B9 star ($T_{\text{eff}} = 10170$\,K) \citep{Gaudi2017}. Its dayside brightness temperature has been measured to be 4600\,K in both the $z$-band \citep{Gaudi2017} and the TESS bandpass \citep{Wong2020}, making the dayside of KELT-9b as hot as the photosphere of a K4 star. This dayside brightness temperature implies a peak blackbody emission wavelength of 0.63\,$\mu$m, which sits in the centre of the CHEOPS bandpass. In \cite{Hooton2018}, they found a near-ultraviolet geometric albedo upper limit of $<0.14$ (3$\sigma$) and in \cite{Sudarsky2000} at wavelengths applicable to CHEOPS, they predict a geometric albedo around 0.05, which corresponds to a percentage of reflected light in the CHEOPS bandpass of $\lesssim 8\%$ of the total dayside flux (thermal emission plus reflected light). These properties imply that the CHEOPS space telescope is well-positioned to monitor the thermal flux of KELT-9b.

The CHaracterising ExOPlanet Satellite (CHEOPS) mission is a 30\,cm space telescope in a low-Earth orbit since December 2019 \citep{Benz2021}. A major component of the guaranteed-time observations planned in the nominal mission involve a thorough atmospheric classification of a wide range of transiting exoplanets, using precise full-phase curve and occultation observations. Published detections include a hot dayside atmosphere for the UHJ WASP-189b \citep{Lendl2020, Deline2022} variations in the phase curve of the super-Earth 55 Cnc e \citep{Morris2021a}, and a hint of dayside reflection for another UHJ, MASCARA-1b \citep{Hooton2021}. In this work, we embark on a comprehensive observational campaign to quantify the climate and variability of KELT-9b. Using the CHEOPS space telescope, we observed 9 secondary eclipses and 4 full phase curves. Additionally, we re-analyse the 4.5\,$\mu$m \Spitzer and TESS phase curves of KELT-9b. We interpret the CHEOPS, TESS, and \Spitzer phase curve variation jointly using \texttt{kelp}, a recently published framework that describes a two-dimensional temperature map using parabolic cylinder functions \citep{Morris2021b}.

In Section \ref{sec:observations} we detail the technical aspects of the CHEOPS, TESS, and \Spitzer observations and include a description of the CHEOPS space telescope and its data reduction pipeline. Section \ref{sec:Analysis} explains the different sections of the data analysis and includes a detailed description of the models used to model the phase curves in each bandpass and the CHEOPS occultations. In Section \ref{sec:results}, we report the results of our phase curve fit and occultation eclipse depths. Within this section, in Section \ref{subsec:thermalmap} and \ref{subsec:daynightinttemps} we show the bandpass-dependent 2D thermal maps derived from the best-fit phase curve parameters and discuss the resulting day and nightside integrated temperatures. The results of the CHEOPS occultation analysis is detailed in Section \ref{subsec:eclipsedepths} and in Section \ref{subsec:albedo} we report the Bond albedo and heat redistribution efficiency. Finally in Section \ref{subsec:litcomparison} we compare our results with previous results using the same datasets. A discussion of the impact and importance of this work for the study of the climate on UHJs and future multi-wavelength JWST phase curves can be found in Section \ref{sec:discussion}.

\section{Observations}
\label{sec:observations}
\subsection{CHEOPS observations}
\label{subsec:CHEOPSobservations}

CHEOPS is an on-axis Ritchey-Chrétien telescope \citep{Benz2021}. Its nadir-locked orientation keeps the bottom of the spacecraft pointed towards Earth throughout each orbit, causing the field of view of CHEOPS to rotate during science observations. This frequently results in systematic noise in phase with the spacecraft roll angle, as neighbouring stars contribute varying levels of contamination into the CHEOPS aperture over one spacecraft orbit \citep[e.g.:][]{Lendl2020}. Detailed systematic investigations have revealed that CHEOPS observations occasionally contain a ramp feature whereby the flux of the first few orbits can show a significant correlation with the temperature fluctuations of the telescope assembly and is most likely caused by the change in temperature of the telescope as it adjusts to a new pointing position \citep{Morris2021a}.

CHEOPS observed nine occultations of KELT-9 between 25 July 2020 and 24 July 2021 and four phase curves; observed on 31 August 2020, 10 September 2020, 31 July 2021 and 22 August 2021, each obversation lasting around 2.4\,days. The individual occultation observations lasted between 5.9 and 6.9\,hours, distributed over 4 to 5 CHEOPS orbits. Further details of the individual observations can be found in Table \ref{tab:obslog}.

CHEOPS data is automatically processed by the CHEOPS Data Reduction Pipeline \citep[DRP,][]{Hoyer2020}.  The CHEOPS DRP performs the basic calibration of the science images (i.e. bias, dark, flat-field corrections) and also performs background correction, cosmic-rays hits removal, correction of bright stars' smear trail contamination and provide estimations of the flux contamination of background stars \citep[see details in][]{Hoyer2020}.  Finally, the DRP extracts the photometric signal of the target in 3 standard apertures called \texttt{RINF}, \texttt{DEFAULT} and \texttt{RSUP} (at radius of R=21.5'', 25'' and 30''), in addition to the \texttt{OPTIMAL} aperture with a radius set to minimise the effect of the contamination by close-by background stars.  In the case of KELT-9 observations this aperture was set at R=40''.  In our analysis we use the light curves obtained with the \texttt{DEFAULT} aperture.  

\begin{table*}[]
\centering
\begin{tabular}{ccccccc}
\hline
Date Start & Date Stop & File Key & Duration & Exposure  & Exposures  & Efficiency \\
$[$UT]& [UT] & & [dd:hh:mm] & Time [s] & per stack & \%\\ 
\hline
2020-07-25 07:15 & 2020-07-25 14:05 & {\small CH\_PR100036\_TG001201\_V0200} & 00:06:50 & 36.7 & 3 ($\times$12.2\,s) & 61 \\
2020-08-04 16:02 & 2020-08-04 21:52 & {\small CH\_PR100036\_TG001202\_V0200} & 00:05:50 & 36.7 & 3 ($\times$12.2\,s) & 66 \\
2020-08-17 23:54 & 2020-08-18 05:49 & {\small CH\_PR100036\_TG001203\_V0200} & 00:05:55 & 36.7 & 3 ($\times$12.2\,s) & 73 \\
2020-08-20 23:28 & 2020-08-21 05:57 & {\small CH\_PR100036\_TG001204\_V0200} & 00:06:28 & 36.7 & 3 ($\times$12.2\,s) & 65 \\
2020-08-23 23:20 & 2020-08-24 05:21 & {\small CH\_PR100036\_TG001205\_V0200} & 00:06:00 & 36.7 & 3 ($\times$12.2\,s) & 72 \\
2020-08-28 10:32 & 2020-08-28 16:30 & {\small CH\_PR100036\_TG001206\_V0200} & 00:05:58 & 36.7 & 3 ($\times$12.2\,s) & 71 \\
2020-09-03 08:19 & 2020-09-03 14:17 & {\small CH\_PR100036\_TG001207\_V0200} & 00:05:58 & 36.7 & 3 ($\times$12.2\,s) & 72 \\
2020-09-13 16:15 & 2020-09-13 22:44 & {\small CH\_PR100036\_TG001208\_V0200} & 00:06:28 & 36.7 & 3 ($\times$12.2\,s) & 50 \\
2021-07-24 15:45 & 2021-07-24 21:37 & {\small CH\_PR100036\_TG001209\_V0200} & 00:05:52 & 36.7 &3 ($\times$12.2\,s)  & 69 \\
\hline
2020-08-31 13:29 & 2020-09-03 00:53 & {\small CH\_PR100036\_TG001001\_V0200} & 02:11:24 & 36.7 & 3 ($\times$12.2\,s) & 61 \\
2020-09-10 03:31 & 2020-09-12 14:11 & {\small CH\_PR100036\_TG001002\_V0200} & 02:10:39 & 36.7 & 3 ($\times$12.2\,s) & 59 \\
2021-07-31 04:25 & 2021-08-02 15:06 & {\small CH\_PR100036\_TG000901\_V0200} & 02:10:40 & 36.7 & 3 ($\times$12.2\,s) & 63 \\
2021-08-22 11:04 & 2021-08-24 22:15 & {\small CH\_PR100036\_TG000902\_V0200} & 02:11:10 & 36.7 & 3 ($\times$12.2\,s) & 62 \\
\hline
\end{tabular}
\caption{CHEOPS observation logs, corresponding to the occultation-only observations in the first 9 rows and the phase curve observations in the last 4 rows. The File Key is useful for uniquely identifying the visits used in this work.}
\label{tab:obslog}
\end{table*}

\subsection{TESS observations}
\label{subsec:tessobs}
The TESS satellite \citep{Ricker2014} observed more than 20 phase curves of KELT-9 during Sector 14 and 15 of the telescope's operation (months of July and August 2019). These observations were first published in \cite{Wong2020} using their own reduction techniques. For our analysis we used the PDCSAP flux measurements at 2 minutes cadence (pre-reduced using the TESS SPOC pipeline to remove long-term trends and systematics) \citep{Jenkins2016}. We downloaded the data and stitched the light curves into a single array using \texttt{lightkurve} \citep{lightkurve}.

\subsection{\Spitzer observations}
\label{subsec:spitzerobs}

In this work, we analyse the \Spitzer archival data of KELT-9\,b that have already been published \citep{Mansfield2020}. We downloaded KELT-9\,b archival IRAC data from the \Spitzer Heritage Archive\footnote{\url{http://sha.ipac.caltech.edu}}. The data consist of one full phase curve at 4.5$\mu$m split in two Astronomical Observation Requests (AORs) obtained under program ID 14059 (PI J. Bean). The reduction and analysis of these datasets are similar to \citet{Demory2016b}. We model the correlated noise associated with IRAC intra-pixel sensitivity \citep{Ingalls2016} using a modified implementation of the BLISS (BiLinearly Interpolated Sub-pixel Sensitivity) mapping algorithm \citep{Stevenson2012}.

In addition to the BLISS mapping, our baseline model includes a linear function of the Point Response Function (PRF) Full Width at Half-Maximum (FWHM) along the $x$ and $y$ axes, which significantly reduces the level of correlated noise as shown in previous studies \citep[see, e.g.:][]{Lanotte2014, Demory2016a, Demory2016b, Gillon2017, Mendonca2018}. Our baseline model does not include time-dependent parameters. Our implementation of this baseline model is included in a Markov Chain Monte Carlo (MCMC) framework already presented in the literature \citep{Gillon2012}. We run two chains of 200,000 steps each to determine the phase curve properties at 4.5 $\mu$m based on the entire dataset described above. From our BLISS mapping+FWHM baseline model, we obtain a median RMS of 723 ppm per 23s integration time for that dataset.

\subsection{Stellar parameters}
\label{sec:stel}

As a key stellar prior in our analysis, we determine the stellar radius of KELT-9 using a Markov-Chain Monte Carlo (MCMC) modified infrared flux method (IRFM; \citealt{Blackwell1977,Schanche2020}). To achieve this, we compute synthetic fluxes from constructed spectral energy distributions (SEDs). These were built from stellar atmospheric models and stellar parameters derived via the spectral analysis in \cite{Borsa2019} that were integrated over bandpasses of interest with the SED attenuated to determine the extinction within the model fitting. We compared these fluxes to observed broadband photometry retrieved from the most recent data releases for the following bandpasses; {\it Gaia} G, G$_{\rm BP}$, and G$_{\rm RP}$, 2MASS J, H, and K, and {\it WISE} W1 and W2 \citep{GaiaCollaboration2021, Skrutskie2006,Wright2010} to calculate the apparent bolometric flux, and hence the stellar angular diameter and effective temperature. In our analysis we use stellar atmospheric models from the \textsc{atlas} Catalogues \citep{Castelli2003}. By converting the angular diameter to the stellar radius using the offset-corrected {\it Gaia} EDR3 parallax \citep{Lindegren2021}, we obtain $R_{\star}=2.379\pm0.038\,R_{\odot}$.

The stellar radius $R_{\star}$ together with the effective temperature $T_{\mathrm{eff}}$ and the metallicity [Fe/H] constitute the input set to determine the isochronal age $t_{\star}$ and mass $M_{\star}$. To make the derivation process more robust we employed two different stellar evolutionary models, namely PARSEC v1.2S\footnote{\textit{PA}dova and T\textit{R}ieste \textit{S}tellar \textit{E}volutionary \textit{C}ode: \url{http://stev.oapd.inaf.it/cgi-bin/cmd}} \citep{marigo17} and CLES \citep[Code Liègeois d'Évolution Stellaire,][]{scuflaire08}. In detail, we used the capability of the isochrone placement technique \citep{bonfanti15,bonfanti16} to fit the input parameters within pre-computed PARSEC grids of tracks and isochrones so to retrieve a first pair of age and mass estimates. A second pair of age and mass values, instead, was directly computed by the CLES code which generates the best evolutionary track that is compatible with the input parameters following a Levenberg-Marquadt minimisation scheme \citep[see][for further details]{salmon21}.
After checking the mutual consistency of the two respective pairs through a $\chi^2$-based criterion, we finally merged our outcomes as described in \citet{bonfanti21} and we obtained $t_{\star}=0.3\pm0.1$\,Gyr and $M_{\star}=2.45_{-0.17}^{+0.19}\,M_{\odot}$.

\begin{table}
\centering
\caption[]{Stellar parameters used to derive the stellar radius and mass used in this paper. This table also includes these calculated parameter values.}
\label{tab:ts3params}
\begin{tabular}{lcc}
\hline \hline
Stellar parameter & Unit & Value\\ 
\hline 
Metallicity & dex & $0.14\pm0.30$ \\
Surface gravity & dex & $4.1\pm0.3$ \\
Radius & $R_{\odot}$ & $2.379\pm0.038$ \\ 
Mass & $M_{\odot}$ & ${2.45}^{+0.19}_{-0.17}$ \\
Age & Gyr & $0.3\pm0.1$ \\
 
\hline 
\end{tabular} 
\end{table}

\section{Analysis}
\label{sec:Analysis}
\subsection{Phase curves} \label{subsec:PhaseCurves}

Over its first and second year, CHEOPS has observed 4 full phase curves of KELT-9; TESS has observed over 20 and \Spitzer has observed 1. We present here a joint analysis of all these observations. 

As already discussed in Section \ref{sec:introduction}, KELT-9b is the hottest know exoplanet with a dayside temperature of $\sim$4600\,K and a thermal emission peak in the CHEOPS bandpass as well as efficient optical absorbers. We therefore assume that all phase curves are completely dominated by thermal emission, an assumption justified in \citet{Schwartz2017} and \citet{Morris2021b}.

We fit each instrument's observations with a self-consistent transit, eclipse and thermal phase curve model. Table \ref{tab:results} shows the parameters common to all phase curves and Table \ref{tab:results2} shows parameters that are instrument-specific. In the following subsections, we detail the different components of the model used in the fitting procedure, along with the detrending and other models specific to each bandpass.

To fit our global model, we used an affine-invariant ensemble sampling to estimate the parameter posterior distributions with 900 walkers, 5000 burn-in steps and then 80000 steps \citep{Foreman-Mackey2013}. We confirmed that the chains converged by analysing the autocorrelation time. The integration length is 50 times the number of samples.

\subsubsection{Gravity-darkened transit model}

Gravity can darken stellar photospheres due to rapid rotation, shaping the star into an oblate spheroid. \cite{vonZeipel1924} showed that this oblateness causes a temperature gradient across the surface of the star and consequently reduces the emitted flux near the equator. Depending on the geometry of the star-planet system, this causes significant deviations from a simple symmetric transit light curve \citep{Barnes2009}. This effect is in addition to limb darkening \citep{Claret2017}.
We must accurately model the transits in our observations in order to calculate the planetary radius and model the phase curve amplitude accurately, which in turn gives us information about the temperature map of the planet.

We model the transit of KELT-9b with \texttt{pytransit} \citep[version 2.5.17, ][]{Parviainen2015}, an open-source Python package which implements a gravity-darkened transit model based on \cite{Barnes2009}. For describing this model we will use the same notation as in \citet{Hooton2021}. The model is characterised by 
\begin{itemize}
    \item $R_*$, the stellar radius
    \item $R_p$, the planetary radius (in units of stellar radii, $R_*$)
    \item $P$, planetary orbital period
    \item $\rho$, the stellar density 
    \item $a$, the semi-major axis in units of $R_*$
    \item $i$, the planetary orbital inclination
    \item $e$, orbital eccentricity
    \item $\omega$, argument of periastron
    \item $P_\mathrm{rot}$, the rotation period of the star
    \item $T_\mathrm{pole}$, the stellar temperature at its pole
    \item $\lambda$, the sky-projected spin orbit angle (the sky-projected angle between the planetary orbital plane and the stellar equatorial plane)
    \item $i^*$, stellar inclination, defined as the angle between the observer's line of sight and the stellar rotation axis. This is related to the stellar obliquity $\phi$, specified in \citet{Ahlers2020}, by $i^* = 90^\circ - \phi $. Note this is \textit{not} the same $\Delta\phi$ that we use to describe the thermal phase curve model.
    \item $\beta$, the gravity-darkening coefficient (defines how strong the temperature gradient is), defined in \cite{vonZeipel1924}.
    \item $u_1$ and $u_2$, the quadratic limb-darkening coefficients, which we reparametrise as $q_1$ and $q_2$ \citep{Kipping2013}
    \item $t_\mathrm{0}$, the time of mid-transit
    \item $f_0$, transit scaling factor (scales the out-of-transit baseline)
    \item $filter$, the telescope's bandpass transmission efficiency
    \item stellar spectrum
\end{itemize}

We fixed $T_\mathrm{pole}= 10170$\,K, $e = 0$ and $\omega = 90^\circ$. We also fixed $R_* = 2.379\,R_\odot$ (see Section \ref{sec:stel}). For CHEOPS and TESS we use a PHOENIX model stellar spectrum \citep{Husser2013}, and for \Spitzer we use a blackbody spectrum. 

Due to the large degeneracy between the limb darkening and gravity darkening parameters, we had to use previous studies to inform our priors, instead of just letting them all float freely - something we tested and found no clear unique solution. Therefore we used results from previous doppler tomography studies \citep{Gaudi2017, Borsa2019} to inform priors on $b$, $\lambda$ and $v\sin{i^*}$ (where $v$ is the rotational velocity of the star), and theoretical limb-darkening tables in \cite{Claret2011, Claret2017, Claret2021} to inform priors on $q_1$ and $q_2$. We also used \cite{Claret2016} to inform priors on $\beta$. See Table \ref{tab:results} and \ref{tab:results2} for the priors used for all the gravity-darkened model parameters. 

Using the information obtained from the transit fit we can calculate the true 3D spin-orbit angle ($\Psi$), given by the equation:
\begin{equation}
\label{eqn:truespinorbit}
    \cos{\Psi} = \sin{i^*} \cos{i} + \cos{i^*} \sin{i} \cos{\lambda}.
\end{equation}

Gravity-darkened transits have also been analysed in other CHEOPS work \citep[see, e.g.:][]{Lendl2020, Hooton2021, Deline2022}.

\subsubsection{Thermal phase curve model}
\label{sec:kelp}

To model the shape of the thermal phase curve we use \texttt{kelp} \citep{Morris2021b}, a Python package that models the surface temperature of the planet as a 2D thermal map constructed by modified spherical harmonics (parabolic cylinder functions) \citep{Heng2014}, given by equation (1) of \cite{Morris2021b}:
\begin{equation}
    T(\theta, \phi) = \bar{T} \left( 1 + \sum_{m, \ell}^{\lmax} h_{m\ell}(\theta, \phi) \right), 
\label{eqn:one}
\end{equation}
where $\theta$ and $\phi$ are latitude and longitude. $\bar{T}$ is a background temperature upon which two-dimensional perturbations exist and are quantified by the ``alphabet" or basis functions $h_{m\ell}$ as defined by equation (258) of \cite{Heng2014}. These functions are also dependent on the dimensionless parameters $\alpha$ and $\omega_\mathrm{drag}$. The dimensionless fluid number $\alpha$ is constructed from the Rossby, Reynolds and Prandtl numbers; it quantifies the competition between stellar heating (forcing) and sources of friction (drag).  The normalised drag frequency $\omega_\mathrm{drag}$ quantifies the strength of friction.  In practice, we find that it is sufficient to truncate the series in the preceding equation at $l_{\rm max}=1$ \citep{Morris2021b}.  The entire temperature map may be shifted back and forth in longitude by $\Delta \phi$, which effectively fits for the hotspot offset if $\omega_{\rm drag} \gtrsim 3$ \citep{Morris2021b}.  Full details of the theoretical formalism may be found in Section 2 of \cite{Morris2021b}, which we will not repeat here.

Overall, our thermal phase curve model is parametrised by the following variables:
\begin{itemize}
    \item $\Delta\phi$, hotspot offset
    \item $\alpha$, dimensionless fluid number
    \item $\omega_\mathrm{drag}$, dimensionless drag frequency
    \item $C_{m\ell}$, the power of individual harmonic modes
    \item $\ell_\mathrm{max}$, describes the highest spherical harmonic mode in the model
    \item planetary parameters including $a$, $R_p$, $T_\mathrm{pole}$
    \item telescope bandpass transmission efficiency
    \item $\bar{T}$, scaling term of the mean temperature field
\end{itemize}

\texttt{kelp} integrates over the passband-weighted thermal map visible at a specific time in its orbit and converts the map into a flux measurement using the following equation from \cite{Cowan2011b}, which can be compared to the observed phase curve:

\begin{equation}
    \frac{F_p}{F_\star} = \frac{1}{\pi I_\star} \left(\frac{R_p}{R_\star}\right)^2 \int_0^\pi \int_{-\xi-\pi/2}^{-\xi+\pi/2} I_p(\theta, \phi) \cos(\phi+\xi) \sin^2(\theta) ~d\phi ~d\theta \label{eqn:diskint}
\end{equation}
where $I$, the intensity, is defined by

\begin{equation}
I = \int \lambda \mathcal{F}_\lambda \mathcal{B}_\lambda(T(\theta, \phi)) d\lambda
\label{eq:intensity}
\end{equation}
where $\mathcal{F}_\lambda$ is the instrument-specific filter response function and $\mathcal{B}_\lambda(T(\theta, \phi))$ is the Planck function of each temperature element $T(\theta, \phi)$.

The initialisation of this model at every time-step required is a bottleneck in the joint MCMC fit for the phase curves. To increase the speed we evaluated the model at 200 orbital phases, spaced equally between -$\pi$ and $\pi$. We then linearly interpolated the model flux values back to the original time resolution. We investigated the effect of increasing the number of samples on the fitted parameters and found that the median parameter values converged to a constant value at around 150, and so we choose 200 samples to increase the speed of the code whilst not decreasing the accuracy of the phase curve fit. 

In the joint fit, we followed \cite{Morris2021b} and set $\alpha = 0.6$, $\omega_{drag} = 4.5$ and $\ell_{max} = 1$, since latitudinal variations in temperature and flux are not constrained by thermal phase curves and it was shown in that study that it is sufficient to describe the temperature map only with the first mode. We also set all $C_{m\ell} = 0$ apart from $C_{11}$ which is a free parameter. See Table \ref{tab:results} for the priors of the phase curve model.

\subsubsection{Secondary eclipse model}

The \texttt{batman} package models the flux decrement as the planet is occulted by the star \citep{Kreidberg2015}. We multiply this normalised model (where the planetary flux is null during the eclipse) by the thermal phase curve model to produce the full phase curve model. 

\subsubsection{Stellar pulsation model}
\label{subsubsec:stellarpulsationmodel}
\begin{figure}
    \centering
    \includegraphics[width=\textwidth/2]{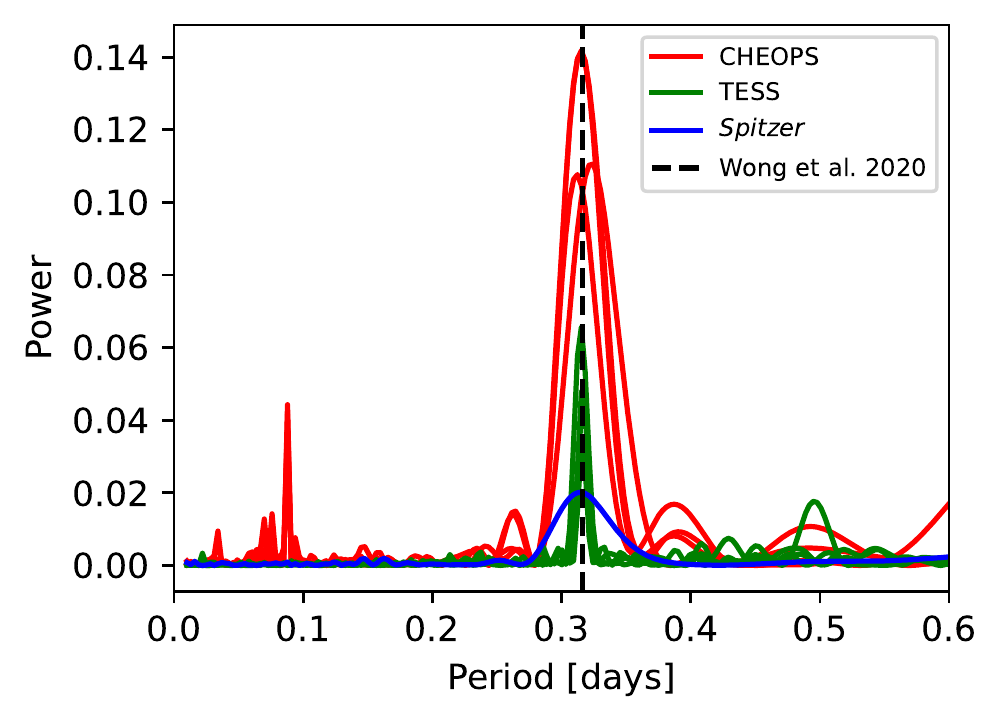}
    \caption{Periodograms reveal the presence of stellar pulsations in the residuals of the phase curve fitting. In red are the four CHEOPS phase curves, in green are the TESS phase curves and in blue is the \Spitzer phase curve. It is clear the stellar pulsation period (near 0.3\,days) is present in all phase curves and matches that signal seen also in \cite{Wong2020} and \cite{Mansfield2020} (dashed black line). The short period peaks are only present in CHEOPS but not TESS or \Spitzer so we assume they are CHEOPS systematics and disregard the signal. }
    \label{fig:periodogram}
\end{figure}

After examining the residuals of an initial CHEOPS phase curve fit, without any stellar pulsation model, it was clear that there was a periodic signal present in the phase curves at a period of around 7.5\,hours (see Figure \ref{fig:periodogram}). This signal was identical to the one observed in \cite{Wong2020} and \cite{Mansfield2020} and has been attributed to stellar pulsations. In \cite{Wyttenbach2020}, they analysed these pulsations and concluded that they are compatible with p-mode oscillations present in a $\delta$ Scuti-type star. 

To correct for this signal we used a Gaussian process with a simple harmonic oscillator (SHO) kernel implemented by \texttt{celerite2} \citep{celerite1, celerite2}. We fixed the amplitude (100\,ppm) and the damping timescale, $\tau$, to 2x the gaussian process period (GP period), which is a fitted parameter. This damping timescale provides coherence of pulsations over several cycles while allowing for evolution of the pulsation signal on longer timescales. Because of this we calculate the Q value of this SHO kernel to be 6.

We found evidence of the same stellar pulsations in the TESS phase curves as well. Although the PDCSAP flux light curves have already been corrected for long-term trends, there were still significant trends present in the KELT-9 light curve which we believe to be of instrument systematic origin. Therefore we implemented the same kernel described above for these TESS phase curves, along with an additional Matérn 3/2 kernel with an amplitude of 200ppm and a timescale of 12 days to remove the long-term trends in this data. 

The pulsations are also detectable in the \Spitzer data, albeit at less than 10\% of the total amplitude of the phase curve (a lower percentage than in the other two bandpasses). After further analysis showed modelling the pulsations within this dataset had a non-trivial impact on the phase curve parameters, we extended the GP to also include the \Spitzer phase curve. The amplitude of the pulsations in the \Spitzer data was also around 100ppm so this hyperparameter could remain the constant. We decided to use the same kernels in both CHEOPS, TESS, and \Spitzer to jointly infer the free kernel hyperparameters.

\subsubsection{CHEOPS}
\label{sec:CHEOPS_phasecurves}
The composite transit, eclipse and thermal phase curve model we use to fit the CHEOPS observations is composed of
\begin{itemize}
  \item gravity-darkened transit model
  \item planet thermal phase curve model
  \item eclipse model 
  \item stellar pulsation model 
  \item systematics model
\end{itemize}

\textbf{Data clipping and systematics model}: For each phase curve, before stitching them together with the other datasets, we sigma-clipped outliers of the centroid position of the target star at 4.5$\sigma$ and masked out individual points with anomalously high levels of background or low temperature readings. As discussed in Section \ref{subsec:CHEOPSobservations}, these telescope assembly temperatures have been shown to coincide with a systematic ``ramp'' feature in the CHEOPS light curves, most likely caused by the change in temperature of the telescope as it adjusts to a new pointing position \citep{Morris2021a}. We also masked out the first few hours of data in the second, third and fourth phase curve, as they coincided with a strong, non-linear increase in the telescope temperature. We mask on average 16\% of datapoints from each phase curve.

We used a linear detrending model including vectors proportional to the temperature, $\text{temperature}^2$ and roll angle in our fit to detrend against the correlated noise sources. We chose these basis vectors based on the analysis performed in \cite{Morris2021a} which used the BIC minimisation to find the parameters that contributed the most to the model without overfitting.

\subsubsection{TESS}
The TESS phase curves model is a linear combination of: 
\begin{itemize}
  \item gravity-darkened transit model
  \item planet thermal phase curve model
  \item eclipse model 
  \item stellar pulsation model 
\end{itemize}

\textbf{Data clipping}: As described in Section \ref{subsec:tessobs}, we use the TESS PDCSAP light curves from the SPOC pipeline. Before the fit we also masked out sections of the phase curves that clearly had strong systematics and that were also removed in the analysis performed in \cite{Wong2020}. These usually affected the data points shortly before or after a gap in the TESS observations. In total we mask out 5\% of the datapoints.

\subsubsection{\Spitzer}
The \Spitzer phase curve model is a linear combination of: 
\begin{itemize}
  \item gravity-darkened transit model
  \item planet thermal phase curve model
  \item eclipse model 
  \item stellar pulsation model 
\end{itemize}

\textbf{Pre-fit conditioning and systematics model}:
See Section \ref{subsec:spitzerobs} for a description of the detrending applied to the \Spitzer dataset. From the BLISS mapping detrending routine, we obtain a detrended lightcurve, which we use for our joint phase curve fit. Observations are also available at 3.6\,$\mu$m, which will be the subject of a forthcoming paper by Beatty et al. (in prep.).

\subsubsection{On ellipsoidal variations}

To include an ellipsoidal variation model \citep[see, e.g.:][]{Welsh2010, Gai2018}, we initially fit the first two CHEOPS phase curves separately using the \citet{Wong2020} sinusoidal model for the stellar pulsations and ellipsoidal variations, leaving the phase and amplitude of both sinusoids as free parameters. The resulting ellipsoidal phase and amplitude were not consistent between each phase curve. We then fit the combined set of CHEOPS phase curves and the chains converged on solutions with ellipsoidal variation amplitudes of $<$10\,ppm. The theoretical expectation is reported as $44\pm6$\,ppm in \cite{Wong2020}, therefore this fitted amplitude is over 5$\sigma$ lower than the theoretical value. We suggest that the variability on the $P/2$ timescale may not be ellipsoidal because it has an unexpected and evolving phase and appears incoherent over 4 orbital timescales (from the first to second phase curve observations). 

Carrying out the same test with only the GP stellar pulsation model instead of the pure sinusoids, we found that both models agreed on the amplitude of the stellar pulsations as well as other planetary parameters. As a result of this analysis, we conclude that the GP stellar pulsation model is flexible enough to capture the stellar signals without absorbing the planetary ones.

\subsection{CHEOPS occultations}
\label{sec:occultations}
As well as the 9 occultation-only CHEOPS observations, 5 eclipses were observed within the 4 CHEOPS phase curves (2 of the 5 eclipses were observed within the same phase curve). To increase our occultation sample size, we added these to the set of occultations, clipping them so that they each had a similar number of CHEOPS orbits to the rest of the occultations. Therefore, there are thirteen occultations in total.

Each occultation was then fitted independently with a model containing a linear combination of:
\begin{itemize}
  \item eclipse model
  \item stellar pulsation model
  \item systematics model
\end{itemize}
We carried out the fit of the occultations with \texttt{PyMC3} \citep{pymc3}, which uses a gradient-based Hamiltonian Monte Carlo (HMC) method to integrate for parameter posteriors.

\textbf{Data clipping and systematics model}: Similarly to the phase curves, the occultations were sigma-clipped to remove outliers in the target centroid-space and flux values (both by $3\sigma$), and points with anomalously large background or temperature readings were masked out. The data points were also flux sigma-clipped. After clipping, 4-14\% of points were removed from each observation.

Since the observation duration for each occultation is short (4-5 CHEOPS orbits), we risk fitting a model that is too complex. To investigate the minimum complexity model needed to reproduce the light curve without over-fitting, we used Leave-One-Out cross-validation \citep{Vehtari2015} to compare the predictive power for each model containing different basis vectors in the systematics model. The basis vectors we tested included a flux constant, $\sin{(\text{roll angle})}$, $\cos{(\text{roll angle})}$, time, $\text{time}^2$, temperature (from the \textit{thermFront2} sensor) and background. These have been shown in a previous study \citep{Morris2021a} to be the most influential basis vectors on the light curves. 

Overall, the preferred systematic model included every basis vector except the background, however other combinations were preferred by nine of the other occulations. After fitting, we investigated whether there was a trend between the fitted eclipse depth and the number of basis vectors chosen for each occultation. We found no correlation.

\subsubsection{Stellar pulsation model}
As detailed in Section \ref{sec:CHEOPS_phasecurves}, We observed a stellar pulsation signal in the phase curves with an amplitude of around 100\,ppm and a timescale of 7.5\,hours. Assuming it is continuous and also present in the occultation observations, the occultations are taken over a long enough timescale that this signal would vary significantly within a single visit. However as the baseline of the occultations is so short, it is impossible to uniquely infer the phase and amplitude of the pulsations in each observation. Therefore we included a sinusoid in the occultation models with a period fixed at 7.5\,hours, with the phase of this signal free to vary independently between each occultation. In order to avoid biasing the prior, we implemented a hierarchical Bayesian technique where the prior of the amplitude was set to $\mathcal{N}(100, \sigma)$ ppm, and truncated at zero so that the amplitude is always positive, and the standard deviation $\sigma$ became a fitted parameter and was allowed to vary as $\mathcal{U}(10, 1000)$ ppm. 

\subsubsection{Eclipse model}
\label{subsubsec:eclipse}

We used the \texttt{batman} package to create a basic secondary eclipse model for the occultations. The out-of-eclipse observations contain a hint of the shape of the phase curve of KELT-9, which peaks near secondary eclipse. We used the best-fit posteriors from the full phase curve fit to produce a phase curve model from \texttt{kelp}. This model is used to scale the out-of-eclipse sections of the basic \texttt{batman} model. This model was characterised by the following parameters:

\begin{itemize}
  \item $t_0$, the time of transit centre*
  \item $R_p$, the planetary radius (in units of stellar radii, $R_*$)
  \item $b$, the impact parameter
  \item $P$, the planetary orbital period*
  \item $\rho$, the stellar density
\end{itemize}
where we fixed starred (*) values and and fit for unstarred ones.

\section{Results}
\label{sec:results}

\begin{figure*}
    \centering
    \begin{subfigure}{\textwidth}
        \centering
        \includegraphics[width=\textwidth*9/10]{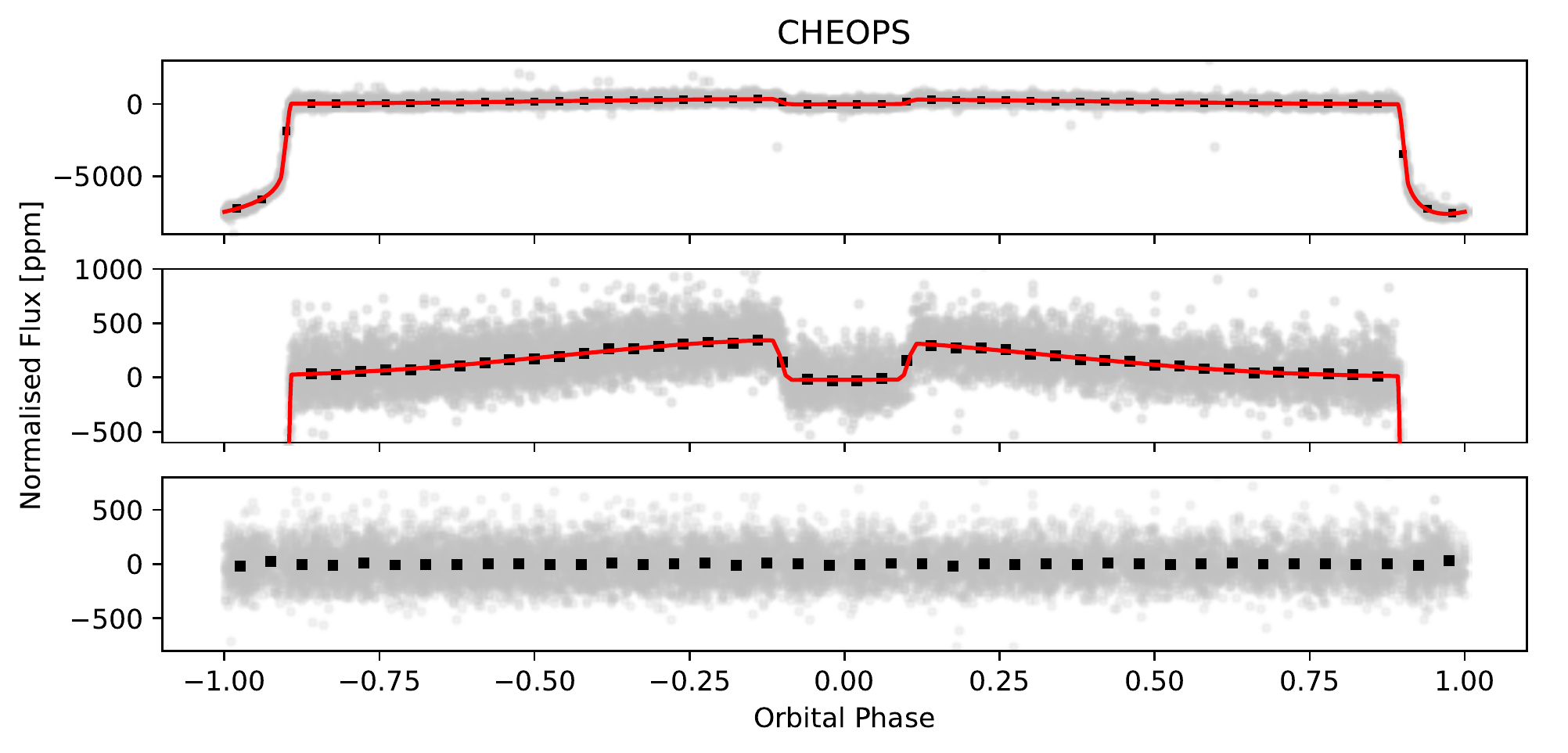}
        \caption{CHEOPS phase-folded and detrended phase curves (first 2 panels, 2nd panel is a zoomed-in view of the 1st). 3rd panel shows residuals of the fit.}
        \label{subfig:cheopsphasecurve}
    \end{subfigure}
    \begin{subfigure}{\textwidth}
        \centering
        \includegraphics[width=\textwidth*9/10]{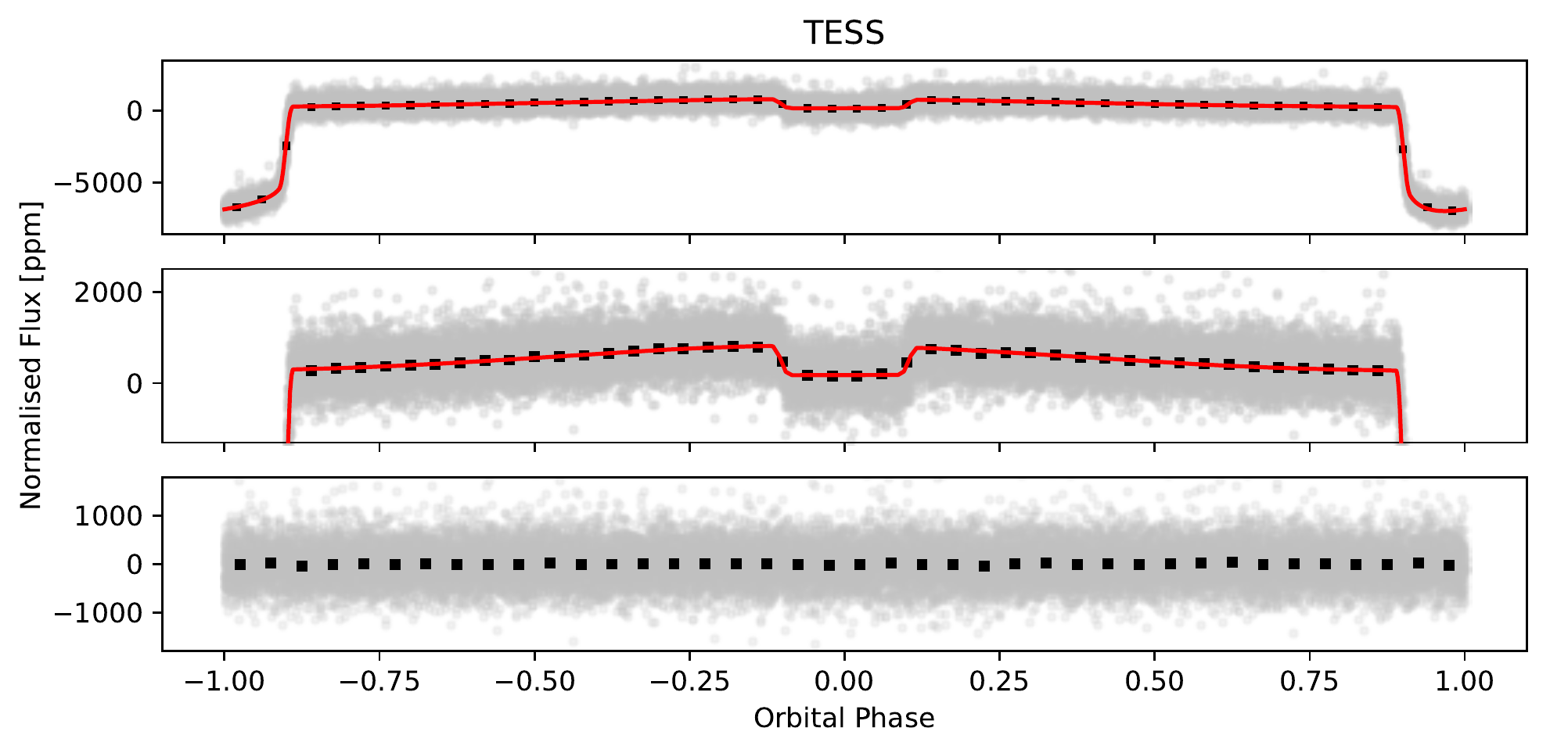}
        \caption{TESS phase-folded and detrended phase curves (first 2 panels, 2nd panel is a zoomed-in view of the 1st). 3rd panel shows residuals of the fit.}
        \label{subfig:tessphasecurve}
    \end{subfigure}
    \begin{subfigure}{\textwidth}
        \centering
        \includegraphics[width=\textwidth*9/10]{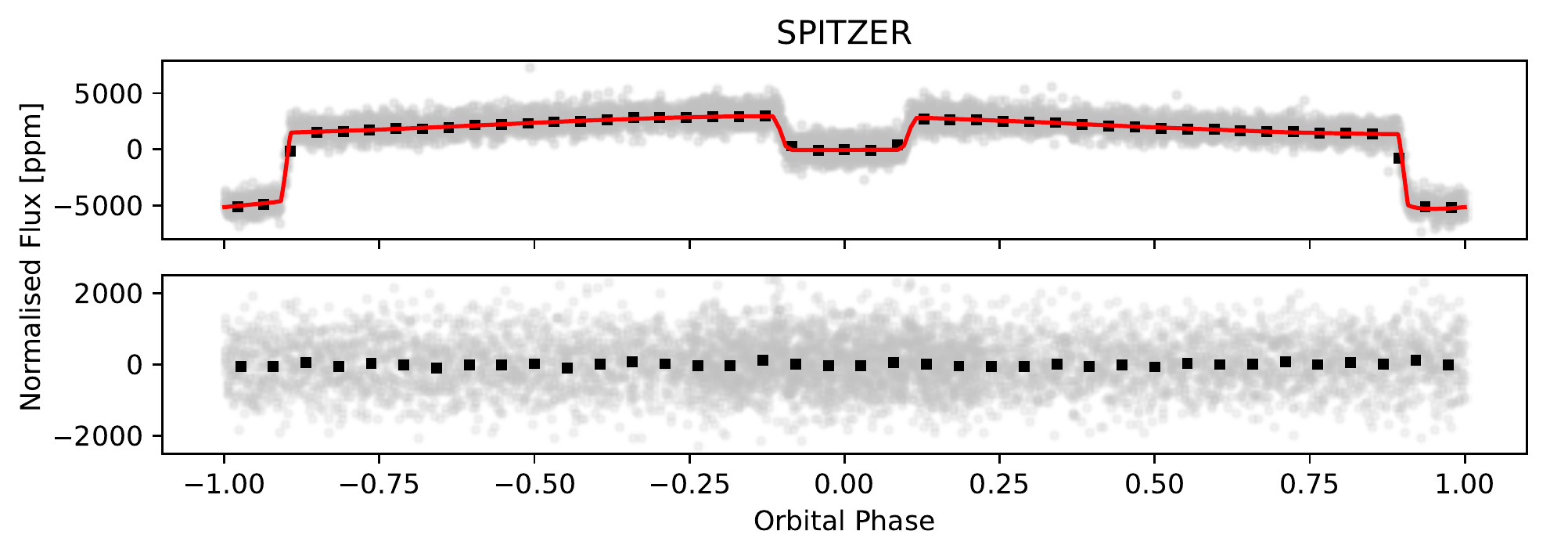}
        \caption{\Spitzer phase-folded and detrended phase curve (first panel). 2nd panel shows residuals of the fit.}
        \label{subfig:spitphasecurve}
    \end{subfigure}
    \caption{CHEOPS, TESS, and \Spitzer phase-folded and detrended (stellar pulsations and other systematic trends removed) phase curves, overplotted with the best-fit phase curve model, transit model and eclipse model (in red). In black are the binned grey datapoints, with error bars that are smaller than the point size in all panels so they are not visible.}
    \label{fig:phasecurves}
\end{figure*}

Figure \ref{fig:phasecurves} shows the phase folded CHEOPS, TESS, and \Spitzer phase curves respectively, with the stellar pulsations and other systematic trends removed, and overplotted with the best-fit phase curve, transit and eclipse model. The fitted parameter posteriors are detailed in Table \ref{tab:results} for global variables and Table \ref{tab:results2} for bandpass-dependent variables. A clearer view of the transit fits can be seen in Appendix \ref{app:transit}. The results of the occultation depth analysis are shown in Figure \ref{fig:occultations}.

In the phase curve analysis we included both year 1 and year 2 CHEOPS phase curves. We analysed these years of phase curves separately and we saw no evidence of variability from one CHEOPS year to the next. The change in parameter values from using only year 1 to including both year 1 and 2 was, for the majority of parameters, within $1\sigma$ and the rest was within $2\sigma$. This justified our decision to use the same parameters in the fitting procedure for all CHEOPS phase curves.

\begin{table*}
\caption[]{Global priors and best-fit values for the model and detrending parameters as described in Section~\ref{sec:Analysis}, along with the derived parameters. The reported errors are the 16$^\textnormal{th}$ and 84$^\textnormal{th}$ percentile interval for every parameter. * is for parameters that are only shared between CHEOPS and TESS.}
\label{tab:results}
\centering
\footnotesize
\begin{tabular}{lcccc}
\hline \hline
\multicolumn{2}{c}{Parameter} & Unit & Prior & Best-fit value\\ 
\hline 
\multicolumn{5}{l}{\textit{\textbf{Global}}} \\ 
\uline{\textit{Fitted parameters}} \\
Period & $P$ & days & $\mathcal{N}(1.4811235, 0.0000011)\tablefootmark{(a)}$ & ${1.48111949}\pm{0.00000034}$\\ 
Transit duration & $\textrm{T}_{14}$ & days & $\mathcal{U}(0.13, 0.2)\tablefootmark{(a)}$ & ${0.16552}^{+0.00016}_{-0.00015}$    \\
Impact parameter & $b$ &  & $\mathcal{N}(0.168, 0.017)\tablefootmark{(b)}$ & ${0.195}^{+0.016}_{-0.015}$\\ 
Sky-projected spin orbit angle & $\lambda$ & deg & $\mathcal{N}(-85.78, 0.46)$\tablefootmark{(b)} & ${-85.67}^{+0.46}_{-0.45}$  \\
Gravity-darkening coefficient & $\beta$ &  & $\mathcal{U}(0.01, \infty)$, $\mathcal{N}(0.237, 0.01)$\tablefootmark{(e)} & ${0.2270}^{+0.0077}_{-0.0080}$ \\
GP periodic timescale$^*$ (SHO)& $\rho_\mathrm{SHO}$ & days & $\mathcal{U}$(0.2, 0.4) & ${0.3386}^{+0.0037}_{-0.0036}$\\
Stellar inclination & $i^*$ & deg & $\mathcal{U}(0, 180)$ & ${47.1}\pm1.1$ \\
\uline{\textit{Derived parameters}} \\
Semi major axis & $a$ & R$_\star$ &  & ${3.0914}^{+0.0090}_{-0.0100}$\\ 
Orbital inclination & $i$ & deg &  & ${86.38}^{0.29}_{0.30}$\\ 
Stellar density & $\rho$ & g cm$^{-3}$ & $\mathcal{N}(0.256, 0.33)\tablefootmark{(d)}$  & ${0.2548}^{+0.0022}_{-0.0025}$\\ 
GP damping timescale$^*$ (SHO)& $\tau_\mathrm{SHO}$ & days & & ${0.6772}^{+0.0074}_{0.0072}$\\
Stellar rotational period & $P_\mathrm{rot}$ & hrs & $\mathcal{U}(4.61, \infty)\tablefootmark{(f)}$ & ${18.96}\pm{0.34}$ \\
True spin orbit angle & $\Psi$ & deg & & ${84.36}^{+0.37}_{-0.38}$ \\
\uline{\textit{Fixed parameters}} & & & \uline{Source} & \uline{Value} \\
Stellar radius & $R_\star$ & $R_\odot$ & This work (Section \ref{sec:stel}) & 2.379\\
Effective/polar stellar temperature & $T_\mathrm{eff}$ & K & \cite{Gaudi2017} & 10170\\
Projected rotational velocity & $v\sin i^*$ & km s$^{-1}$ & \cite{Borsa2019} & $111.8$ \\
Eccentricity & $e$ & & assumed & 0\\
Argument of periastron & $\omega$ & deg & assumed & 90 \\
Fluid number & $\alpha$ & & \cite{Morris2021b} & 0.6 \\
Drag frequency & $\omega_\mathrm{drag}$ & & \cite{Morris2021b} & 4.5 \\
Highest present spherical mode & $l_\mathrm{max}$ & & \cite{Morris2021b} & 1 \\
GP Amplitude$^*$ (SHO) & $\sigma_\mathrm{SHO}$ & ppm & & 100 \\
GP Amplitude$^*$ (Mat\'ern) & $\sigma_\mathrm{Matern}$ & ppm & & 200 \\
GP timescale$^*$ (Mat\'ern) & $\rho_\mathrm{Matern}$ & days & & 12 \\
\end{tabular} 
\tablefoot{Priors based on previous results from \tablefoottext{a}{\cite{Wong2020}} \tablefoottext{b}{\cite{Borsa2019}} \tablefoottext{c}{\cite{Ahlers2020}} \tablefoottext{d}{This work (See Section \ref{sec:stel})} \tablefoottext{e}{\cite{Claret2016}} \tablefoottext{f}{Restricted by break-up velocity}}
\end{table*}

\begin{table*}
\caption[]{Bandpass-specific priors and best-fit values for the model and detrending parameters as described in Section~\ref{sec:Analysis}, along with the derived parameters. The reported errors are the 16$^\textnormal{th}$ and 84$^\textnormal{th}$ percentile interval for every parameter.}
\label{tab:results2}
\footnotesize
\centering
\begin{tabular}{lcccc}
\hline \hline
\multicolumn{2}{c}{Parameter} & Unit & Prior & Best-fit value\\ 
\hline
\multicolumn{5}{l}{\textit{\textbf{CHEOPS}}} \\ 
\uline{\textit{Fitted parameters}} \\
Zero transit epoch & $t_\mathrm{0}$ & BJD time - 2459095.2 & $\mathcal{U}$(-0.01, 0.01) & ${-0.003751}^{+0.000060}_{-0.000059}$ \\
Transit depth & \textit{depth} & & $\mathcal{U}$(0.0049, 1) & ${0.006212}^{+0.000019}_{-0.000020}$\\ 
Transit scaling factor & $f_0$ &  & $\mathcal{U}$(0.9, 1.1) &  ${0.99998}^{+0.00012}_{-0.00011}$ \\
$1^{st}$ quadratic limb darkening component & $q_{1}$ &  & $\mathcal{N}$(0.34052, 0.01)\tablefootmark{(a)} & ${0.3299}^{+0.0078}_{-0.0074}$\\ 
$2^{nd}$ quadratic limb darkening component & $q_{2}$ &  & $\mathcal{N}$(0.22030, 0.01)\tablefootmark{(a)} & ${0.2210}^{+0.0086}_{-0.0087}$\\ 
Hotspot offset & $\Delta\phi$ & deg & $\mathcal{U}$(-180, 29) & ${-14.1}\pm{2.4}$\\ 
$C_{m\ell}$ power coefficient & $C_\mathrm{1,1}$ &  & $\mathcal{U}$(0, 1) & ${0.205}^{+0.039}_{-0.028}$ \\ 
Mean background temperature & $\bar{T}$ & K & $\mathcal{U}$(2890, 5780)\tablefootmark{(d)} & ${4060}^{+120}_{-150}$\\ 
\uline{\textit{Derived parameters}} \\
Planetary radius & $R_\mathrm{p}$ & $R_\star$ &  & ${0.07882}^{+0.00012}_{-0.00013}$\\ 
$1^{st}$ quadratic limb darkening component & $u_{1}$ & & & ${0.2541}^{+0.0094}_{-0.0097}$\\ 
$2^{nd}$ quadratic limb darkening component & $u_{2}$ & & & ${0.320}\pm{0.012}$ \\ 

\hline 
\multicolumn{5}{l}{\textit{\textbf{TESS}}} \\ 
\uline{\textit{Fitted parameters}} \\
Zero transit epoch & $t_\mathrm{0}$ & BJD time - 2458693.8 & $\mathcal{U}$(-0.01, 0.01) & ${0.013185}^{+0.000044}_{-0.000045}$\\
Transit depth & \textit{depth} & & $\mathcal{U}$(0.0049, 1) & ${0.006263}\pm0.000016$ \\ 
Transit scaling factor & $f_0$ &  & $\mathcal{U}$(0.9, 1.1) & ${1.00017}\pm0.00012$ \\
$1^{st}$ quadratic limb darkening component & $q_{1}$ &  & $\mathcal{N}$(0.1690, 0.01)\tablefootmark{(b)} & ${0.1541}^{+0.0062}_{-0.0064}$ \\ 
$2^{nd}$ quadratic limb darkening component & $q_{2}$ &  & $\mathcal{N}$(0.2082, 0.01)\tablefootmark{(b)} & ${0.2175}^{+0.0094}_{-0.0097}$\\ 
Hotspot offset & $\Delta\phi$ & deg & $\mathcal{U}$(-180, 29) & ${-12.32}\pm{0.97}$\\ 
$C_{m\ell}$ power coefficient & $C_\mathrm{1,1}$ &  & $\mathcal{U}$(0, 1) & ${0.1884}^{+0.0088}_{-0.0083}$ \\ 
Mean background temperature & $\bar{T}$ & K & $\mathcal{U}$(2890, 5780)\tablefootmark{(d)} & ${3955}^{+40}_{-42}$\\ 
\uline{\textit{Derived parameters}} \\
Planetary radius & $R_\mathrm{p}$ & $R_\star$ &  & ${0.079142}\pm{0.000099}$ \\
$1^{st}$ quadratic limb darkening component & $u_{1}$ &  &  & ${0.1705}^{+0.0076}_{-0.0075}$\\ 
$2^{nd}$ quadratic limb darkening component & $u_{2}$ &  &  & ${0.2216}\pm{0.0096}$\\ 

\hline 
\multicolumn{5}{l}{\textit{\textbf{\Spitzer}}} \\ 
\uline{\textit{Fitted parameters}} \\
Zero transit epoch & $t_\mathrm{0}$ & BJD time - 2458415.36 & $\mathcal{U}$(-0.01, 0.01) & ${0.00206}\pm{0.00015}$\\
Transit depth & \textit{depth} & & $\mathcal{U}$(0.0049, 1) &  ${0.006152}\pm{0.000068}$ \\ 
Transit scaling factor & $f_0$ &  & $\mathcal{U}$(0.9, 1.1) &  ${0.99993}\pm{0.00020}$\\
$1^{st}$ quadratic limb darkening component & $q_{1}$ &  & $\mathcal{N}$(0.0097, 0.01)\tablefootmark{(c)} & ${0.0118}^{+0.0065}_{-0.0054}$  \\ 
$2^{nd}$ quadratic limb darkening component & $q_{2}$ &  & $\mathcal{N}$(0.1463, 0.01)\tablefootmark{(c)} & ${0.1464}^{+0.0099}_{-0.0101}$ \\ 
Hotspot offset & $\Delta\phi$ & deg & $\mathcal{U}$(-180, 29) & ${-18.2}^{+1.7}_{-1.6}$\\ 
$C_{m\ell}$ power coefficient & $C_\mathrm{1,1}$ &  & $\mathcal{U}$(0, 1) & ${0.2453}\pm{0.0092}$\\ 
Mean background temperature & $\bar{T}$ & K & $\mathcal{U}$(2890, 5780)\tablefootmark{(d)} & $3934^{+54}_{-58}$ \\ 
\uline{\textit{Derived parameters}} \\
Planetary radius & $R_\mathrm{p}$ & $R_\star$ &  & ${0.07843}\pm{0.00043}$\\
$1^{st}$ quadratic limb darkening component & $u_{1}$ &  &  & ${0.0318}^{+0.0082}_{-0.0084}$\\ 
$2^{nd}$ quadratic limb darkening component & $u_{2}$ &  &  & ${0.077}^{+0.019}_{-0.020}$ \\ 
\end{tabular} 
\tablefoot{Priors based on previous results from \tablefoottext{a}{\cite{Claret2021}} \tablefoottext{b}{\cite{Claret2017}} \tablefoottext{c}{\cite{Claret2011}} \tablefoottext{d}{\cite{Morris2021b}}}
\end{table*} 

\subsection{Thermal map of KELT-9b}
\label{subsec:thermalmap}
\begin{figure}
    \includegraphics[width=\textwidth/2]{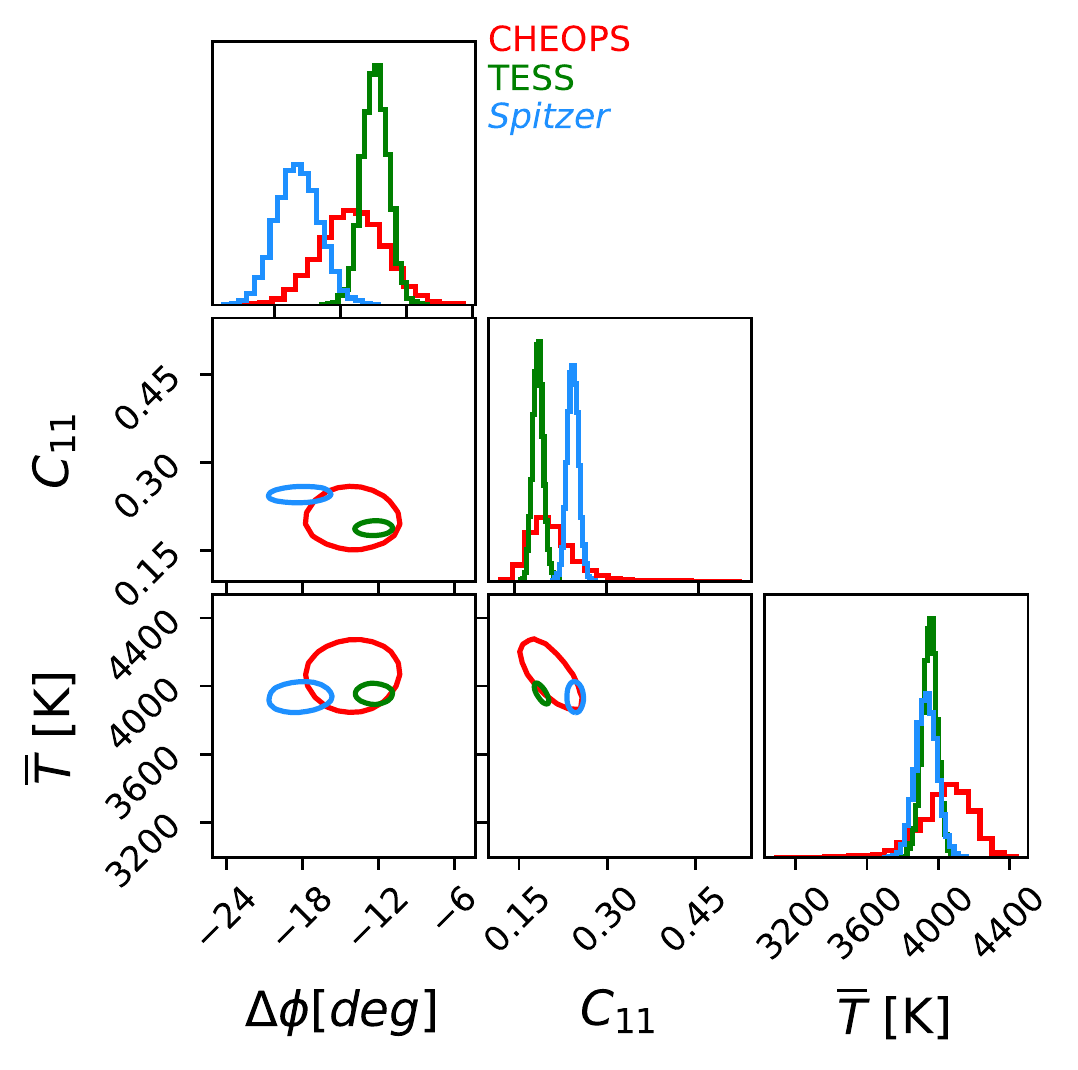}
    \caption{Posteriors of the phase curve parameters across the three bandpasses showing $1\sigma$-confidence contours. Red is CHEOPS, green is TESS and blue is \Spitzer.}
    \label{fig:phasecorner}
\end{figure}

\begin{figure*}
    \includegraphics[width=\textwidth]{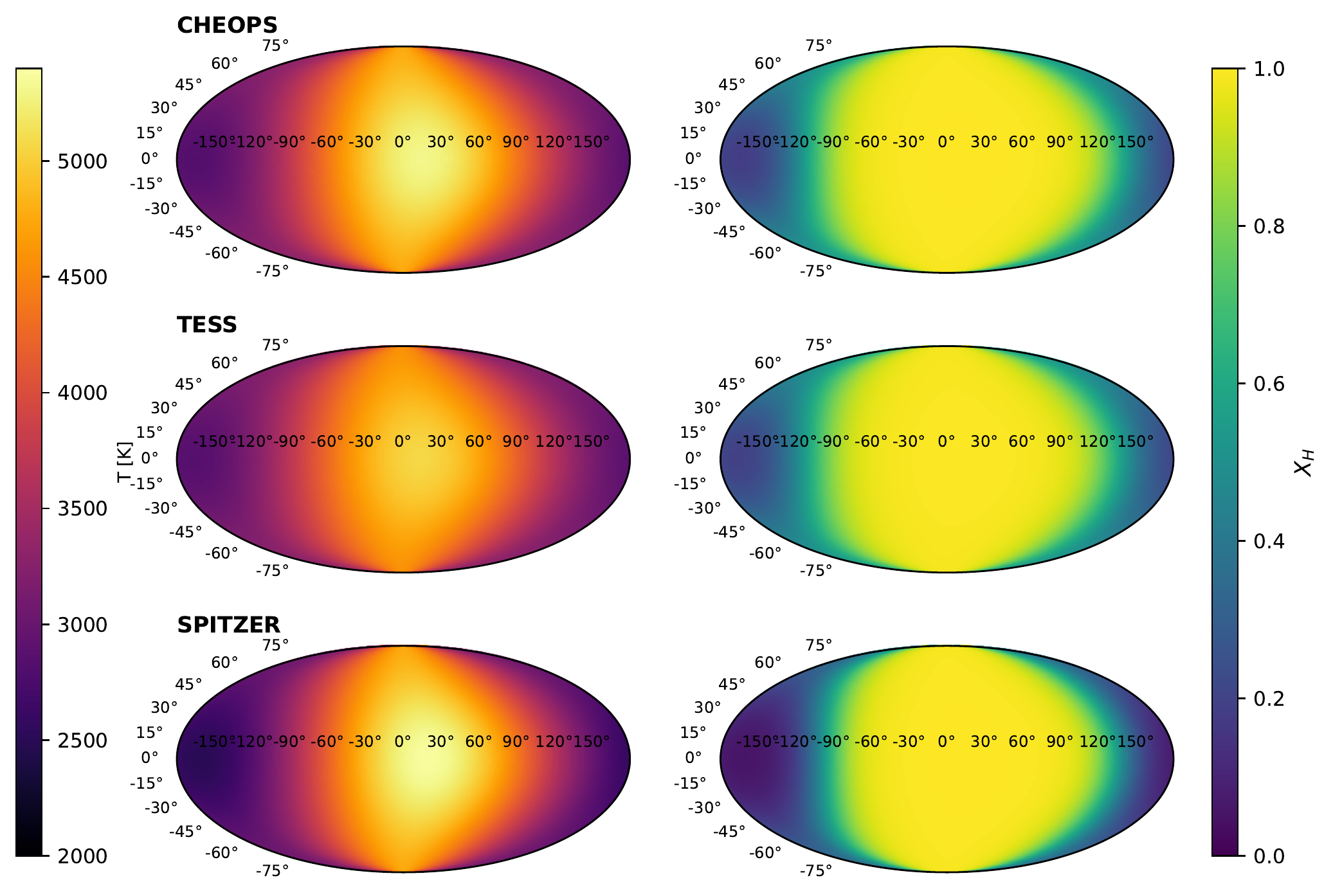}
    \caption{2D thermal temperature maps (left column and colour bar) and maps showing ratio of atomic to molecular hydrogen, assuming chemical equilibrium, that would be present given the temperature maps (right column and colour bar) of KELT-9b in each bandpass. The hottest temperature is reached in the CHEOPS bandpass near the substellar point and is just over 5000 K whereas the lowest temperature is reached in the \Spitzer bandpass at just under 2500 K. On all three maps, the dayside is mostly atomic hydrogen-saturated while across the terminator and over to the nightside, the ratio falls to near zero at the antistellar point.}
    \label{fig:tempmap}
\end{figure*}

As described in Section \ref{sec:kelp}, we use a generalised spherical harmonic temperature map to fit the thermal phase curve variation in the light curves. To a good first-approximation, there are only three free variables that can describe most hot-Jupiter phase curves: $\Delta\phi$ (hotspot offset), $C_\mathrm{1,1}$ ($C_{m\ell}$ power coefficient) and $\bar{T}$ (mean background temperature). As each instrument observes in a different bandpass, they are probing different atmospheric depths of KELT-9b and so we used an independent phase curve model for each bandpass. Although \cite{Hooton2021} used spherical harmonic temperature maps in a similar way to fit the phase variations in CHEOPS and \Spitzer light curves for MASCARA-1b, they used a single set of the same three parameters for bandpasses due to the limited phase coverage of the CHEOPS observations. Figure \ref{fig:tempmap} shows the temperature maps derived from the best-fit phase curve models for each instrument. Similar to other hot Jupiters, the result in all three bandpasses is a hotspot slightly offset to the east. Furthermore, the minimum nightside temperature remains above 2000\,K in all bandpasses, further evidence for significant heat-redistribution in the atmosphere. It is also apparent that in the TESS data, there is a smaller temperature contrast between the maximum and minimum temperature than the other datasets.

In Figure \ref{fig:phasecorner} we illustrate the fitted posteriors of the three free variables in these phase curve models. Table \ref{tab:results2} details the best-fit values of these parameters along with the other bandpass-dependent parameters. It is clear that these parameters change significantly as we probe different depths of KELT-9b's atmosphere.

The CHEOPS nightside posteriors are considerably broader than those of TESS and \Spitzer. The TESS data has around 50 days of photometry compared to the $\sim$10 days of CHEOPS, which contributes to narrower TESS posteriors. The CHEOPS nightside flux (in ppm) is also lower in the CHEOPS observations and so is detected at at lower significance than the other instruments. Furthermore, \Spitzer observes a larger planet-to-star contrast than the other two instruments and also other factors such as a larger collecting area and the longer observing durations all contribute to the narrower posteriors. This may also be related to the fact that we detrend the CHEOPS data simultaneously in the fitting procedure but not the TESS and Spitzer data. However, we try to account for this by fitting for an additional white noise term to each individual data set, modifying the uncertainties based on the standard deviation of the flux values.

In the second column of Figure \ref{fig:tempmap} we show the ratio of atomic hydrogen to molecular hydrogen in the atmosphere of KELT-9b, assuming equilibrium chemistry and using the method and equations described in \cite{Heng2016a}. This supports the theory that the dissociation from atomic to molecular hydrogen occurs near the day-night terminator for the wavelengths observed. 

\subsection{Dayside and nightside brightness temperatures}
\label{subsec:daynightinttemps}

\begin{figure}
    \includegraphics[width=\textwidth/2]{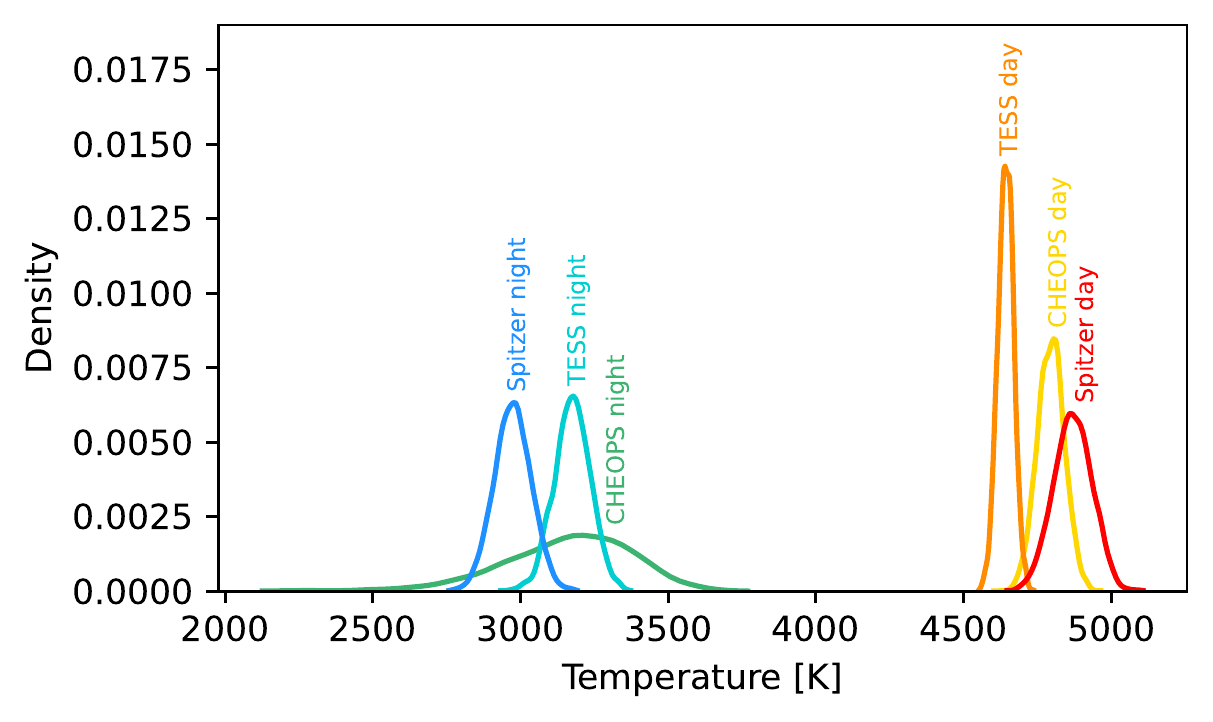}
    \caption{Posteriors of the brightness day and nightside temperatures from the different bandpasses.}
    \label{fig:daynighthist}
\end{figure}

\begin{table*}
\centering
\caption[]{Fitted eclipse depths and night fluxes derived directly from the phase curve fitting procedure for each of the three different bandpasses. The last two columns show the corresponding dayside and nightside brightness temperatures, derived using a PHOENIX model for KELT-9 and the respective bandpass filter functions. }
\label{tab:daynighttemp}
\begin{tabular}{lcccc}
\hline \hline
  & Eclipse Depth [ppm] & Night Flux [ppm] & Dayside Temp [K] & Nightside Temp [K] \\ 
\hline 
CHEOPS & ${367}\pm{17}$ & ${37}^{+16}_{-14}$ & ${4796}\pm{46}$ & ${3180}^{+190}_{-230}$ \\ 
TESS & ${645}\pm{15}$ & ${105}^{+11}_{-11}$ & ${4643}\pm26$ & ${3177}^{+60}_{-61}$\\ 
\Spitzer &  ${3007}^{+57}_{-55}$ & ${1440}^{+53}_{-49}$ & ${4870}^{+67}_{-65}$ & ${2973}^{+66}_{-62}$\\
\hline 
\end{tabular} 
\end{table*} 

Although our work reveals 2-dimensional information regarding the planet's temperature profile, we believe it is useful to still report the dayside and nightside brightness temperatures that we infer directly from two single ('1D') measurements: the eclipse depth and the flux at half an orbital period away from the centre of the eclipse. Assuming a model spectrum for the star, a black body for the planet and by using each instrument's filter function, these brightness temperatures can be calculated. Although this is not the main focus of this paper, it is useful to report these temperatures due to the wide understanding of this temperature statistic in the community, along with the opportunity for direct comparison to previous work. 

We obtain a dayside brightness temperature of CHEOPS, TESS, and \Spitzer of ${4796}\pm{46}$\,K, ${4643}\pm26$\,K and ${4870}^{+67}_{-65}$\,K respectively. We found the nightside brightness temperatures of the three bandpasses (in the same order) to be ${3180}^{+190}_{-230}$\,K, ${3177}^{+60}_{-61}$\,K and ${2973}^{+66}_{-62}$\,K. Figure \ref{fig:daynighthist} shows the posteriors of these wavelength-dependent temperatures (derived from the posteriors of the thermal phase curve parameters). It is clear that the highest average dayside temperature is observed in the \Spitzer bandpass, followed by CHEOPS and then TESS. Generally, the nightside temperatures also have wider posteriors as the lower nightside flux is detected at a lower significance.

Our TESS dayside brightness temperature is consistent with the \cite{Wong2020} value of ${4600}\pm100$\,K and our TESS nightside temperature is also consistent with their nightside temperature of ${3040}\pm100$\,K. This is an encouraging test of our full phase curve model as both of these studies used the same TESS data. For our \Spitzer dayside temperature, our results are consistent within $2\sigma$ to the dayside brightness temperature reported in \cite{Mansfield2020}, ${4566}^{+140}_{-136}$\,K. However our nightside temperature is not consistent with their nightside brightness temperature of ${2556}^{+101}_{-97}$\,K (difference of over $3\sigma$). This could in part be due to the different detrending method applied to \Spitzer data in this study compared to the \cite{Mansfield2020} study.

\begin{figure}
    \includegraphics[width=\textwidth/2]{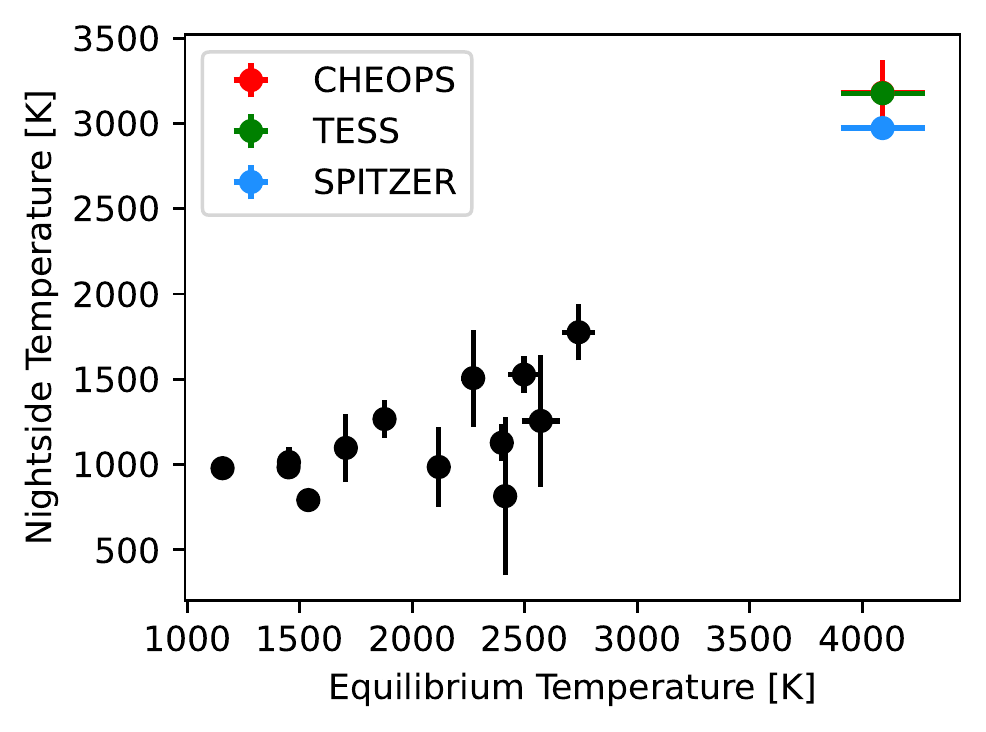}
    \caption{Nightside integrated temperatures of hot Jupiters plotted against their equilibrium temperatures \citep{Keating2019}. The coloured points show the KELT-9b temperatures derived in this study.}
    \label{fig:teqtn}
\end{figure}

Figure \ref{fig:teqtn} shows the nightside temperatures of the most extreme hot Jupiters plotted against their equilibrium temperatures. We define the equilibrium temperature as $T_{eq} = T_* \sqrt{R_*/2a}$. It has been reported in \cite{Keating2019} that due to the presence of clouds the nightside temperature of most hot Jupiters is around 1100 K, with the caveat that clouds would disperse for hotter planets, and therefore the nightside temperature may increase again proportionally to the amount of stellar irradiation. KELT-9b supports this caveat as it is clearly an exception to the pattern of constant nightside temperature, having a nightside temperature exceeding 2900\,K.

\begin{figure}
    \includegraphics[width=\textwidth/2]{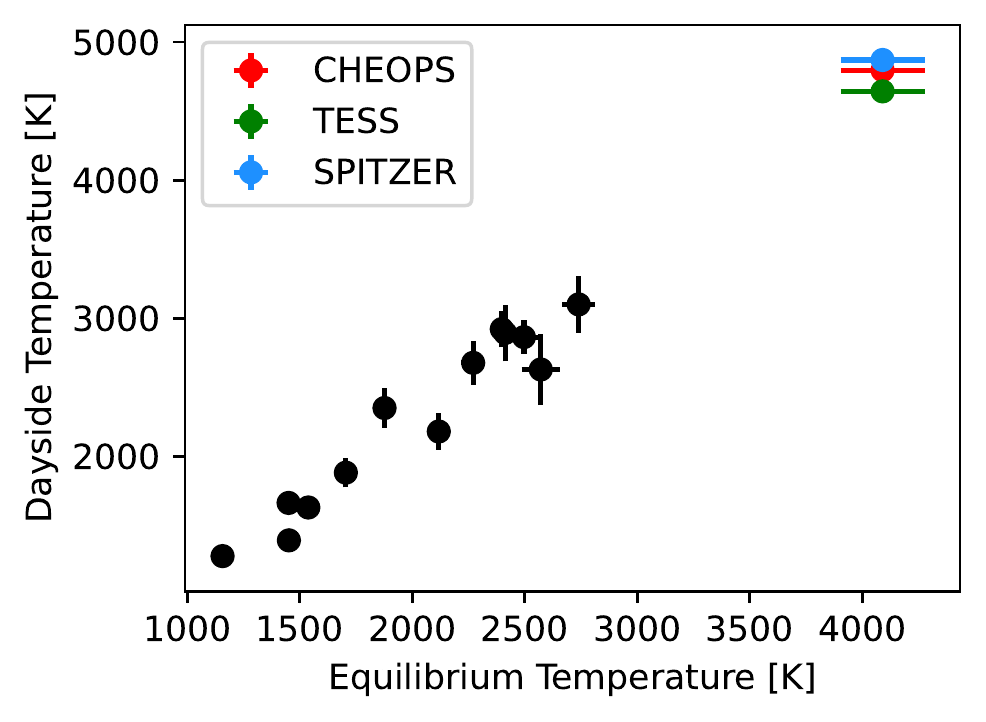}
    \caption{Dayside integrated temperatures of hot Jupiters plotted against their equilibrium temperatures \citep{Keating2019}. The coloured points show the KELT-9b temperatures derived in this study.}
    \label{fig:teqtd}
\end{figure}

\subsection{Eclipse depths}
\label{subsec:eclipsedepths}

\begin{figure*}
    \includegraphics[width=\textwidth]{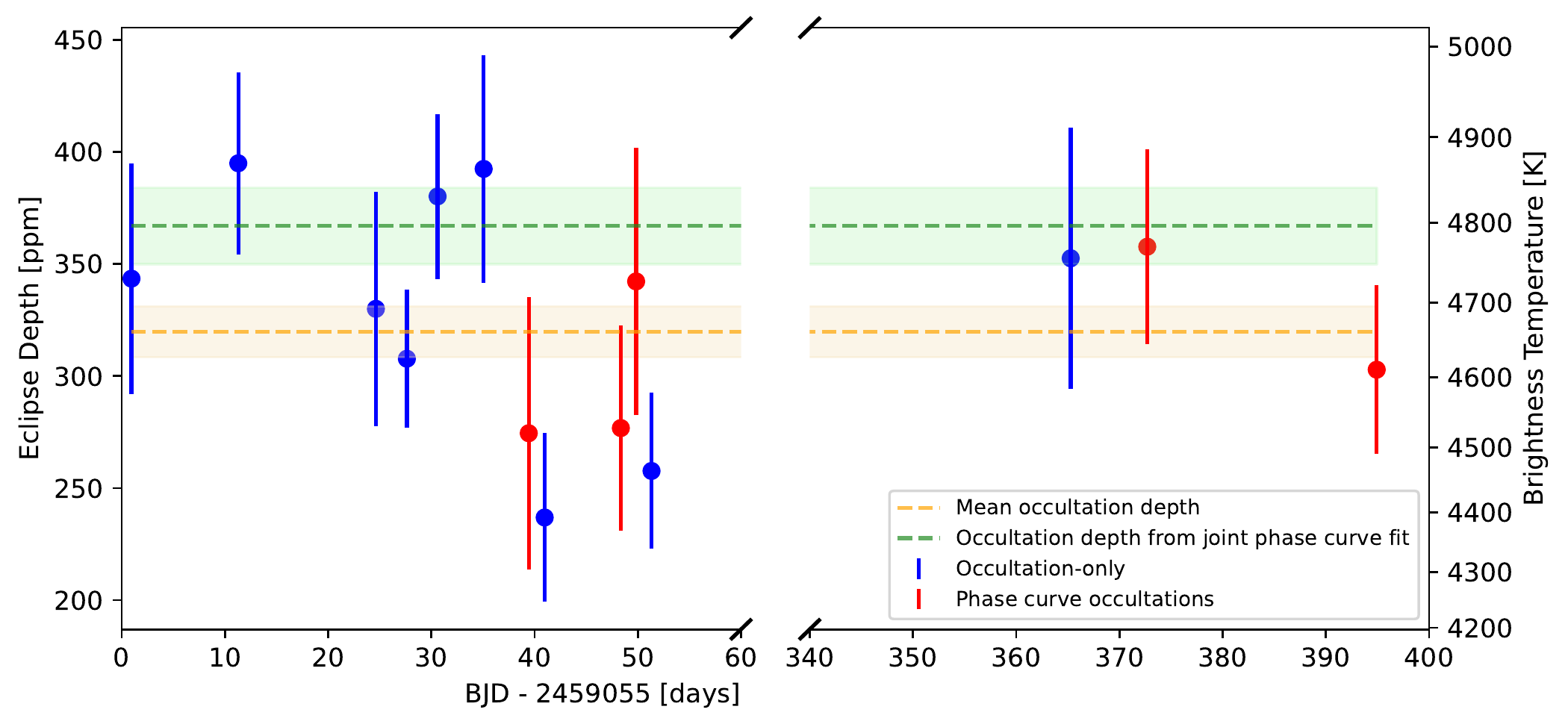}
    \caption{Fitted eclipse depths from the 9 CHEOPS occultation observations (in blue) and 5 CHEOPS occultations extracted from the phase curves (in red). The eclipse depth derived from the joint CHEOPS phase curve fit is shown in green (367 $\pm$ 17 ppm, with $1\sigma$ shaded) and the mean value of these 14 occultations (320 $\pm$ 11 ppm) is shown in orange (with $1\sigma$ shaded). These two mean eclipse depths are consistent with each other by just over $2\sigma$. The right-hand y-axis shows how the eclipse depths convert to dayside brightness temperatures, assuming the dayside hemisphere is a blackbody and radiates at a uniform temperature.}
    \label{fig:occultations}
\end{figure*}

Figure \ref{fig:occultations} shows the eclipse depths fitted from the CHEOPS occultation observations. As explained in Section \ref{sec:occultations}, this dataset includes 9 occultation-only observations and four occultations cut-out from the four CHEOPS phase curves. Together they have a mean eclipse depth of $320 \pm 11$\,ppm. This is consistent within $2.3\sigma$ of the eclipse depth found from the full phase curve fit of the four CHEOPS phase curves which was ${367}\pm{17}$\,ppm. The mean eclipse depth variation (from the occultations) corresponds to a brightness temperature change of around $\pm 33$\,K.

To validate our eclipse model uncertainties, we fit the same size error bar to each eclipse depth datapoint at the same time as performing a straight-line fit to the eclipse depths. This fitted error bar amplitude was similar to the average of our occultation fit error bars, therefore we suggest that the error bars are appropriate to justify our results.

As described in Section \ref{sec:occultations}, fitting the eclipses is particularly challenging due to lack of data out-of-eclipse. This makes finding the correct phase and amplitude of the stellar variations, which we know are present from the CHEOPS, TESS, and \Spitzer phase curves, very difficult. As the amplitude and half-period of these variations are on the same order as the eclipse depth and duration, this inflates the uncertainty in the eclipse depth measurements, and including them in the eclipse depth fit allows the reported error bars to reflect this uncertainty. This may also be an explanation as to why the mean occultation depth is less than the estimated occultation depth from the full phase curve-only fits, as the stellar pulsations being fitted can mimic the dip of the eclipse and produce a good fit with an anomalously low eclipse depth. It is worth noting that this difference is probably not due to the difference in modelling of the out-of-eclipse flux between the occultation-only observations and the occultations within the phase curves as we used a phase curve model in the occultation-only analysis as well (see Section \ref{subsubsec:eclipse}). In models without this stellar pulsation model, the eclipse depths we retrieve are very different from one another and from the analysis with the pulsation model, and the errors reported were considerably smaller than the scatter in the depths. In this case it is clear that phase curve observations have been vital for informing the priors and model of the occultation-only observations. In future CHEOPS projects, one must be extremely cautious when working with occultations from variable and pulsating stars. Phase curves are essential in this case to constrain this stellar source of variability, due to the very limited baseline of occultation observations.

These eclipse depths, especially with the addition of the three occultations observed a year later, suggests a lack of significant variability in the atmosphere of KELT-9b. The variation in eclipse depths is roughly consistent with the error bar of the mean depth. Therefore we set an upper limit of temperature variability of KELT-9b at 1$\%$ of the mean brightness temperature ($\sim1\sigma$). This observation is consistent with the lack of variability observed by \Spitzer for HD 189733 b by \citet{Agol2010}, for HD 189733 b and HD 209458 b by \citet{Kilpatrick2020}, and with theoretical expectations by, for example, \citet{Showman2009} and \citet{Komacek2020}.

\subsection{Albedo and heat redistribution}
\label{subsec:albedo}

A thermal phase curve constrains the global dayside and nightside brightness temperatures. If the atmosphere radiates like a perfect blackbody, then the brightness temperature is equal to the real temperature, despite the thermal phase curve being observed only within a limited range of wavelengths. The traditional approach is to then use a 0D 'box model' to convert these temperatures into the Bond albedo and heat redistribution efficiency, e.g., equations (4) and (5) of \cite{Cowan2011b}. We do not expect KELT-9b to radiate like a perfect blackbody, but our approach for converting the temperature map to fluxes assumes a Planck function \citep{Morris2021b}. Generally, the CHEOPS, TESS, and \Spitzer thermal phase curves are probing different atmospheric layers (across radial distance or atmospheric pressure), which are described by different temperature maps.

We utilise the method described in \cite{Morris2021b} to use the entire 2D temperature maps derived from the phase curve fitting to calculate these values. Following this, the Bond albedo is defined as
\begin{equation}
    A_{\rm B} = 1 - \left( \frac{a}{R_\star} \right)^2 \frac{\int^{\pi}_{-\pi} \int^{\pi}_0 F_p(\theta,\phi) \sin\theta ~d\theta ~d\phi}{\pi \sigma T_\star^4},
\label{eq:Bond_albedo}
\end{equation}
where $\sigma$ is the Stefan-Boltzmann constant and $F_p(\theta,\phi)$ is the flux from the planet derived from the temperature map. We must note here that the Bond albedo is a wavelength-independent quantity, however in this paper we quote three Bond albedos, one derived from each temperature map. We do this by assuming that each temperature element from the temperature map behaves like a blackbody. For each temperature element the flux of that element is estimated to be equal to $F_p(\theta,\phi) = \sigma T_p^4(\theta,\phi)$. This is not a perfect assumption, and so the Bond albedos reported from each temperature map may vary slightly (and can go negative) due to non-blackbody behaviour of the atmosphere. For example, if, at a certain wavelength, we are observing a strong absorption feature, then the estimated bolometric flux and brightness temperature may be less than the actual bolometric flux, as we assume that the small wavelength-band we observe is representative of the entire spectrum. The Bond albedos reported using the temperature map from each bandpass is reported in Table \ref{tab:albedo}. 

Considering the Bond albedo reported from each temperature map, there is strong evidence to suggest it is consistent with zero. This result is expected due to the extreme level of irradiation on the planet, which would contribute to a highly extended atmosphere where light entering from the direction of the star would have a very low probability of escaping before being absorbed. This therefore implies KELT-9b predictably, behaves similar to a blackbody. 

The heat redistribution parameter of the atmosphere is defined as the ratio of the nightside flux to the dayside flux and describes the extent to which heat is transported around the planet. As KELT-9b is tidally locked, this parameter is entirely dependent on the dynamics and chemistry of the atmosphere. It is also derived using information from the entire temperature map using the equation from \cite{Morris2021b}:
\begin{equation}
\varepsilon = \frac{\int^{\pi}_{\pi/2} \int^{\pi}_{0} F_p \sin\theta ~d\theta ~d\phi + \int^{-\pi/2}_{-\pi} \int^{\pi}_{0} F_p \sin\theta ~d\theta ~d\phi}{\int^{\pi/2}_{-\pi/2} \int^{\pi}_0 F_p \sin\theta ~d\theta ~d\phi}.
\end{equation}

\begin{figure}
    \includegraphics[width=\textwidth/2]{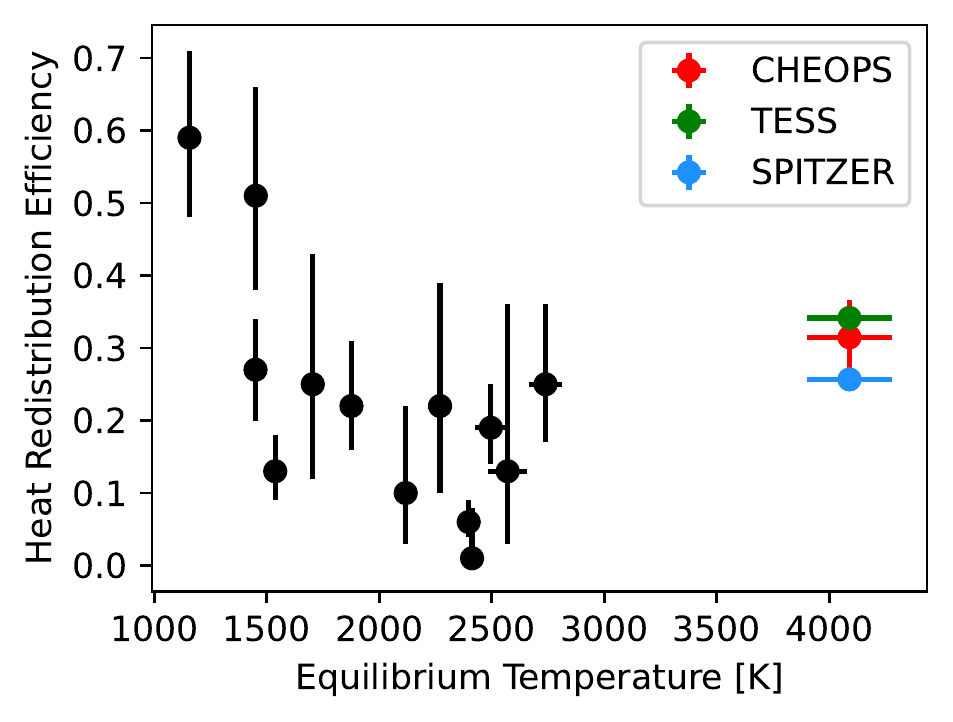}
    \caption{Heat redistribution efficiency of hot Jupiters plotted against their equilibrium temperatures. The coloured points show the KELT-9b efficiency, $\varepsilon$, derived in this study.}
    \label{fig:teqeps}
\end{figure}

Figure \ref{fig:teqeps} shows the heat redistribution ($\varepsilon$) of KELT-9b plotted along with other hot-Jupiters. In previous papers \citep{Wallack2021, Komacek2016}, it has been shown that for hot Jupiters, the incident stellar flux is the primary decider of the level of $\varepsilon$. However in \cite{Bell2018}, they predict a rising heat redistribution for UHJ due to the $H_2$ dissociation and recombination increasing the heat transport around the planet. Our work agrees with this theory and in Figure \ref{fig:teqeps}, it appears that $\varepsilon$ does indeed fall with planet equilibrium temperatures up to around 2500\,K, however after that the $\varepsilon$ rises again with temperature and KELT-9b's $\varepsilon$ calculated in this paper supports this trend.

\begin{table}
\centering
\caption[]{Bond albedo and heat redistribution in the three bandpasses.}
\label{tab:albedo}
\begin{tabular}{lcc}
\hline \hline
Bandpass & $A_B$ & $\varepsilon$ \\ 
\hline 
CHEOPS & ${-0.168}^{+0.096}_{-0.092}$ & ${0.314}^{+0.051}_{-0.058}$ \\ 
TESS & ${-0.025}\pm{0.032}$ & ${0.342}^{+0.016}_{-0.017}$ \\ 
\Spitzer & ${-0.109}\pm0.068$ & ${0.257}\pm{0.013}$\\ 
\hline 
\end{tabular} 
\end{table}

\subsection{Comparison with \Spitzer/TESS literature}
\label{subsec:litcomparison}

\subsubsection{TESS}
We find that for the phase curve parameters, our TESS hotspot offset of $-12.32\pm0.97^\circ$ is inconsistent with the offset reported in \cite{Wong2020} (hereafter \citetalias{Wong2020}) ($5.2\pm0.9^\circ$) by over 5$\sigma$ (N.B. this error is a combination of the uncertainty in this paper's value and \citetalias{Wong2020}). However we obtain larger values for the offset in all passbands, and the \Spitzer result is consistent with previous analyses. This lends credibility to our TESS measurement. For the other orbital parameters our planetary radius ($R_p = 0.079142\pm{0.000099}$\,$R_\star$) in the TESS bandpass is consistent with \citetalias{Wong2020}.

For some of the global parameters, our best-fit impact parameter ($b = 0.195^{+0.016}_{-0.015}$) differs from the \citetalias{Wong2020} value by 1.4$\sigma$. Our best-fit semi-major axis ($a = 3.0914^{+0.0090}_{-0.0100}$\,$R_\star$) is different from the \citetalias{Wong2020} value by around 4$\sigma$ and our best-fit period ($P = 1.48111949\pm{0.00000034}$\,days) is different to \citetalias{Wong2020}'s value by 3.5$\sigma$. Finally, our best-fit orbital inclination ($i = {86.38^{+0.29}_{-0.30}}^\circ$) is just over 1.5$\sigma$ away from the inclination reported in \citetalias{Wong2020} and 1.6$\sigma$ away from \cite{Ahlers2020}. 

We find that for the TESS phase curves, the eclipse depth reported in this paper from the phase curve fit ($645\pm{15}$\,ppm) is consistent with the eclipse depth found in \citetalias{Wong2020} ($650^{+14}_{-15}$\,ppm). As discussed in Section \ref{subsec:daynightinttemps}, from our fitted eclipse depth and nightside flux we derived a TESS dayside brightness temperature of $4643\pm{26}$\,K which is also consistent with \citetalias{Wong2020} ($4600\pm100$\,K). Our TESS nightside brightness temperature of $3177^{+60}_{-61}$\,K is consistent within $1.2\sigma$ of \citetalias{Wong2020}'s nightside temperature ($3040\pm100$\,K). Following from this, in this work we have used the method in \citep{Morris2021b} to derive the Bond albedo and heat redistribution efficiency. Our values differ with the Bond albedo in \citetalias{Wong2020} by 2.3$\sigma$ and the heat redistribution parameter is consistent within 1$\sigma$. The gravity-darkening parameters will be discussed in Section \ref{subsec:GravDark}.

\subsubsection{\Spitzer}

For the orbital and system parameters, the planetary radius and transit duration differ with the same parameters reported in \cite{Mansfield2020} (hereafter \citetalias{Mansfield2020}) by $2.1\sigma$ and $6\sigma$ respectively, indicative of the different reduction methods influencing the derived parameters.

For the phase curve parameters, the \Spitzer hotspot offset reported in this paper (${-18.2^{+1.7}_{-1.6}}^\circ$ eastwards) is consistent the offset reported in \citetalias{Mansfield2020} (${-18.7^{+2.1}_{-2.3}}^\circ$). However as mentioned in Section \ref{subsec:daynightinttemps}, our nightside brightness temperature is inconsistent with the nightside brightness temperature reported in \citetalias{Mansfield2020} by 3.5$\sigma$. The dayside brightness temperature reported in this paper is consistent with \citetalias{Mansfield2020}'s value by just under 2$\sigma$.

\subsubsection{Gravity-darkened transits}
\label{subsec:GravDark}

We find a sky-projected spin orbit angle of $-85.67^{+0.46}_{-0.45}$\,degrees which is consistent with the spin orbit angle reported in \cite{Ahlers2020} (hereafter \citetalias{Ahlers2020}), and \cite{Gaudi2017}. Our value of the gravity-darkening coefficient, $\beta$,  ($0.2270^{+0.0077}_{-0.0080}$) is not consistent with \citetalias{Ahlers2020}, but is consistent with the \cite{Claret2016} theoretical value. Our best-fit stellar inclination, $i^* = 47.1\pm{1.1}^\circ$ (note this is related to the stellar obliquity $\phi$, specified in \citetalias{Ahlers2020}, by $i^* = 90^\circ - \phi$), and stellar rotation period ($P_{\text{rot}}= 18.96\pm{0.34}$\,hours) are consistent with the value reported in \citetalias{Ahlers2020}.

Using the fitted values for $i^*$, $i$ and $\lambda$ we used Equation \ref{eqn:truespinorbit} to calculate a value for the true spin orbit angle of ${{84.36}^{+0.37}_{-0.38}}^\circ$. This is within 1$\sigma$ of the value \citetalias{Ahlers2020} reported.

\section{Discussion}
\label{sec:discussion}
\subsection{Challenges of simulating UHJs}

Tidally locked, highly irradiated exoplanets, including UHJs, are complex, three-dimensional objects. General circulation models (GCMs), which are numerical solvers of the three-dimensional fluid equations, have been adapted to study hot Jupiters \citep[see, e.g.:][]{Showman2009, Rauscher2010, Heng2011, Kataria2013, Mayne2014}. Recently, GCMs have been used to study UHJs \citep[e.g.:][]{Tan2019}.

The higher temperatures of UHJs present additional technical challenges for GCMs. As we have shown in Figure \ref{fig:tempmap}, the atmosphere of KELT-9b transitions from temperatures where it is dominated by atomic hydrogen on its dayside to being dominated by molecular hydrogen on its nightside, verifying the prediction of \cite{Bell2018}. Since the specific heat capacity at constant pressure, $c_P$, varies as the reciprocal of the mean molecular mass $m$, i.e., $c_P \propto 1/m$, it implies that $c_P$ changes by a factor of 2 within the atmosphere of an UHJ, unlike for a regular hot Jupiter where it is roughly constant. The transformation from atomic to molecular hydrogen (and back) means that an additional cooling/heating term needs to be inserted into the governing equations \citep{Bell2018, Tan2019}. KELT-9b is an extreme example of these processes occurring in UHJs and correctly reproducing the observed dayside-to-nightside flux redistribution will require them to be simulated correctly. The CHEOPS, TESS, and \Spitzer phase curves presented in the current study, as well as their associated temperature maps, will provide valuable constraints for future GCM studies of KELT-9b.

Another important constraint provided by the current work is that the climate of KELT-9b is somewhat stable: over around 270 orbital periods, the globally averaged dayside temperature varies by less than 50 K (see Section \ref{subsec:eclipsedepths} on results of occultation analysis). These results are consistent with theoretical expectations produced from GCM analysis in works such as \cite{Showman2009} and \cite{Komacek2020}. Simulating variability accurately using GCMs is challenging, as it depends on the choice of governing equations \citep[e.g.:][]{Cho2008}, numerical dissipation \citep[e.g.:][]{Heng2011, Thrastarson2011} and choice of bottom boundary condition \citep[e.g.:][]{Liu2013}. In particular, our inability to specify numerical dissipation, which is often required to numerically stabilise GCM runs, from first principles \citep{Heng2011} implies that energy and momentum conservation, and therefore our ability to accurately predict wind speeds and variability, is limited \citep{Goodman2009}.

In these ways, observations of UHJs lead our current ability to simulate them, with KELT-9b presenting the most extreme case study. The empirical constraints derived here therefore provide important checks on future GCMs of UHJs.

\subsection{Anticipating JWST multi-wavelength phase curves}

Formally, the spherical and Bond albedos are monochromatic and bolometric quantities, respectively.  The spherical albedo is the monochromatic version of the Bond albedo (monochromatic version of equation \ref{eq:Bond_albedo}). One of the key limitations of the current study is that the measured CHEOPS, TESS, and \Spitzer phase curves are neither monochromatic nor bolometric.  Essentially, our ability to extract temperatures and Bond albedos is based on an extrapolation: the assumption that the spectral energy distribution (SED) of KELT-9b follows a perfect blackbody.  Generally, a non-blackbody SED sampled in the CHEOPS versus \Spitzer bandpasses will return different brightness temperatures.  These differences translate into differences in the inferred Bond albedos that we report in Table \ref{tab:albedo}.

In the era of the James Webb Space Telescope (JWST), multi-wavelength phase curves, which will be monochromatic to a good approximation, will become the norm.  Two questions concerning the data analysis and interpretation of JWST thermal phase curves arise.

\subsubsection{May we still use box models to extract Bond albedos from JWST thermal phase curves?}

Generally, the heating of an atmosphere by starlight occurs in the near-ultraviolet to optical range of wavelengths, which is processed and re-emitted in the infrared range of wavelengths as thermal emission.  (As already mentioned in the introduction, KELT-9b is a special case where its dayside thermal emission radiates in the optical.)  While it is possible to quantify the wavelength-dependent flux of starlight incident upon the \textit{top} of the atmosphere, it is much more challenging to describe how much starlight penetrates to each atmospheric layer as a function of wavelength.  Any such attempt will be model-dependent and probabilistic (in a Bayesian sense).  

If one integrates over all wavelengths, the bolometric flux of starlight is simply given by the Stefan-Boltzmann law.  If the set of JWST thermal phase curves covers the entire wavelength range of the SED, then one may empirically derive the bolometric thermal flux emitted by the atmosphere.  The Bond albedo may then be inferred without any assumption on the nature of the SED of the exoplanetary atmosphere, i.e., it is not necessary to assume a blackbody SED.  Essentially, one bypasses the need for a 0D box model.

Alternatively, one could still perform an analysis like we have done in this paper and obtain different Bond albedos for every JWST thermal phase curve, assuming a blackbody SED for each surface element of the planet. However, as previously explained, this assumption will not be perfect for a planet with non-blackbody behaviour. Therefore the former approach, assuming the JWST phase curves cover the entire planetary SED, would be much more accurate to determine the total bolometric flux of the planet, and therefore a more accurate (single value) of the Bond albedo.

\subsubsection{Is it possible to extract spherical albedos from JWST thermal phase curves?}

As mentioned, the spherical albedo is the monochromatic version of the Bond albedo. At first thought, if JWST can produce monochromatic phase curves, it is easy to suggest this could translate into a set of spherical albedos for the planet. However this would be incorrect. In principle, if one knew exactly how much starlight was deposited in a specific atmospheric layer, then one could compare that to the thermal emission from that layer and derive the spherical albedo.  However, as previously described this would be a model-dependent exercise that involves some assumption on the model atmosphere in order to perform radiative transfer from the top of the atmosphere to the layer in question.

Specifically, equation (\ref{eq:Bond_albedo}) describes our approach for deriving the Bond albedo from a 2D temperature map.  Despite the CHEOPS, TESS, and \Spitzer thermal phase curves being neither monochromatic nor bolometric, this equation allows one to work with temperatures because of the assumption of a blackbody SED.  When JWST multi-wavelength thermal phase curves are available, the numerator of equation (\ref{eq:Bond_albedo}) may be generalised such that $F_p$ is empirically derived from the data with no need to assume a blackbody SED.  However, the denominator cannot be generalised in a straightforward way as $\sigma T^4_\star$ needs to be replaced by the flux of starlight deposited in the same atmospheric layer.  In principle, it is possible to solve for these wavelength-dependent fluxes within a holistic framework that simultaneously interprets phase-dependent emission spectra and wavelength-dependent phase curves.  Such a framework would have to account for scattered starlight versus thermal emission as functions of wavelength. However, as explained in the previous subsection, calculating the Bond albedo would still be possible from the JWST thermal phase curves alone, provided they sufficiently spanned the planetary SED. As the Bond albedo describes the planet's input and output energy bolometrically, it removes the need to undertake the difficult task of probing the individual atmospheric layers.

\section{Conclusion}
In this work we have simultaneously analysed CHEOPS phase curves as well as public phase curves from TESS and \Spitzer to infer joint constraints on the phase curve variation, gravity-darkened transits and occultation depth in the three bandpasses (Figure \ref{fig:phasecurves}). From this analysis we find the following results:

\begin{itemize}
    \item We derive 2D temperature maps of the atmosphere at three different depths, and calculate dayside and nightside brightness temperatures of the planet in each bandpass (Figure \ref{fig:tempmap}, Table \ref{tab:daynighttemp} and Sections \ref{subsec:thermalmap} and \ref{subsec:daynightinttemps}).
    \item The day-night heat redistribution of $\sim\,$0.3 confirms theoretical expectations of enhanced energy transfer to the planetary nightside due to dissociation and recombination of molecular hydrogen in ultra-hot Jupiters (Figure \ref{fig:teqeps}, Table \ref{tab:albedo} and Section \ref{subsec:albedo}).
    \item We also find a Bond albedo consistent with zero (Table \ref{tab:albedo} and Section \ref{subsec:albedo}).
    \item We also analyse 9 CHEOPS occultations of KELT-9 and find no evidence of variability of the brightness temperature of the planet, excluding variability greater than 1\% (1$\sigma$) (Figure \ref{fig:occultations} and Section \ref{subsec:eclipsedepths}).
\end{itemize}

\begin{acknowledgements}
    We gratefully acknowledge the open source software which made this work possible: \texttt{astropy} \citep{Astropy2013, Astropy2018}, \texttt{ipython} \citep{ipython}, \texttt{numpy} \citep{numpy}, \texttt{scipy} \citep{scipy}, \texttt{matplotlib} \citep{matplotlib}, \texttt{emcee} \citep{Foreman-Mackey2013}, \texttt{batman} \citep{Kreidberg2015}, \texttt{PyMC3} \citep{pymc3}. This research has made use of the SVO Filter Profile Service (http://svo2.cab.inta-csic.es/theory/fps/) supported from the Spanish MINECO through grant AYA2017-84089.
    CHEOPS is an ESA mission in partnership with Switzerland with important contributions to the payload and the ground segment from Austria, Belgium, France, Germany, Hungary, Italy, Portugal, Spain, Sweden, and the United Kingdom. The CHEOPS Consortium would like to gratefully acknowledge the support received by all the agencies, offices, universities, and industries involved. Their flexibility and willingness to explore new approaches were essential to the success of this mission. 
    B.-O.D. acknowledges support from the Swiss National Science Foundation (PP00P2-190080). 
    This project has received funding from the European Research Council (ERC) under the European Union’s Horizon 2020 research and innovation programme (project {\sc Four Aces}. 
    grant agreement No 724427). It has also been carried out in the frame of the National Centre for Competence in Research PlanetS supported by the Swiss National Science Foundation (SNSF). DE acknowledges financial support from the Swiss National Science Foundation for project 200021\_200726. 
    SH gratefully acknowledges CNES funding through the grant 837319. 
    ML acknowledges support of the Swiss National Science Foundation under grant number PCEFP2\_194576. 
    This work was supported by FCT - Fundação para a Ciência e a Tecnologia through national funds and by FEDER through COMPETE2020 - Programa Operacional Competitividade e Internacionalizacão by these grants: UID/FIS/04434/2019, UIDB/04434/2020, UIDP/04434/2020, PTDC/FIS-AST/32113/2017 \& POCI-01-0145-FEDER- 032113, PTDC/FIS-AST/28953/2017 \& POCI-01-0145-FEDER-028953, PTDC/FIS-AST/28987/2017 \& POCI-01-0145-FEDER-028987, O.D.S.D. is supported in the form of work contract (DL 57/2016/CP1364/CT0004) funded by national funds through FCT. 
    S.G.S. acknowledge support from FCT through FCT contract nr. CEECIND/00826/2018 and POPH/FSE (EC). 
    ACC and TW acknowledge support from STFC consolidated grant numbers ST/R000824/1 and ST/V000861/1, and UKSA grant number ST/R003203/1. 
    YA and MJH acknowledge the support of the Swiss National Fund under grant 200020\_172746. 
    We acknowledge support from the Spanish Ministry of Science and Innovation and the European Regional Development Fund through grants ESP2016-80435-C2-1-R, ESP2016-80435-C2-2-R, PGC2018-098153-B-C33, PGC2018-098153-B-C31, ESP2017-87676-C5-1-R, MDM-2017-0737 Unidad de Excelencia Maria de Maeztu-Centro de Astrobiologí­a (INTA-CSIC), as well as the support of the Generalitat de Catalunya/CERCA programme. The MOC activities have been supported by the ESA contract No. 4000124370. 
    S.C.C.B. acknowledges support from FCT through FCT contracts nr. IF/01312/2014/CP1215/CT0004. 
    XB, SC, DG, MF and JL acknowledge their role as ESA-appointed CHEOPS science team members. 
    ABr was supported by the SNSA. 
    ACC acknowledges support from STFC consolidated grant numbers ST/R000824/1 and ST/V000861/1, and UKSA grant number ST/R003203/1. 
    This project was supported by the CNES. 
    The Belgian participation to CHEOPS has been supported by the Belgian Federal Science Policy Office (BELSPO) in the framework of the PRODEX Program, and by the University of Liège through an ARC grant for Concerted Research Actions financed by the Wallonia-Brussels Federation. 
    L.D. is an F.R.S.-FNRS Postdoctoral Researcher. 
    MF and CMP gratefully acknowledge the support of the Swedish National Space Agency (DNR 65/19, 174/18). 
    DG gratefully acknowledges financial support from the CRT foundation under Grant No. 2018.2323 ``Gaseousor rocky? Unveiling the nature of small worlds''. 
    M.G. is an F.R.S.-FNRS Senior Research Associate. 
    KGI is the ESA CHEOPS Project Scientist and is responsible for the ESA CHEOPS Guest Observers Programme. She does not participate in, or contribute to, the definition of the Guaranteed Time Programme of the CHEOPS mission through which observations described in this paper have been taken, nor to any aspect of target selection for the programme. 
    This work was granted access to the HPC resources of MesoPSL financed by the Region Ile de France and the project Equip@Meso (reference ANR-10-EQPX-29-01) of the programme Investissements d'Avenir supervised by the Agence Nationale pour la Recherche. 
    PM acknowledges support from STFC research grant number ST/M001040/1. 
    GSc, GPi, IPa, LBo, VNa, RRa and GBr acknowledge support from CHEOPS ASI-INAF agreement n. 2019-29-HH.0. 
    This work was also partially supported by a grant from the Simons Foundation (PI Queloz, grant number 327127). 
    IRI acknowledges support from the Spanish Ministry of Science and Innovation and the European Regional Development Fund through grant PGC2018-098153-B- C33, as well as the support of the Generalitat de Catalunya/CERCA programme. 
    GyMSz acknowledges the support of the Hungarian National Research, Development and Innovation Office (NKFIH) grant K-125015, a a PRODEX Experiment Agreement No. 4000137122, the Lend\"ulet LP2018-7/2021 grant of the Hungarian Academy of Science and the support of the city of Szombathely. 
    V.V.G. is an F.R.S-FNRS Research Associate. 
    NAW acknowledges UKSA grant ST/R004838/1. 
 
\end{acknowledgements}

\bibliographystyle{aa} 
\bibliography{bibliography} 

\begin{thebibliography}{99}
\expandafter\ifx\csname natexlab\endcsname\relax\def\natexlab#1{#1}\fi

\bibitem[{{Agol} {et~al.}(2010){Agol}, {Cowan}, {Knutson}, {Deming}, {Steffen},
  {Henry}, \& {Charbonneau}}]{Agol2010}
{Agol}, E., {Cowan}, N.~B., {Knutson}, H.~A., {et~al.} 2010, \apj, 721, 1861

\bibitem[{{Ahlers} {et~al.}(2020){Ahlers}, {Johnson}, {Stassun}, {Col{\'o}n},
  {Barnes}, {Stevens}, {Beatty}, {Gaudi}, {Collins}, {Rodriguez}, {Ricker},
  {Vanderspek}, {Latham}, {Seager}, {Winn}, {Jenkins}, {Caldwell}, {Goeke},
  {Osborn}, {Paegert}, {Rowden}, \& {Tenenbaum}}]{Ahlers2020}
{Ahlers}, J.~P., {Johnson}, M.~C., {Stassun}, K.~G., {et~al.} 2020, The
  Astronomical Journal, 160, 4

\bibitem[{{Arcangeli} {et~al.}(2018){Arcangeli}, {D{\'e}sert}, {Line}, {Bean},
  {Parmentier}, {Stevenson}, {Kreidberg}, {Fortney}, {Mansfield}, \&
  {Showman}}]{Arcangeli2018}
{Arcangeli}, J., {D{\'e}sert}, J.-M., {Line}, M.~R., {et~al.} 2018, The
  Astrophysical Journal, 855, L30

\bibitem[{{Arcangeli} {et~al.}(2019){Arcangeli}, {D{\'e}sert}, {Parmentier},
  {Stevenson}, {Bean}, {Line}, {Kreidberg}, {Fortney}, \&
  {Showman}}]{Arcangeli2019}
{Arcangeli}, J., {D{\'e}sert}, J.-M., {Parmentier}, V., {et~al.} 2019,
  Astronomy and Astrophysics, 625, A136

\bibitem[{{Armstrong} {et~al.}(2016){Armstrong}, {de Mooij}, {Barstow},
  {Osborn}, {Blake}, \& {Saniee}}]{Armstrong2016}
{Armstrong}, D.~J., {de Mooij}, E., {Barstow}, J., {et~al.} 2016, Nature
  Astronomy, 1, 0004

\bibitem[{{Astropy Collaboration} {et~al.}(2018){Astropy Collaboration},
  {Price-Whelan}, {Sip{\H{o}}cz}, {G{\"u}nther}, {Lim}, {Crawford}, {Conseil},
  {Shupe}, {Craig}, {Dencheva}, {Ginsburg}, {Vand erPlas}, {Bradley},
  {P{\'e}rez-Su{\'a}rez}, {de Val-Borro}, {Aldcroft}, {Cruz}, {Robitaille},
  {Tollerud}, {Ardelean}, {Babej}, {Bach}, {Bachetti}, {Bakanov}, {Bamford},
  {Barentsen}, {Barmby}, {Baumbach}, {Berry}, {Biscani}, {Boquien}, {Bostroem},
  {Bouma}, {Brammer}, {Bray}, {Breytenbach}, {Buddelmeijer}, {Burke},
  {Calderone}, {Cano Rodr{\'\i}guez}, {Cara}, {Cardoso}, {Cheedella}, {Copin},
  {Corrales}, {Crichton}, {D'Avella}, {Deil}, {Depagne}, {Dietrich}, {Donath},
  {Droettboom}, {Earl}, {Erben}, {Fabbro}, {Ferreira}, {Finethy}, {Fox},
  {Garrison}, {Gibbons}, {Goldstein}, {Gommers}, {Greco}, {Greenfield},
  {Groener}, {Grollier}, {Hagen}, {Hirst}, {Homeier}, {Horton}, {Hosseinzadeh},
  {Hu}, {Hunkeler}, {Ivezi{\'c}}, {Jain}, {Jenness}, {Kanarek}, {Kendrew},
  {Kern}, {Kerzendorf}, {Khvalko}, {King}, {Kirkby}, {Kulkarni}, {Kumar},
  {Lee}, {Lenz}, {Littlefair}, {Ma}, {Macleod}, {Mastropietro}, {McCully},
  {Montagnac}, {Morris}, {Mueller}, {Mumford}, {Muna}, {Murphy}, {Nelson},
  {Nguyen}, {Ninan}, {N{\"o}the}, {Ogaz}, {Oh}, {Parejko}, {Parley}, {Pascual},
  {Patil}, {Patil}, {Plunkett}, {Prochaska}, {Rastogi}, {Reddy Janga},
  {Sabater}, {Sakurikar}, {Seifert}, {Sherbert}, {Sherwood-Taylor}, {Shih},
  {Sick}, {Silbiger}, {Singanamalla}, {Singer}, {Sladen}, {Sooley},
  {Sornarajah}, {Streicher}, {Teuben}, {Thomas}, {Tremblay}, {Turner},
  {Terr{\'o}n}, {van Kerkwijk}, {de la Vega}, {Watkins}, {Weaver}, {Whitmore},
  {Woillez}, {Zabalza}, \& {Astropy Contributors}}]{Astropy2018}
{Astropy Collaboration}, {Price-Whelan}, A.~M., {Sip{\H{o}}cz}, B.~M., {et~al.}
  2018, \aj, 156, 123

\bibitem[{{Astropy Collaboration} {et~al.}(2013){Astropy Collaboration},
  {Robitaille}, {Tollerud}, {Greenfield}, {Droettboom}, {Bray}, {Aldcroft},
  {Davis}, {Ginsburg}, {Price-Whelan}, {Kerzendorf}, {Conley}, {Crighton},
  {Barbary}, {Muna}, {Ferguson}, {Grollier}, {Parikh}, {Nair}, {Unther},
  {Deil}, {Woillez}, {Conseil}, {Kramer}, {Turner}, {Singer}, {Fox}, {Weaver},
  {Zabalza}, {Edwards}, {Azalee Bostroem}, {Burke}, {Casey}, {Crawford},
  {Dencheva}, {Ely}, {Jenness}, {Labrie}, {Lim}, {Pierfederici}, {Pontzen},
  {Ptak}, {Refsdal}, {Servillat}, \& {Streicher}}]{Astropy2013}
{Astropy Collaboration}, {Robitaille}, T.~P., {Tollerud}, E.~J., {et~al.} 2013,
  \aap, 558, A33

\bibitem[{{Barnes}(2009)}]{Barnes2009}
{Barnes}, J.~W. 2009, The Astrophysical Journal, 705, 683

\bibitem[{{Bell} \& {Cowan}(2018)}]{Bell2018}
{Bell}, T.~J. \& {Cowan}, N.~B. 2018, The Astrophysical Journal, 857, L20

\bibitem[{{Bello-Arufe} {et~al.}(2022){Bello-Arufe}, {Buchhave},
  {Mendon{\c{c}}a}, {Tronsgaard}, {Heng}, {Hoeijmakers}, \&
  {Mayo}}]{Bello-Arufe2022}
{Bello-Arufe}, A., {Buchhave}, L.~A., {Mendon{\c{c}}a}, J.~M., {et~al.} 2022,
  arXiv e-prints, arXiv:2203.04969

\bibitem[{{Benz} {et~al.}(2021){Benz}, {Broeg}, {Fortier}, {Rando}, {Beck},
  {Beck}, {Queloz}, {Ehrenreich}, {Maxted}, {Isaak}, {Billot}, {Alibert},
  {Alonso}, {Ant{\'o}nio}, {Asquier}, {Bandy}, {B{\'a}rczy}, {Barrado},
  {Barros}, {Baumjohann}, {Bekkelien}, {Bergomi}, {Biondi}, {Bonfils},
  {Borsato}, {Brandeker}, {Busch}, {Cabrera}, {Cessa}, {Charnoz}, {Chazelas},
  {Collier Cameron}, {Corral Van Damme}, {Cortes}, {Davies}, {Deleuil},
  {Deline}, {Delrez}, {Demangeon}, {Demory}, {Erikson}, {Farinato}, {Fossati},
  {Fridlund}, {Futyan}, {Gandolfi}, {Garcia Munoz}, {Gillon}, {Guterman},
  {Gutierrez}, {Hasiba}, {Heng}, {Hernandez}, {Hoyer}, {Kiss}, {Kovacs},
  {Kuntzer}, {Laskar}, {Lecavelier des Etangs}, {Lendl}, {L{\'o}pez}, {Lora},
  {Lovis}, {L{\"u}ftinger}, {Magrin}, {Malvasio}, {Marafatto}, {Michaelis}, {de
  Miguel}, {Modrego}, {Munari}, {Nascimbeni}, {Olofsson}, {Ottacher},
  {Ottensamer}, {Pagano}, {Palacios}, {Pall{\'e}}, {Peter}, {Piazza}, {Piotto},
  {Pizarro}, {Pollaco}, {Ragazzoni}, {Ratti}, {Rauer}, {Ribas}, {Rieder},
  {Rohlfs}, {Safa}, {Salatti}, {Santos}, {Scandariato}, {S{\'e}gransan},
  {Simon}, {Smith}, {Sordet}, {Sousa}, {Steller}, {Szab{\'o}}, {Szoke},
  {Thomas}, {Tschentscher}, {Udry}, {Van Grootel}, {Viotto}, {Walter},
  {Walton}, {Wildi}, \& {Wolter}}]{Benz2021}
{Benz}, W., {Broeg}, C., {Fortier}, A., {et~al.} 2021, Experimental Astronomy,
  51, 109

\bibitem[{{Blackwell} \& {Shallis}(1977)}]{Blackwell1977}
{Blackwell}, D.~E. \& {Shallis}, M.~J. 1977, \mnras, 180, 177

\bibitem[{{Bonfanti} {et~al.}(2021){Bonfanti}, {Delrez}, {Hooton}, {Wilson},
  {Fossati}, {Alibert}, {Hoyer}, {Mustill}, {Osborn}, {Adibekyan}, {Gandolfi},
  {Salmon}, {Sousa}, {Tuson}, {Van Grootel}, {Cabrera}, {Nascimbeni}, {Maxted},
  {Barros}, {Billot}, {Bonfils}, {Borsato}, {Broeg}, {Davies}, {Deleuil},
  {Demangeon}, {Fridlund}, {Lacedelli}, {Lendl}, {Persson}, {Santos},
  {Scandariato}, {Szab{\'o}}, {Collier Cameron}, {Udry}, {Benz}, {Beck},
  {Ehrenreich}, {Fortier}, {Isaak}, {Queloz}, {Alonso}, {Asquier}, {Bandy},
  {B{\'a}rczy}, {Barrado}, {Barrag{\'a}n}, {Baumjohann}, {Beck}, {Bekkelien},
  {Bergomi}, {Brandeker}, {Busch}, {Cessa}, {Charnoz}, {Chazelas}, {Corral Van
  Damme}, {Demory}, {Erikson}, {Farinato}, {Futyan}, {Garcia Mu{\~n}oz},
  {Gillon}, {Guedel}, {Guterman}, {Hasiba}, {Heng}, {Hernandez}, {Kiss},
  {Kuntzer}, {Laskar}, {Lecavelier des Etangs}, {Lovis}, {Magrin}, {Malvasio},
  {Marafatto}, {Michaelis}, {Munari}, {Olofsson}, {Ottacher}, {Ottensamer},
  {Pagano}, {Pall{\'e}}, {Peter}, {Piazza}, {Piotto}, {Pollacco}, {Ragazzoni},
  {Rando}, {Ratti}, {Rauer}, {Ribas}, {Rieder}, {Rohlfs}, {Safa}, {Salatti},
  {S{\'e}gransan}, {Simon}, {Smith}, {Sordet}, {Steller}, {Thomas},
  {Tschentscher}, {Van Eylen}, {Viotto}, {Walter}, {Walton}, {Wildi}, \&
  {Wolter}}]{bonfanti21}
{Bonfanti}, A., {Delrez}, L., {Hooton}, M.~J., {et~al.} 2021, \aap, 646, A157

\bibitem[{{Bonfanti} {et~al.}(2016){Bonfanti}, {Ortolani}, \&
  {Nascimbeni}}]{bonfanti16}
{Bonfanti}, A., {Ortolani}, S., \& {Nascimbeni}, V. 2016, \aap, 585, A5

\bibitem[{{Bonfanti} {et~al.}(2015){Bonfanti}, {Ortolani}, {Piotto}, \&
  {Nascimbeni}}]{bonfanti15}
{Bonfanti}, A., {Ortolani}, S., {Piotto}, G., \& {Nascimbeni}, V. 2015, \aap,
  575, A18

\bibitem[{{Borsa} {et~al.}(2019){Borsa}, {Rainer}, {Bonomo}, {Barbato},
  {Fossati}, {Malavolta}, {Nascimbeni}, {Lanza}, {Esposito}, {Affer},
  {Andreuzzi}, {Benatti}, {Biazzo}, {Bignamini}, {Brogi}, {Carleo}, {Claudi},
  {Cosentino}, {Covino}, {Damasso}, {Desidera}, {Garrido Rubio}, {Giacobbe},
  {Gonz{\'a}lez-{\'A}lvarez}, {Harutyunyan}, {Knapic}, {Leto}, {Ligi},
  {Maggio}, {Maldonado}, {Mancini}, {Fiorenzano}, {Masiero}, {Micela},
  {Molinari}, {Pagano}, {Pedani}, {Piotto}, {Pino}, {Poretti}, {Scandariato},
  {Smareglia}, \& {Sozzetti}}]{Borsa2019}
{Borsa}, F., {Rainer}, M., {Bonomo}, A.~S., {et~al.} 2019, Astronomy and
  Astrophysics, 631, A34

\bibitem[{{Castelli} \& {Kurucz}(2003)}]{Castelli2003}
{Castelli}, F. \& {Kurucz}, R.~L. 2003, in IAU Symposium, Vol. 210, Modelling
  of Stellar Atmospheres, ed. N.~{Piskunov}, W.~W. {Weiss}, \& D.~F. {Gray},
  A20

\bibitem[{{Cho} {et~al.}(2008){Cho}, {Menou}, {Hansen}, \& {Seager}}]{Cho2008}
{Cho}, J. Y.~K., {Menou}, K., {Hansen}, B. M.~S., \& {Seager}, S. 2008, The
  Astrophysical Journal, 675, 817

\bibitem[{{Claret}(2016)}]{Claret2016}
{Claret}, A. 2016, Astronomy and Astrophysics, 588, A15

\bibitem[{{Claret}(2017)}]{Claret2017}
{Claret}, A. 2017, Astronomy and Astrophysics, 600, A30

\bibitem[{{Claret}(2021)}]{Claret2021}
{Claret}, A. 2021, VizieR Online Data Catalog (other), 0690, J/other/RNAAS/5

\bibitem[{{Claret} \& {Bloemen}(2011)}]{Claret2011}
{Claret}, A. \& {Bloemen}, S. 2011, Astronomy and Astrophysics, 529, A75

\bibitem[{{Cowan} \& {Agol}(2011)}]{Cowan2011b}
{Cowan}, N.~B. \& {Agol}, E. 2011, The Astrophysical Journal, 729, 54

\bibitem[{{Deline} {et~al.}(2022){Deline}, {Hooton}, {Lendl}, {Morris},
  {Salmon}, {Olofsson}, {Broeg}, {Ehrenreich}, {Beck}, {Brandeker}, {Hoyer},
  {Sulis}, {Van Grootel}, {Bourrier}, {Demangeon}, {Demory}, {Heng},
  {Parviainen}, {Serrano}, {Singh}, {Bonfanti}, {Fossati}, {Kitzmann}, {Sousa},
  {Wilson}, {Alibert}, {Alonso}, {Anglada}, {B{\'a}rczy}, {Barrado Navascues},
  {Barros}, {Baumjohann}, {Beck}, {Bekkelien}, {Benz}, {Billot}, {Bonfils},
  {Cabrera}, {Charnoz}, {Collier Cameron}, {Corral van Damme}, {Csizmadia},
  {Davies}, {Deleuil}, {Delrez}, {de Roche}, {Erikson}, {Fortier}, {Fridlund},
  {Futyan}, {Gandolfi}, {Gillon}, {G{\"u}del}, {Gutermann}, {Hasiba}, {Isaak},
  {Kiss}, {Laskar}, {Lecavelier des Etangs}, {Lovis}, {Magrin}, {Maxted},
  {Munari}, {Nascimbeni}, {Ottensamer}, {Pagano}, {Pall{\'e}}, {Peter},
  {Piotto}, {Pollacco}, {Queloz}, {Ragazzoni}, {Rando}, {Rauer}, {Ribas},
  {Santos}, {Scandariato}, {S{\'e}gransan}, {Simon}, {Smith}, {Steller},
  {Szab{\'o}}, {Thomas}, {Udry}, {Walter}, \& {Walton}}]{Deline2022}
{Deline}, A., {Hooton}, M.~J., {Lendl}, M., {et~al.} 2022, arXiv e-prints,
  arXiv:2201.04518

\bibitem[{{Demory} {et~al.}(2016{\natexlab{a}}){Demory}, {Gillon}, {de Wit},
  {Madhusudhan}, {Bolmont}, {Heng}, {Kataria}, {Lewis}, {Hu}, {Krick},
  {Stamenkovi{\'c}}, {Benneke}, {Kane}, \& {Queloz}}]{Demory2016b}
{Demory}, B.-O., {Gillon}, M., {de Wit}, J., {et~al.} 2016{\natexlab{a}}, \nat,
  532, 207

\bibitem[{{Demory} {et~al.}(2016{\natexlab{b}}){Demory}, {Gillon},
  {Madhusudhan}, \& {Queloz}}]{Demory2016a}
{Demory}, B.-O., {Gillon}, M., {Madhusudhan}, N., \& {Queloz}, D.
  2016{\natexlab{b}}, \mnras, 455, 2018

\bibitem[{{Foreman-Mackey}(2018)}]{celerite2}
{Foreman-Mackey}, D. 2018, Research Notes of the American Astronomical Society,
  2, 31

\bibitem[{{Foreman-Mackey} {et~al.}(2017){Foreman-Mackey}, {Agol},
  {Ambikasaran}, \& {Angus}}]{celerite1}
{Foreman-Mackey}, D., {Agol}, E., {Ambikasaran}, S., \& {Angus}, R. 2017, \aj,
  154, 220

\bibitem[{{Foreman-Mackey} {et~al.}(2013){Foreman-Mackey}, {Hogg}, {Lang}, \&
  {Goodman}}]{Foreman-Mackey2013}
{Foreman-Mackey}, D., {Hogg}, D.~W., {Lang}, D., \& {Goodman}, J. 2013,
  Publications of the Astronomical Society of the Pacific, 125, 306

\bibitem[{{Gai} \& {Knuth}(2018)}]{Gai2018}
{Gai}, A.~D. \& {Knuth}, K.~H. 2018, The Astrophysical Journal, 853, 49

\bibitem[{{Gaia Collaboration} {et~al.}(2021){Gaia Collaboration}, {Brown},
  {Vallenari}, {Prusti}, {de Bruijne}, {Babusiaux}, {Biermann}, {Creevey},
  {Evans}, {Eyer}, {Hutton}, {Jansen}, {Jordi}, {Klioner}, {Lammers},
  {Lindegren}, {Luri}, {Mignard}, {Panem}, {Pourbaix}, {Randich}, {Sartoretti},
  {Soubiran}, {Walton}, {Arenou}, {Bailer-Jones}, {Bastian}, {Cropper},
  {Drimmel}, {Katz}, {Lattanzi}, {van Leeuwen}, {Bakker}, {Cacciari},
  {Casta{\~n}eda}, {De Angeli}, {Ducourant}, {Fabricius}, {Fouesneau},
  {Fr{\'e}mat}, {Guerra}, {Guerrier}, {Guiraud}, {Jean-Antoine Piccolo},
  {Masana}, {Messineo}, {Mowlavi}, {Nicolas}, {Nienartowicz}, {Pailler},
  {Panuzzo}, {Riclet}, {Roux}, {Seabroke}, {Sordo}, {Tanga}, {Th{\'e}venin},
  {Gracia-Abril}, {Portell}, {Teyssier}, {Altmann}, {Andrae}, {Bellas-Velidis},
  {Benson}, {Berthier}, {Blomme}, {Brugaletta}, {Burgess}, {Busso}, {Carry},
  {Cellino}, {Cheek}, {Clementini}, {Damerdji}, {Davidson}, {Delchambre},
  {Dell'Oro}, {Fern{\'a}ndez-Hern{\'a}ndez}, {Galluccio}, {Garc{\'\i}a-Lario},
  {Garcia-Reinaldos}, {Gonz{\'a}lez-N{\'u}{\~n}ez}, {Gosset}, {Haigron},
  {Halbwachs}, {Hambly}, {Harrison}, {Hatzidimitriou}, {Heiter},
  {Hern{\'a}ndez}, {Hestroffer}, {Hodgkin}, {Holl}, {Jan{\ss}en}, {Jevardat de
  Fombelle}, {Jordan}, {Krone-Martins}, {Lanzafame}, {L{\"o}ffler}, {Lorca},
  {Manteiga}, {Marchal}, {Marrese}, {Moitinho}, {Mora}, {Muinonen}, {Osborne},
  {Pancino}, {Pauwels}, {Petit}, {Recio-Blanco}, {Richards}, {Riello},
  {Rimoldini}, {Robin}, {Roegiers}, {Rybizki}, {Sarro}, {Siopis}, {Smith},
  {Sozzetti}, {Ulla}, {Utrilla}, {van Leeuwen}, {van Reeven}, {Abbas}, {Abreu
  Aramburu}, {Accart}, {Aerts}, {Aguado}, {Ajaj}, {Altavilla}, {{\'A}lvarez},
  {{\'A}lvarez Cid-Fuentes}, {Alves}, {Anderson}, {Anglada Varela}, {Antoja},
  {Audard}, {Baines}, {Baker}, {Balaguer-N{\'u}{\~n}ez}, {Balbinot}, {Balog},
  {Barache}, {Barbato}, {Barros}, {Barstow}, {Bartolom{\'e}}, {Bassilana},
  {Bauchet}, {Baudesson-Stella}, {Becciani}, {Bellazzini}, {Bernet}, {Bertone},
  {Bianchi}, {Blanco-Cuaresma}, {Boch}, {Bombrun}, {Bossini}, {Bouquillon},
  {Bragaglia}, {Bramante}, {Breedt}, {Bressan}, {Brouillet}, {Bucciarelli},
  {Burlacu}, {Busonero}, {Butkevich}, {Buzzi}, {Caffau}, {Cancelliere},
  {C{\'a}novas}, {Cantat-Gaudin}, {Carballo}, {Carlucci}, {Carnerero},
  {Carrasco}, {Casamiquela}, {Castellani}, {Castro-Ginard}, {Castro Sampol},
  {Chaoul}, {Charlot}, {Chemin}, {Chiavassa}, {Cioni}, {Comoretto}, {Cooper},
  {Cornez}, {Cowell}, {Crifo}, {Crosta}, {Crowley}, {Dafonte}, {Dapergolas},
  {David}, {David}, {de Laverny}, {De Luise}, {De March}, {De Ridder}, {de
  Souza}, {de Teodoro}, {de Torres}, {del Peloso}, {del Pozo}, {Delbo},
  {Delgado}, {Delgado}, {Delisle}, {Di Matteo}, {Diakite}, {Diener},
  {Distefano}, {Dolding}, {Eappachen}, {Edvardsson}, {Enke}, {Esquej}, {Fabre},
  {Fabrizio}, {Faigler}, {Fedorets}, {Fernique}, {Fienga}, {Figueras},
  {Fouron}, {Fragkoudi}, {Fraile}, {Franke}, {Gai}, {Garabato},
  {Garcia-Gutierrez}, {Garc{\'\i}a-Torres}, {Garofalo}, {Gavras}, {Gerlach},
  {Geyer}, {Giacobbe}, {Gilmore}, {Girona}, {Giuffrida}, {Gomel}, {Gomez},
  {Gonzalez-Santamaria}, {Gonz{\'a}lez-Vidal}, {Granvik},
  {Guti{\'e}rrez-S{\'a}nchez}, {Guy}, {Hauser}, {Haywood}, {Helmi}, {Hidalgo},
  {Hilger}, {H{\l}adczuk}, {Hobbs}, {Holland}, {Huckle}, {Jasniewicz},
  {Jonker}, {Juaristi Campillo}, {Julbe}, {Karbevska}, {Kervella}, {Khanna},
  {Kochoska}, {Kontizas}, {Kordopatis}, {Korn}, {Kostrzewa-Rutkowska},
  {Kruszy{\'n}ska}, {Lambert}, {Lanza}, {Lasne}, {Le Campion}, {Le Fustec},
  {Lebreton}, {Lebzelter}, {Leccia}, {Leclerc}, {Lecoeur-Taibi}, {Liao},
  {Licata}, {Lindstr{\o}m}, {Lister}, {Livanou}, {Lobel}, {Madrero Pardo},
  {Managau}, {Mann}, {Marchant}, {Marconi}, {Marcos Santos}, {Marinoni},
  {Marocco}, {Marshall}, {Martin Polo}, {Mart{\'\i}n-Fleitas}, {Masip},
  {Massari}, {Mastrobuono-Battisti}, {Mazeh}, {McMillan}, {Messina},
  {Michalik}, {Millar}, {Mints}, {Molina}, {Molinaro}, {Moln{\'a}r},
  {Montegriffo}, {Mor}, {Morbidelli}, {Morel}, {Morris}, {Mulone}, {Munoz},
  {Muraveva}, {Murphy}, {Musella}, {Noval}, {Ord{\'e}novic}, {Orr{\`u}},
  {Osinde}, {Pagani}, {Pagano}, {Palaversa}, {Palicio}, {Panahi}, {Pawlak},
  {Pe{\~n}alosa Esteller}, {Penttil{\"a}}, {Piersimoni}, {Pineau}, {Plachy},
  {Plum}, {Poggio}, {Poretti}, {Poujoulet}, {Pr{\v{s}}a}, {Pulone}, {Racero},
  {Ragaini}, {Rainer}, {Raiteri}, {Rambaux}, {Ramos}, {Ramos-Lerate}, {Re
  Fiorentin}, {Regibo}, {Reyl{\'e}}, {Ripepi}, {Riva}, {Rixon}, {Robichon},
  {Robin}, {Roelens}, {Rohrbasser}, {Romero-G{\'o}mez}, {Rowell}, {Royer},
  {Rybicki}, {Sadowski}, {Sagrist{\`a} Sell{\'e}s}, {Sahlmann}, {Salgado},
  {Salguero}, {Samaras}, {Sanchez Gimenez}, {Sanna}, {Santove{\~n}a},
  {Sarasso}, {Schultheis}, {Sciacca}, {Segol}, {Segovia}, {S{\'e}gransan},
  {Semeux}, {Shahaf}, {Siddiqui}, {Siebert}, {Siltala}, {Slezak}, {Smart},
  {Solano}, {Solitro}, {Souami}, {Souchay}, {Spagna}, {Spoto}, {Steele},
  {Steidelm{\"u}ller}, {Stephenson}, {S{\"u}veges}, {Szabados}, {Szegedi-Elek},
  {Taris}, {Tauran}, {Taylor}, {Teixeira}, {Thuillot}, {Tonello}, {Torra},
  {Torra}, {Turon}, {Unger}, {Vaillant}, {van Dillen}, {Vanel}, {Vecchiato},
  {Viala}, {Vicente}, {Voutsinas}, {Weiler}, {Wevers}, {Wyrzykowski}, {Yoldas},
  {Yvard}, {Zhao}, {Zorec}, {Zucker}, {Zurbach}, \&
  {Zwitter}}]{GaiaCollaboration2021}
{Gaia Collaboration}, {Brown}, A.~G.~A., {Vallenari}, A., {et~al.} 2021, \aap,
  649, A1

\bibitem[{{Gaudi} {et~al.}(2017){Gaudi}, {Stassun}, {Collins}, {Beatty},
  {Zhou}, {Latham}, {Bieryla}, {Eastman}, {Siverd}, {Crepp}, {Gonzales},
  {Stevens}, {Buchhave}, {Pepper}, {Johnson}, {Colon}, {Jensen}, {Rodriguez},
  {Bozza}, {Novati}, {D'Ago}, {Dumont}, {Ellis}, {Gaillard}, {Jang-Condell},
  {Kasper}, {Fukui}, {Gregorio}, {Ito}, {Kielkopf}, {Manner}, {Matt}, {Narita},
  {Oberst}, {Reed}, {Scarpetta}, {Stephens}, {Yeigh}, {Zambelli}, {Fulton},
  {Howard}, {James}, {Penny}, {Bayliss}, {Curtis}, {Depoy}, {Esquerdo},
  {Gould}, {Joner}, {Kuhn}, {Labadie-Bartz}, {Lund}, {Marshall}, {McLeod},
  {Pogge}, {Relles}, {Stockdale}, {Tan}, {Trueblood}, \&
  {Trueblood}}]{Gaudi2017}
{Gaudi}, B.~S., {Stassun}, K.~G., {Collins}, K.~A., {et~al.} 2017, Nature, 546,
  514

\bibitem[{{Gillon} {et~al.}(2012){Gillon}, {Demory}, {Benneke}, {Valencia},
  {Deming}, {Seager}, {Lovis}, {Mayor}, {Pepe}, {Queloz}, {S{\'e}gransan}, \&
  {Udry}}]{Gillon2012}
{Gillon}, M., {Demory}, B.~O., {Benneke}, B., {et~al.} 2012, Astronomy and
  Astrophysics, 539, A28

\bibitem[{{Gillon} {et~al.}(2017){Gillon}, {Triaud}, {Demory}, {Jehin}, {Agol},
  {Deck}, {Lederer}, {de Wit}, {Burdanov}, {Ingalls}, {Bolmont}, {Leconte},
  {Raymond}, {Selsis}, {Turbet}, {Barkaoui}, {Burgasser}, {Burleigh}, {Carey},
  {Chaushev}, {Copperwheat}, {Delrez}, {Fernandes}, {Holdsworth}, {Kotze}, {Van
  Grootel}, {Almleaky}, {Benkhaldoun}, {Magain}, \& {Queloz}}]{Gillon2017}
{Gillon}, M., {Triaud}, A.~H.~M.~J., {Demory}, B.-O., {et~al.} 2017, \nat, 542,
  456

\bibitem[{{Goodman}(2009)}]{Goodman2009}
{Goodman}, J. 2009, The Astrophysical Journal, 693, 1645

\bibitem[{Harris {et~al.}(2020)Harris, Millman, van~der Walt, Gommers,
  Virtanen, Cournapeau, Wieser, Taylor, Berg, Smith, Kern, Picus, Hoyer, van
  Kerkwijk, Brett, Haldane, del R{\'{i}}o, Wiebe, Peterson,
  G{\'{e}}rard-Marchant, Sheppard, Reddy, Weckesser, Abbasi, Gohlke, \&
  Oliphant}]{numpy}
Harris, C.~R., Millman, K.~J., van~der Walt, S.~J., {et~al.} 2020, Nature, 585,
  357

\bibitem[{{Heng} {et~al.}(2016){Heng}, {Lyons}, \& {Tsai}}]{Heng2016a}
{Heng}, K., {Lyons}, J.~R., \& {Tsai}, S.-M. 2016, The Astrophysical Journal,
  816, 96

\bibitem[{{Heng} {et~al.}(2011){Heng}, {Menou}, \& {Phillipps}}]{Heng2011}
{Heng}, K., {Menou}, K., \& {Phillipps}, P.~J. 2011, Monthly Notices of the
  Royal Astronomical Society, 413, 2380

\bibitem[{{Heng} \& {Showman}(2015)}]{Heng2015}
{Heng}, K. \& {Showman}, A.~P. 2015, Annual Review of Earth and Planetary
  Sciences, 43, 509

\bibitem[{{Heng} \& {Workman}(2014)}]{Heng2014}
{Heng}, K. \& {Workman}, J. 2014, The Astrophysical Journal Supplement Series,
  213, 27

\bibitem[{{Hoeijmakers} {et~al.}(2019){Hoeijmakers}, {Ehrenreich}, {Kitzmann},
  {Allart}, {Grimm}, {Seidel}, {Wyttenbach}, {Pino}, {Nielsen}, {Fisher},
  {Rimmer}, {Bourrier}, {Cegla}, {Lavie}, {Lovis}, {Patzer}, {Stock}, {Pepe},
  \& {Heng}}]{Hoeijmakers2019}
{Hoeijmakers}, H.~J., {Ehrenreich}, D., {Kitzmann}, D., {et~al.} 2019, \aap,
  627, A165

\bibitem[{{Hoeijmakers} {et~al.}(2018){Hoeijmakers}, {Snellen}, \& {van
  Terwisga}}]{Hoeijmakers2018}
{Hoeijmakers}, H.~J., {Snellen}, I.~A.~G., \& {van Terwisga}, S.~E. 2018, \aap,
  610, A47

\bibitem[{{Hooton} {et~al.}(2021){Hooton}, {Hoyer}, {Kitzmann}, {Morris},
  {Smith}, {Collier Cameron}, {Futyan}, {Maxted}, {Queloz}, {Demory}, {Heng},
  {Lendl}, {Cabrera}, {Csizmadia}, {Deline}, {Parviainen}, {Salmon}, {Sulis},
  {Wilson}, {Bonfanti}, {Brandeker}, {Demangeon}, {Oshagh}, {Persson},
  {Scandariato}, {Alibert}, {Alonso}, {Anglada Escud{\'e}}, {B{\'a}rczy},
  {Barrado}, {Barros}, {Baumjohann}, {Beck}, {Beck}, {Benz}, {Billot},
  {Bonfils}, {Bourrier}, {Broeg}, {Busch}, {Charnoz}, {Davies}, {Deleuil},
  {Delrez}, {Ehrenreich}, {Erikson}, {Farinato}, {Fortier}, {Fossati},
  {Fridlund}, {Gandolfi}, {Gillon}, {G{\"u}del}, {Isaak}, {Jones}, {Kiss},
  {Laskar}, {Lecavelier des Etangs}, {Lovis}, {Luntzer}, {Magrin},
  {Nascimbeni}, {Olofsson}, {Ottensamer}, {Pagano}, {Pall{\'e}}, {Peter},
  {Piotto}, {Pollacco}, {Ragazzoni}, {Rando}, {Ratti}, {Rauer}, {Ribas},
  {Santos}, {S{\'e}gransan}, {Simon}, {Sousa}, {Steller}, {Szab{\'o}},
  {Thomas}, {Udry}, {Ulmer}, {Van Grootel}, \& {Walton}}]{Hooton2021}
{Hooton}, M.~J., {Hoyer}, S., {Kitzmann}, D., {et~al.} 2021, arXiv e-prints,
  arXiv:2109.05031

\bibitem[{{Hooton} {et~al.}(2018){Hooton}, {Watson}, {de Mooij}, {Gibson}, \&
  {Kitzmann}}]{Hooton2018}
{Hooton}, M.~J., {Watson}, C.~A., {de Mooij}, E. J.~W., {Gibson}, N.~P., \&
  {Kitzmann}, D. 2018, The Astrophysical Journal, 869, L25

\bibitem[{{Hoyer} {et~al.}(2020){Hoyer}, {Guterman}, {Demangeon}, {Sousa},
  {Deleuil}, {Meunier}, \& {Benz}}]{Hoyer2020}
{Hoyer}, S., {Guterman}, P., {Demangeon}, O., {et~al.} 2020, \aap, 635, A24

\bibitem[{{Hunter}(2007)}]{matplotlib}
{Hunter}, J.~D. 2007, Computing in Science and Engineering, 9, 90

\bibitem[{{Husser} {et~al.}(2013){Husser}, {Wende-von Berg}, {Dreizler},
  {Homeier}, {Reiners}, {Barman}, \& {Hauschildt}}]{Husser2013}
{Husser}, T.-O., {Wende-von Berg}, S., {Dreizler}, S., {et~al.} 2013, \aap,
  553, A6

\bibitem[{{Ingalls} {et~al.}(2016){Ingalls}, {Krick}, {Carey}, {Stauffer},
  {Lowrance}, {Grillmair}, {Buzasi}, {Deming}, {Diamond-Lowe}, {Evans},
  {Morello}, {Stevenson}, {Wong}, {Capak}, {Glaccum}, {Laine}, {Surace}, \&
  {Storrie-Lombardi}}]{Ingalls2016}
{Ingalls}, J.~G., {Krick}, J.~E., {Carey}, S.~J., {et~al.} 2016, The
  Astronomical Journal, 152, 44

\bibitem[{{Jenkins} {et~al.}(2016){Jenkins}, {Twicken}, {McCauliff},
  {Campbell}, {Sanderfer}, {Lung}, {Mansouri-Samani}, {Girouard}, {Tenenbaum},
  {Klaus}, {Smith}, {Caldwell}, {Chacon}, {Henze}, {Heiges}, {Latham},
  {Morgan}, {Swade}, {Rinehart}, \& {Vanderspek}}]{Jenkins2016}
{Jenkins}, J.~M., {Twicken}, J.~D., {McCauliff}, S., {et~al.} 2016, in Society
  of Photo-Optical Instrumentation Engineers (SPIE) Conference Series, Vol.
  9913, Software and Cyberinfrastructure for Astronomy IV, ed. G.~{Chiozzi} \&
  J.~C. {Guzman}, 99133E

\bibitem[{{Kataria} {et~al.}(2013){Kataria}, {Showman}, {Lewis}, {Fortney},
  {Marley}, \& {Freedman}}]{Kataria2013}
{Kataria}, T., {Showman}, A.~P., {Lewis}, N.~K., {et~al.} 2013, The
  Astrophysical Journal, 767, 76

\bibitem[{{Keating} {et~al.}(2019){Keating}, {Cowan}, \& {Dang}}]{Keating2019}
{Keating}, D., {Cowan}, N.~B., \& {Dang}, L. 2019, Nature Astronomy, 3, 1092

\bibitem[{{Kilpatrick} {et~al.}(2020){Kilpatrick}, {Kataria}, {Lewis},
  {Zellem}, {Henry}, {Cowan}, {de Wit}, {Fortney}, {Knutson}, {Seager},
  {Showman}, \& {Tucker}}]{Kilpatrick2020}
{Kilpatrick}, B.~M., {Kataria}, T., {Lewis}, N.~K., {et~al.} 2020, The
  Astronomical Journal, 159, 51

\bibitem[{{Kipping}(2013)}]{Kipping2013}
{Kipping}, D.~M. 2013, \mnras, 435, 2152

\bibitem[{{Kitzmann} {et~al.}(2018){Kitzmann}, {Heng}, {Rimmer}, {Hoeijmakers},
  {Tsai}, {Malik}, {Lendl}, {Deitrick}, \& {Demory}}]{Kitzmann2018b}
{Kitzmann}, D., {Heng}, K., {Rimmer}, P.~B., {et~al.} 2018, The Astrophysical
  Journal, 863, 183

\bibitem[{{Komacek} \& {Showman}(2016)}]{Komacek2016}
{Komacek}, T.~D. \& {Showman}, A.~P. 2016, The Astrophysical Journal, 821, 16

\bibitem[{{Komacek} \& {Showman}(2020)}]{Komacek2020}
{Komacek}, T.~D. \& {Showman}, A.~P. 2020, The Astrophysical Journal, 888, 2

\bibitem[{{Kreidberg}(2015)}]{Kreidberg2015}
{Kreidberg}, L. 2015, \pasp, 127, 1161

\bibitem[{{Kreidberg} {et~al.}(2018){Kreidberg}, {Line}, {Parmentier},
  {Stevenson}, {Louden}, {Bonnefoy}, {Faherty}, {Henry}, {Williamson},
  {Stassun}, {Beatty}, {Bean}, {Fortney}, {Showman}, {D{\'e}sert}, \&
  {Arcangeli}}]{Kreidberg2018}
{Kreidberg}, L., {Line}, M.~R., {Parmentier}, V., {et~al.} 2018, The
  Astronomical Journal, 156, 17

\bibitem[{{Lally} \& {Vanderburg}(2022)}]{Lally2022}
{Lally}, M. \& {Vanderburg}, A. 2022, The Astronomical Journal, 163, 181

\bibitem[{{Lanotte} {et~al.}(2014){Lanotte}, {Gillon}, {Demory}, {Fortney},
  {Astudillo}, {Bonfils}, {Magain}, {Delfosse}, {Forveille}, {Lovis}, {Mayor},
  {Neves}, {Pepe}, {Queloz}, {Santos}, \& {Udry}}]{Lanotte2014}
{Lanotte}, A.~A., {Gillon}, M., {Demory}, B.~O., {et~al.} 2014, Astronomy and
  Astrophysics, 572, A73

\bibitem[{{Lendl} {et~al.}(2020){Lendl}, {Csizmadia}, {Deline}, {Fossati},
  {Kitzmann}, {Heng}, {Hoyer}, {Salmon}, {Benz}, {Broeg}, {Ehrenreich},
  {Fortier}, {Queloz}, {Bonfanti}, {Brandeker}, {Collier Cameron}, {Delrez},
  {Garcia Mu{\~n}oz}, {Hooton}, {Maxted}, {Morris}, {Van Grootel}, {Wilson},
  {Alibert}, {Alonso}, {Asquier}, {Bandy}, {B{\'a}rczy}, {Barrado}, {Barros},
  {Baumjohann}, {Beck}, {Beck}, {Bekkelien}, {Bergomi}, {Billot}, {Biondi},
  {Bonfils}, {Bourrier}, {Busch}, {Cabrera}, {Cessa}, {Charnoz}, {Chazelas},
  {Corral Van Damme}, {Davies}, {Deleuil}, {Demangeon}, {Demory}, {Erikson},
  {Farinato}, {Fridlund}, {Futyan}, {Gandolfi}, {Gillon}, {Guterman}, {Hasiba},
  {Hernandez}, {Isaak}, {Kiss}, {Kuntzer}, {Lecavelier des Etangs},
  {L{\"u}ftinger}, {Laskar}, {Lovis}, {Magrin}, {Malvasio}, {Marafatto},
  {Michaelis}, {Munari}, {Nascimbeni}, {Olofsson}, {Ottacher}, {Ottensamer},
  {Pagano}, {Pall{\'e}}, {Peter}, {Piazza}, {Piotto}, {Pollacco}, {Ratti},
  {Rauer}, {Ragazzoni}, {Rando}, {Ribas}, {Rieder}, {Rohlfs}, {Safa}, {Santos},
  {Scandariato}, {S{\'e}gransan}, {Simon}, {Singh}, {Smith}, {Sordet}, {Sousa},
  {Steller}, {Szab{\'o}}, {Thomas}, {Tschentscher}, {Udry}, {Viotto}, {Walter},
  {Walton}, {Wildi}, \& {Wolter}}]{Lendl2020}
{Lendl}, M., {Csizmadia}, S., {Deline}, A., {et~al.} 2020, Astronomy and
  Astrophysics, 643, A94

\bibitem[{{Lightkurve Collaboration} {et~al.}(2018){Lightkurve Collaboration},
  {Cardoso}, {Hedges}, {Gully-Santiago}, {Saunders}, {Cody}, {Barclay}, {Hall},
  {Sagear}, {Turtelboom}, {Zhang}, {Tzanidakis}, {Mighell}, {Coughlin}, {Bell},
  {Berta-Thompson}, {Williams}, {Dotson}, \& {Barentsen}}]{lightkurve}
{Lightkurve Collaboration}, {Cardoso}, J. V. d. M.~a., {Hedges}, C., {et~al.}
  2018, {Lightkurve: Kepler and TESS time series analysis in Python}

\bibitem[{{Lindegren} {et~al.}(2021){Lindegren}, {Bastian}, {Biermann},
  {Bombrun}, {de Torres}, {Gerlach}, {Geyer}, {Hern{\'a}ndez}, {Hilger},
  {Hobbs}, {Klioner}, {Lammers}, {McMillan}, {Ramos-Lerate},
  {Steidelm{\"u}ller}, {Stephenson}, \& {van Leeuwen}}]{Lindegren2021}
{Lindegren}, L., {Bastian}, U., {Biermann}, M., {et~al.} 2021, \aap, 649, A4

\bibitem[{{Liu} \& {Showman}(2013)}]{Liu2013}
{Liu}, B. \& {Showman}, A.~P. 2013, The Astrophysical Journal, 770, 42

\bibitem[{{Lothringer} {et~al.}(2018){Lothringer}, {Barman}, \&
  {Koskinen}}]{Lothringer2018}
{Lothringer}, J.~D., {Barman}, T., \& {Koskinen}, T. 2018, The Astrophysical
  Journal, 866, 27

\bibitem[{{Mansfield} {et~al.}(2020){Mansfield}, {Bean}, {Stevenson},
  {Komacek}, {Bell}, {Tan}, {Malik}, {Beatty}, {Wong}, {Cowan}, {Dang},
  {D{\'e}sert}, {Fortney}, {Gaudi}, {Keating}, {Kempton}, {Kreidberg}, {Line},
  {Parmentier}, {Stassun}, {Swain}, \& {Zellem}}]{Mansfield2020}
{Mansfield}, M., {Bean}, J.~L., {Stevenson}, K.~B., {et~al.} 2020, The
  Astrophysical Journal, 888, L15

\bibitem[{{Marigo} {et~al.}(2017){Marigo}, {Girardi}, {Bressan}, {Rosenfield},
  {Aringer}, {Chen}, {Dussin}, {Nanni}, {Pastorelli}, {Rodrigues}, {Trabucchi},
  {Bladh}, {Dalcanton}, {Groenewegen}, {Montalb{\'a}n}, \& {Wood}}]{marigo17}
{Marigo}, P., {Girardi}, L., {Bressan}, A., {et~al.} 2017, \apj, 835, 77

\bibitem[{{Mayne} {et~al.}(2014){Mayne}, {Baraffe}, {Acreman}, {Smith},
  {Browning}, {Sk{\r{a}}lid Amundsen}, {Wood}, {Thuburn}, \&
  {Jackson}}]{Mayne2014}
{Mayne}, N.~J., {Baraffe}, I., {Acreman}, D.~M., {et~al.} 2014, Astronomy and
  Astrophysics, 561, A1

\bibitem[{{Mendon{\c{c}}a} {et~al.}(2018){Mendon{\c{c}}a}, {Malik}, {Demory},
  \& {Heng}}]{Mendonca2018}
{Mendon{\c{c}}a}, J.~M., {Malik}, M., {Demory}, B.-O., \& {Heng}, K. 2018, The
  Astronomical Journal, 155, 150

\bibitem[{{Morris} {et~al.}(2021{\natexlab{a}}){Morris}, {Delrez}, {Brandeker},
  {Cameron}, {Simon}, {Futyan}, {Olofsson}, {Hoyer}, {Fortier}, {Demory},
  {Lendl}, {Wilson}, {Oshagh}, {Heng}, {Ehrenreich}, {Sulis}, {Alibert},
  {Alonso}, {Anglada Escud{\'e}}, {Barrado}, {Barros}, {Baumjohann}, {Beck},
  {Beck}, {Bekkelien}, {Benz}, {Bergomi}, {Billot}, {Bonfils}, {Bourrier},
  {Broeg}, {B{\'a}rczy}, {Cabrera}, {Charnoz}, {Davies}, {Ferreras}, {Deleuil},
  {Deline}, {Demangeon}, {Erikson}, {Floren}, {Fossati}, {Fridlund},
  {Gandolfi}, {Garc{\'\i}a Mu{\~n}oz}, {Gillon}, {Guedel}, {Guterman}, {Isaak},
  {Kiss}, {Laskar}, {Lecavelier des Etangs}, {Lieder}, {Lovis}, {Magrin},
  {Maxted}, {Nascimbeni}, {Ottensamer}, {Pagano}, {Pall{\'e}}, {Peter},
  {Piotto}, {Pizarro Rubio}, {Pollacco}, {Pozuelos}, {Queloz}, {Ragazzoni},
  {Rando}, {Rauer}, {Ribas}, {Santos}, {Scandariato}, {Smith}, {Sousa},
  {Steller}, {Szab{\'o}}, {S{\'e}gransan}, {Thomas}, {Udry}, {Ulmer}, {Van
  Grootel}, \& {Walton}}]{Morris2021a}
{Morris}, B.~M., {Delrez}, L., {Brandeker}, A., {et~al.} 2021{\natexlab{a}},
  arXiv e-prints, arXiv:2106.07443

\bibitem[{{Morris} {et~al.}(2021{\natexlab{b}}){Morris}, {Heng}, {Jones},
  {Piaulet}, {Demory}, {Kitzmann}, \& {Hoeijmakers}}]{Morris2021b}
{Morris}, B.~M., {Heng}, K., {Jones}, K., {et~al.} 2021{\natexlab{b}}, arXiv
  e-prints, arXiv:2110.11837

\bibitem[{{Owens} {et~al.}(2021){Owens}, {de Mooij}, {Watson}, \&
  {Hooton}}]{Owens2021}
{Owens}, N., {de Mooij}, E.~J.~W., {Watson}, C.~A., \& {Hooton}, M.~J. 2021,
  Monthly Notices of the Royal Astronomical Society, 503, L38

\bibitem[{{Parmentier} {et~al.}(2018){Parmentier}, {Line}, {Bean}, {Mansfield},
  {Kreidberg}, {Lupu}, {Visscher}, {D{\'e}sert}, {Fortney}, {Deleuil},
  {Arcangeli}, {Showman}, \& {Marley}}]{Parmentier2018}
{Parmentier}, V., {Line}, M.~R., {Bean}, J.~L., {et~al.} 2018, Astronomy and
  Astrophysics, 617, A110

\bibitem[{{Parviainen}(2015)}]{Parviainen2015}
{Parviainen}, H. 2015, Monthly Notices of the Royal Astronomical Society, 450,
  3233

\bibitem[{Peix\'oto \& Oort(1984)}]{Peixoto1984}
Peix\'oto, J.~P. \& Oort, A.~H. 1984, Rev. Mod. Phys., 56, 365

\bibitem[{P\'erez \& Granger(2007)}]{ipython}
P\'erez, F. \& Granger, B.~E. 2007, Computing in Science and Engineering, 9, 21

\bibitem[{{Pino} {et~al.}(2020){Pino}, {D{\'e}sert}, {Brogi}, {Malavolta},
  {Wyttenbach}, {Line}, {Hoeijmakers}, {Fossati}, {Bonomo}, {Nascimbeni},
  {Panwar}, {Affer}, {Benatti}, {Biazzo}, {Bignamini}, {Borsa}, {Carleo},
  {Claudi}, {Cosentino}, {Covino}, {Damasso}, {Desidera}, {Giacobbe},
  {Harutyunyan}, {Lanza}, {Leto}, {Maggio}, {Maldonado}, {Mancini}, {Micela},
  {Molinari}, {Pagano}, {Piotto}, {Poretti}, {Rainer}, {Scandariato},
  {Sozzetti}, {Allart}, {Borsato}, {Bruno}, {Di Fabrizio}, {Ehrenreich},
  {Fiorenzano}, {Frustagli}, {Lavie}, {Lovis}, {Magazz{\`u}}, {Nardiello},
  {Pedani}, \& {Smareglia}}]{Pino2020}
{Pino}, L., {D{\'e}sert}, J.-M., {Brogi}, M., {et~al.} 2020, The Astrophysical
  Journal, 894, L27

\bibitem[{{Rauscher} \& {Menou}(2010)}]{Rauscher2010}
{Rauscher}, E. \& {Menou}, K. 2010, The Astrophysical Journal, 714, 1334

\bibitem[{{Ricker} {et~al.}(2014){Ricker}, {Winn}, {Vanderspek}, {Latham},
  {Bakos}, {Bean}, {Berta-Thompson}, {Brown}, {Buchhave}, {Butler}, {Butler},
  {Chaplin}, {Charbonneau}, {Christensen-Dalsgaard}, {Clampin}, {Deming},
  {Doty}, {De Lee}, {Dressing}, {Dunham}, {Endl}, {Fressin}, {Ge}, {Henning},
  {Holman}, {Howard}, {Ida}, {Jenkins}, {Jernigan}, {Johnson}, {Kaltenegger},
  {Kawai}, {Kjeldsen}, {Laughlin}, {Levine}, {Lin}, {Lissauer}, {MacQueen},
  {Marcy}, {McCullough}, {Morton}, {Narita}, {Paegert}, {Palle}, {Pepe},
  {Pepper}, {Quirrenbach}, {Rinehart}, {Sasselov}, {Sato}, {Seager},
  {Sozzetti}, {Stassun}, {Sullivan}, {Szentgyorgyi}, {Torres}, {Udry}, \&
  {Villasenor}}]{Ricker2014}
{Ricker}, G.~R., {Winn}, J.~N., {Vanderspek}, R., {et~al.} 2014, in \procspie,
  Vol. 9143, Space Telescopes and Instrumentation 2014: Optical, Infrared, and
  Millimeter Wave, 914320

\bibitem[{{Salmon} {et~al.}(2021){Salmon}, {Van Grootel}, {Buldgen}, {Dupret},
  \& {Eggenberger}}]{salmon21}
{Salmon}, S.~J.~A.~J., {Van Grootel}, V., {Buldgen}, G., {Dupret}, M.~A., \&
  {Eggenberger}, P. 2021, \aap, 646, A7

\bibitem[{{Salvatier} {et~al.}(2016){Salvatier}, {Wiecki{\^a}}, \&
  {Fonnesbeck}}]{pymc3}
{Salvatier}, J., {Wiecki{\^a}}, T.~V., \& {Fonnesbeck}, C. 2016, {PyMC3: Python
  probabilistic programming framework}

\bibitem[{{Schanche} {et~al.}(2020){Schanche}, {H{\'e}brard}, {Collier
  Cameron}, {Dalal}, {Smalley}, {Wilson}, {Boisse}, {Bouchy}, {Brown},
  {Demangeon}, {Haswell}, {Hellier}, {Kolb}, {Lopez}, {Maxted}, {Pollacco},
  {West}, \& {Wheatley}}]{Schanche2020}
{Schanche}, N., {H{\'e}brard}, G., {Collier Cameron}, A., {et~al.} 2020,
  \mnras, 499, 428

\bibitem[{{Schwartz} {et~al.}(2017){Schwartz}, {Kashner}, {Jovmir}, \&
  {Cowan}}]{Schwartz2017}
{Schwartz}, J.~C., {Kashner}, Z., {Jovmir}, D., \& {Cowan}, N.~B. 2017, The
  Astrophysical Journal, 850, 154

\bibitem[{{Scuflaire} {et~al.}(2008){Scuflaire}, {Th{\'e}ado}, {Montalb{\'a}n},
  {Miglio}, {Bourge}, {Godart}, {Thoul}, \& {Noels}}]{scuflaire08}
{Scuflaire}, R., {Th{\'e}ado}, S., {Montalb{\'a}n}, J., {et~al.} 2008, \apss,
  316, 83

\bibitem[{{Showman} {et~al.}(2009){Showman}, {Fortney}, {Lian}, {Marley},
  {Freedman}, {Knutson}, \& {Charbonneau}}]{Showman2009}
{Showman}, A.~P., {Fortney}, J.~J., {Lian}, Y., {et~al.} 2009, The
  Astrophysical Journal, 699, 564

\bibitem[{{Skrutskie} {et~al.}(2006){Skrutskie}, {Cutri}, {Stiening},
  {Weinberg}, {Schneider}, {Carpenter}, {Beichman}, {Capps}, {Chester},
  {Elias}, {Huchra}, {Liebert}, {Lonsdale}, {Monet}, {Price}, {Seitzer},
  {Jarrett}, {Kirkpatrick}, {Gizis}, {Howard}, {Evans}, {Fowler}, {Fullmer},
  {Hurt}, {Light}, {Kopan}, {Marsh}, {McCallon}, {Tam}, {Van Dyk}, \&
  {Wheelock}}]{Skrutskie2006}
{Skrutskie}, M.~F., {Cutri}, R.~M., {Stiening}, R., {et~al.} 2006, \aj, 131,
  1163

\bibitem[{{Stevenson} {et~al.}(2012){Stevenson}, {Harrington}, {Fortney},
  {Loredo}, {Hardy}, {Nymeyer}, {Bowman}, {Cubillos}, {Bowman}, \&
  {Hardin}}]{Stevenson2012}
{Stevenson}, K.~B., {Harrington}, J., {Fortney}, J.~J., {et~al.} 2012, The
  Astrophysical Journal, 754, 136

\bibitem[{{Sudarsky} {et~al.}(2000){Sudarsky}, {Burrows}, \&
  {Pinto}}]{Sudarsky2000}
{Sudarsky}, D., {Burrows}, A., \& {Pinto}, P. 2000, The Astrophysical Journal,
  538, 885

\bibitem[{{Tan} \& {Komacek}(2019)}]{Tan2019}
{Tan}, X. \& {Komacek}, T.~D. 2019, The Astrophysical Journal, 886, 26

\bibitem[{{Thrastarson} \& {Cho}(2011)}]{Thrastarson2011}
{Thrastarson}, H.~T. \& {Cho}, J. Y.-K. 2011, The Astrophysical Journal, 729,
  117

\bibitem[{{Vehtari} {et~al.}(2015){Vehtari}, {Gelman}, \&
  {Gabry}}]{Vehtari2015}
{Vehtari}, A., {Gelman}, A., \& {Gabry}, J. 2015, arXiv e-prints,
  arXiv:1507.04544

\bibitem[{Virtanen {et~al.}(2020)Virtanen, Gommers, Oliphant, Haberland, Reddy,
  Cournapeau, Burovski, Peterson, Weckesser, Bright, {van der Walt}, Brett,
  Wilson, Millman, Mayorov, Nelson, Jones, Kern, Larson, Carey, Polat, Feng,
  Moore, {VanderPlas}, Laxalde, Perktold, Cimrman, Henriksen, Quintero, Harris,
  Archibald, Ribeiro, Pedregosa, {van Mulbregt}, \& {SciPy 1.0
  Contributors}}]{scipy}
Virtanen, P., Gommers, R., Oliphant, T.~E., {et~al.} 2020, Nature Methods, 17,
  261

\bibitem[{{von Zeipel}(1924)}]{vonZeipel1924}
{von Zeipel}, H. 1924, Monthly Notices of the Royal Astronomical Society, 84,
  665

\bibitem[{{Wallack} {et~al.}(2021){Wallack}, {Knutson}, \&
  {Deming}}]{Wallack2021}
{Wallack}, N.~L., {Knutson}, H.~A., \& {Deming}, D. 2021, arXiv e-prints,
  arXiv:2103.15833

\bibitem[{{Welsh} {et~al.}(2010){Welsh}, {Orosz}, {Seager}, {Fortney},
  {Jenkins}, {Rowe}, {Koch}, \& {Borucki}}]{Welsh2010}
{Welsh}, W.~F., {Orosz}, J.~A., {Seager}, S., {et~al.} 2010, The Astrophysical
  Journal, 713, L145

\bibitem[{{Wong} {et~al.}(2020){Wong}, {Shporer}, {Kitzmann}, {Morris}, {Heng},
  {Hoeijmakers}, {Demory}, {Ahlers}, {Mansfield}, {Bean}, {Daylan},
  {Fetherolf}, {Rodriguez}, {Benneke}, {Ricker}, {Latham}, {Vanderspek},
  {Seager}, {Winn}, {Jenkins}, {Burke}, {Christiansen}, {Essack}, {Rose},
  {Smith}, {Tenenbaum}, \& {Yahalomi}}]{Wong2020}
{Wong}, I., {Shporer}, A., {Kitzmann}, D., {et~al.} 2020, The Astronomical
  Journal, 160, 88

\bibitem[{{Wright} {et~al.}(2010){Wright}, {Eisenhardt}, {Mainzer}, {Ressler},
  {Cutri}, {Jarrett}, {Kirkpatrick}, {Padgett}, {McMillan}, {Skrutskie},
  {Stanford}, {Cohen}, {Walker}, {Mather}, {Leisawitz}, {Gautier}, {McLean},
  {Benford}, {Lonsdale}, {Blain}, {Mendez}, {Irace}, {Duval}, {Liu}, {Royer},
  {Heinrichsen}, {Howard}, {Shannon}, {Kendall}, {Walsh}, {Larsen}, {Cardon},
  {Schick}, {Schwalm}, {Abid}, {Fabinsky}, {Naes}, \& {Tsai}}]{Wright2010}
{Wright}, E.~L., {Eisenhardt}, P. R.~M., {Mainzer}, A.~K., {et~al.} 2010, \aj,
  140, 1868

\bibitem[{{Wyttenbach} {et~al.}(2020){Wyttenbach}, {Molli{\`e}re},
  {Ehrenreich}, {Cegla}, {Bourrier}, {Lovis}, {Pino}, {Allart}, {Seidel},
  {Hoeijmakers}, {Nielsen}, {Lavie}, {Pepe}, {Bonfils}, \&
  {Snellen}}]{Wyttenbach2020}
{Wyttenbach}, A., {Molli{\`e}re}, P., {Ehrenreich}, D., {et~al.} 2020,
  Astronomy and Astrophysics, 638, A87

\bibitem[{{Yan} \& {Henning}(2018)}]{Yan2018}
{Yan}, F. \& {Henning}, T. 2018, Nature Astronomy, 2, 714

\end{thebibliography}

\appendix

\section{Transit best-fits}
\label{app:transit}

\begin{figure*}
    \centering
    \includegraphics[width=\textwidth* 5/6]{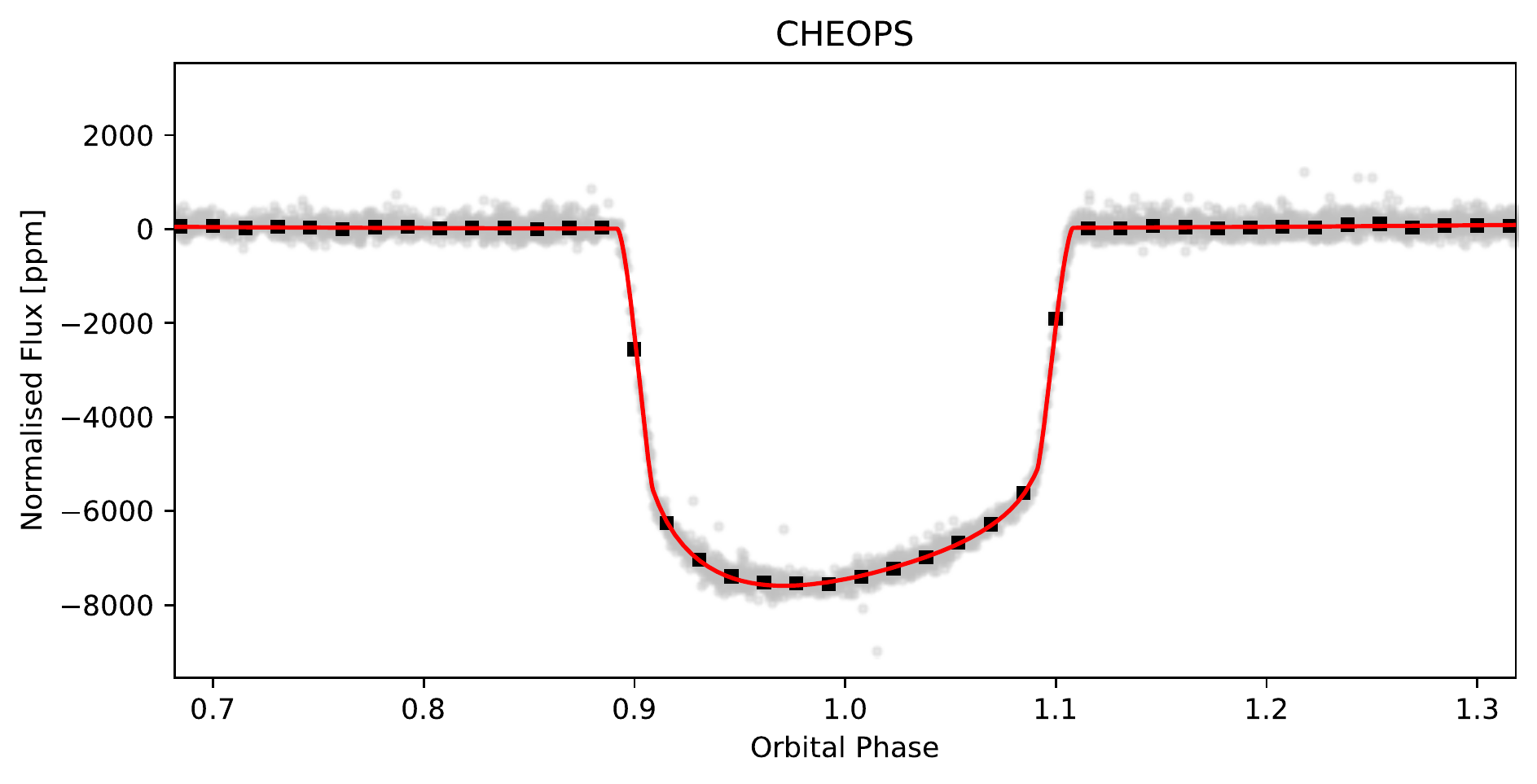}
    \includegraphics[width=\textwidth* 5/6]{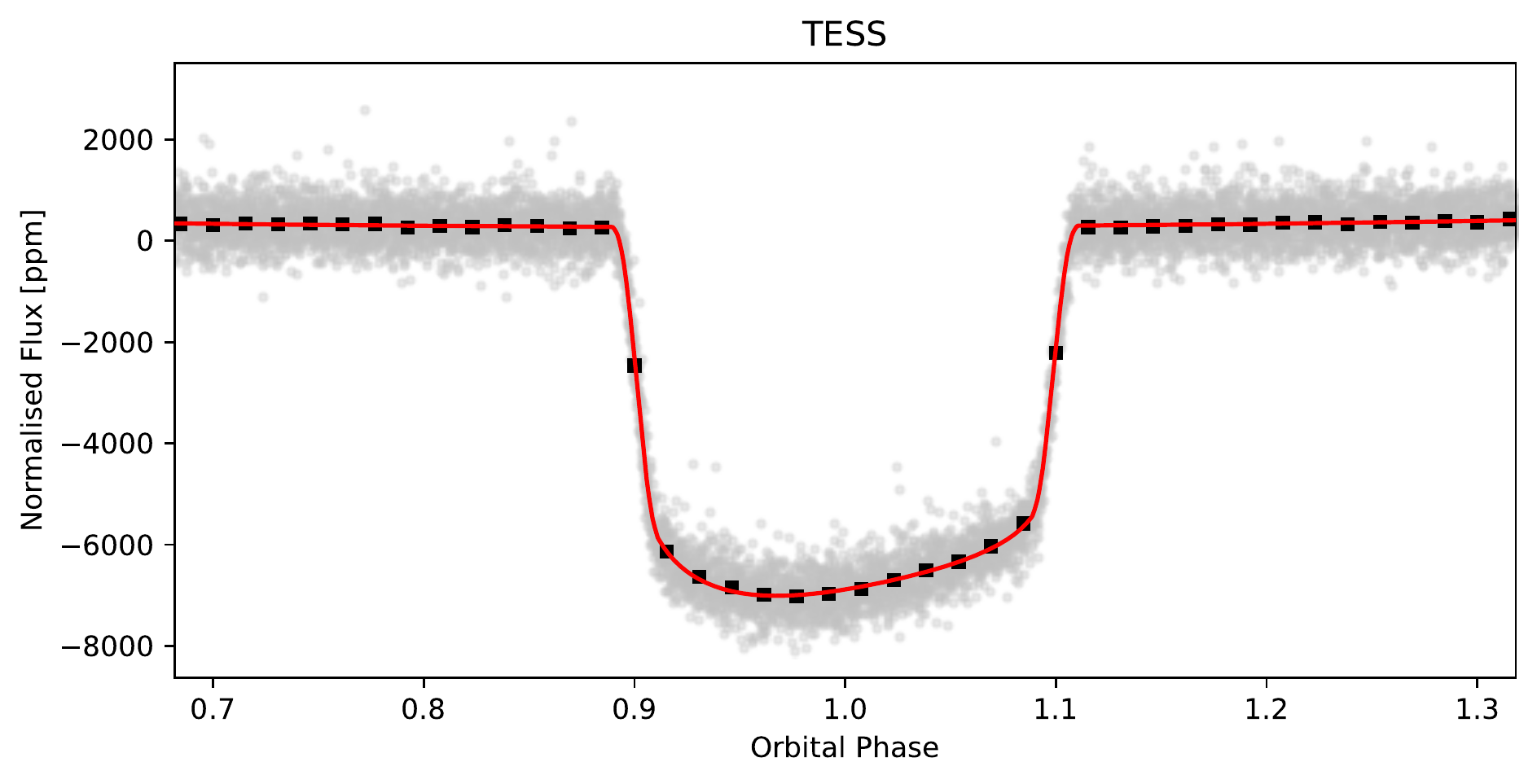}
    \includegraphics[width=\textwidth* 5/6]{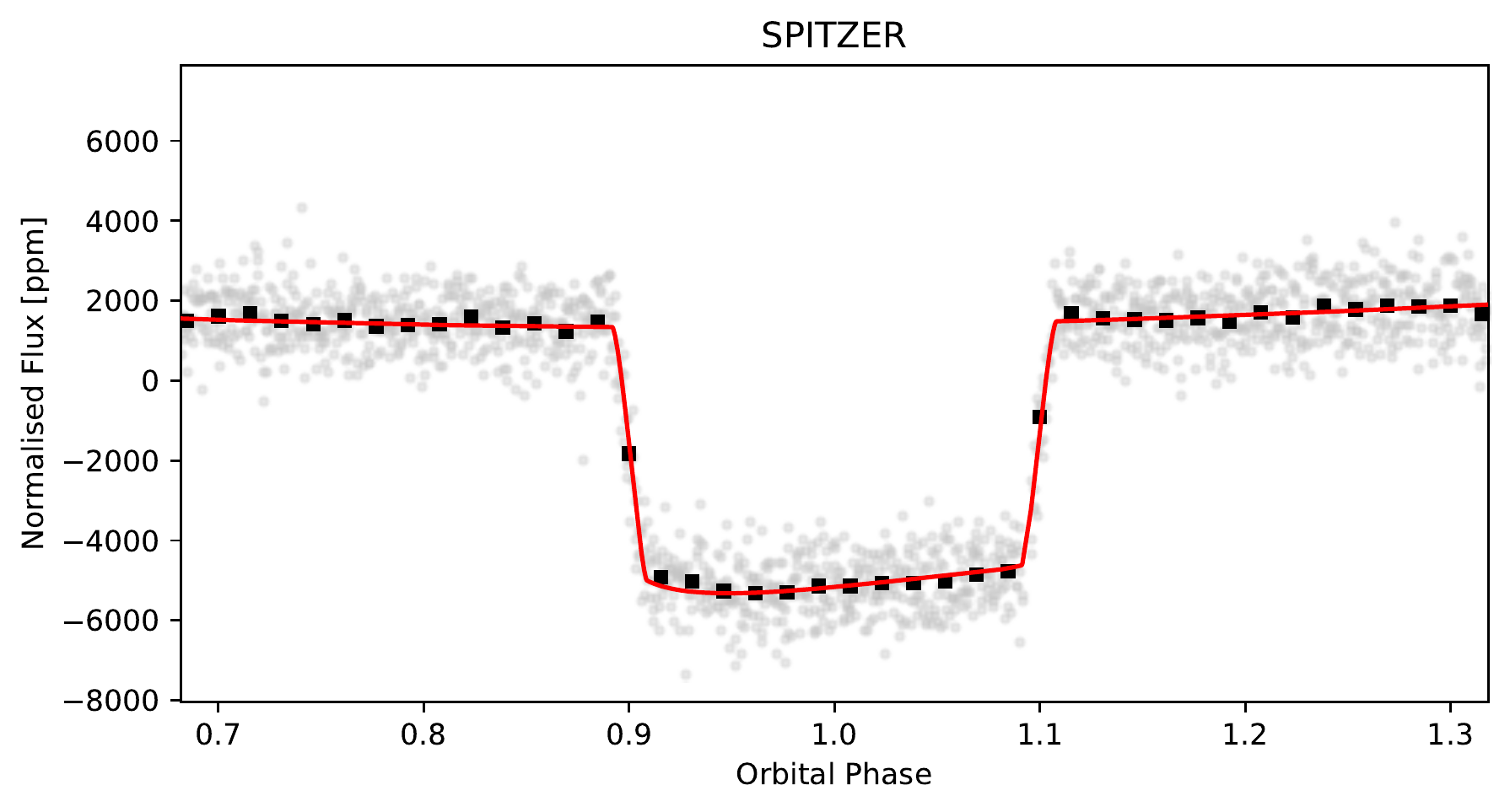}
    \caption{CHEOPS, TESS, and \Spitzer phase-folded and detrended transits overplotted with the best-fit transit model (in red), with all other models and systematics removed. In black are the binned grey datapoints, with error bars that are smaller than the point size in all panels so they are not visible.}
    \label{fig:All_transits}
\end{figure*}

Figure \ref{fig:All_transits} shows the transit models and photometry.

\section{Posterior distributions}
\label{app:corner}

\begin{figure*}
    \centering
    \includegraphics[width=\textwidth]{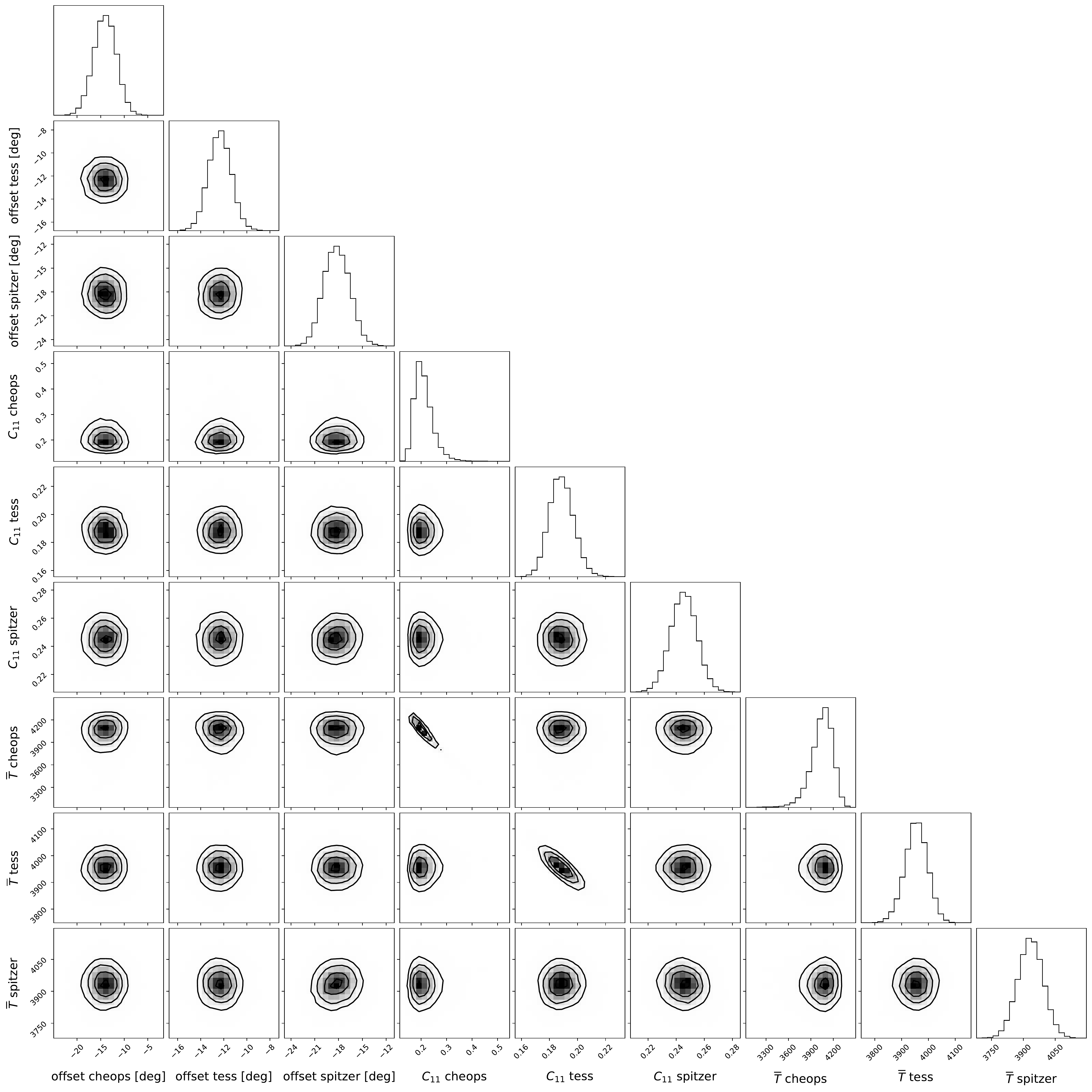}
    \caption{Corner plot showing posteriors of the phase curve parameters in the joint fit. The three contour lines in each subplot refer to the 1-, 2- and 3-sigma contour levels.}
    \label{fig:fullphasecorner}
    
\end{figure*}

Figure \ref{fig:fullphasecorner} shows the posterior distributions for the phase curve parameters.

\end{document}